\documentclass[twocolumn,bbm, trackchanges]{aastex631}

\newcommand{\redshift}{\ensuremath{z_r}}
\newcommand{\rco}{\ensuremath{R_{\mathrm{co}}}}
\newcommand{\Qgas}{\ensuremath{Q_{\mathrm{gas}}}}

\newcommand{\Msat}{\ensuremath{M_{\mathrm{sat}}}}
\newcommand{\Mhost}{\ensuremath{M_{\mathrm{host}}}}
\newcommand{\dhost}{\ensuremath{D_{\mathrm{H}}}}
\newcommand{\rhost}{\ensuremath{R_{\mathrm{H}}}}

\newcommand{\upenn}{Department of Physics \& Astronomy, University of Pennsylvania, 209 S 33rd St., Philadelphia, PA 19104, USA}
\newcommand{\cca}{Center for Computational Astrophysics, Flatiron Institute, 162 5th Ave, 21st Street, New York City, NY 10010, USA}

\shorttitle{Galactic bars in FIRE-2}
\shortauthors{Ansar et al.}

\graphicspath{{./}{figures/}}

\begin{document}

\title{Bar formation and destruction in the FIRE-2 simulations\footnote{Released on March, 1st, 2021}}

\correspondingauthor{Sioree Ansar}
\email{sioreeansar@gmail.com}

\author[0000-0002-3353-6421]{Sioree Ansar}
\affiliation{\cca}
\affiliation{Indian Institute of Astrophysics, 2nd Block, Koramangala, Bangalore, Karnataka 560034, India}
\affiliation{Pondicherry University, R.V. Nagar, Kalapet 605014, Puducherry, India}

\author[0000-0003-0256-5446]{Sarah Pearson}\thanks{Hubble Fellow}
\affiliation{The Center for Cosmology and Particle Physics, New York University, 726 Broadway, New York, NY 10003}
\affiliation{Niels Bohr International Academy \& DARK, Niels Bohr Institute, University of Copenhagen, Blegdamsvej 17, 2100 Copenhagen,  Denmark}

\author[0000-0003-3939-3297]{Robyn E. Sanderson}
\affiliation{\upenn}
\affiliation{\cca}

\author[0000-0002-8354-7356]{Arpit Arora}
\affiliation{\upenn}

\author[0000-0003-3729-1684]{Philip F. Hopkins}
\affiliation{TAPIR, Mailcode 350-17, California Institute of Technology, Pasadena, CA 91125, USA}

\author[0000-0003-0603-8942]{Andrew Wetzel}
\affiliation{Department of Physics \& Astronomy, University of California, Davis, CA 95616, USA}

\author[0000-0002-6993-0826]{Emily C. Cunningham}\thanks{Hubble Fellow}
\affiliation{Department of Astronomy, Columbia University, 550 West 120th Street, New York, NY 10027, USA}
\affiliation{\cca}

\author[0000-0001-6086-9873]{Jamie Quinn}
\affiliation{Department of Physics, University of California, Merced, 5200 N. Lake Road, Merced, CA 95343}

\begin{abstract}
The physical mechanisms responsible for bar formation and destruction in galaxies remain a subject of debate. While we have gained valuable insight into how bars form and evolve from isolated idealized simulations, in the cosmological domain, galactic bars evolve in complex environments with mergers, gas accretion events, in presence of turbulent Inter Stellar Medium (ISM) with multiple star formation episodes, in addition to coupling to their host galaxies’ dark matter halos. We investigate bar formation in 13 Milky Way-mass galaxies from the FIRE-2 (Feedback in Realistic Environments) cosmological zoom-in simulations. 8 of the 13 simulated galaxies form bars at some point during their history: three from tidal interactions and five from internal evolution of the disk. The bars in FIRE-2 are generally shorter than the corotation radius (mean bar radius $\sim 1.53$ kpc), have a wide range of pattern speeds (36--97 km s$^{-1}$kpc$^{-1}$), and live for a wide range of dynamical times (2--160 bar rotations). We find that bar formation in FIRE-2 galaxies is influenced by satellite interactions and the stellar-to-dark matter mass ratio in the inner galaxy, but neither is a sufficient condition for bar formation. Bar formation is more likely to occur, and the bars formed are stronger and longer-lived, if the disks are kinematically cold; galaxies with high central gas fractions and/or vigorous star formation, on the other hand, tend to form weaker bars. In the case of the FIRE-2 galaxies these properties combine to produce ellipsoidal bars with strengths $A_2/A_0 \sim $ 0.1--0.2.
\end{abstract}


\section{Introduction} \label{sec:intro}
Bars are triaxial structures at the centers of galaxy disks that form through global disk instabilities. Stars in barred potentials follow elongated orbits. 
Studies using isolated galaxy simulations have explored different formation mechanisms in controlled simulations with fine-tuned initial conditions \citep{Hohl.1971,Julian.and.Toomre.1966, Ostriker.and.Peebles.1973,Sellwood.and.Wilkinson1993,Sellwood.2014}. These studies have shown that bars can form during isolated evolution of disk galaxies \citep{Gadotti.2011, Sellwood.2014} and during encounters with other galaxies or satellite galaxies \citep{Gerin.et.al.1990,Ghosh.et.al.2021}. Some studies have also shown that bars can be destroyed and re-generated  \citep{Cavanagh.et.al.2022}. Recently, observations of bars using the James Webb Space Telescope (JWST) have generated interest in bar formation at high redshifts ($z_r\sim 1\text{--}4$) \citep{Guo2023, LeConte2024, Guo2024}. Studies have provided insights into the formation of bars in thick disks, which may exhibit similar properties to disk galaxies observed at high redshifts \citep{Klypin2009, Aumer2017, Reddish.et.al.2022, Ghosh2023}.

Detailed studies of the bar in the Milky Way (MW) have played a crucial role in understanding its dynamics, formation and evolution using voluminous new survey data \citep[e.g.][]{2015MNRAS.450.4050W,2019MNRAS.485.3296W, Bovy.et.al.2019,lucey2022}. Other studies estimate the MW bar pattern speed using action space \citep{Trick.et.al.2021, Trick.2022} and transverse velocities \citep{Sanders.et.al.2019}, and \cite{Grion_Filho.2021} have studied high resolution numerical simulations of galaxy interactions (analogous to MW--Sagittarius dwarf galaxy system) that can result in bar formation. Bars are thought to play a major role in star formation \citep[e.g.][]{Aguerri.1999,Masters.et.al.2011,Masters.et.al.2012}, internal evolution of galaxies \citep[e.g.][]{Gadotti.2011,Sellwood.2014}, and stellar and gas dynamics \citep{Sheth.et.al.2005, Khoperskov.et.al.2018, Seo.et.al.2019, Lokas.2020}. Formation of bars during tidal interactions with satellite galaxies have been studied in details using N-body simulations \citep{Lang.et.al.2014, Lokas.et.al.2014, Lokas.et.al.2016, Lokas.2018} and in cosmological simulations \citep{Zana2018a, Zana2018b}. The bar-halo connection has been explored in detail using isolated galaxy evolution simulations \citep[e.g.][]{Athanassoula.2002, Athanassoula.and.Misiriotis.2002}, where bars slow down through the transfer of angular momentum to the dark matter halos and the rest of the disk in the host galaxy due to angular momentum exchange \citep{Debattista.and.Sellwood.1998, Debattista.and.Sellwood.2000, Athanassoula.et.al.2013MNRAS}. Multiple studies have highlighted the connection between dark matter halo angular momentum, disk angular momentum and bar structure \citep{Romeo2023, Ansar2023}.

However, bars in the Universe evolve in a complex environment: their host galaxies undergo satellite interactions, mergers, gas accretion, and star formation while the bars simultaneously interact with the host's dark matter halo. Multiple cosmological simulations have studied bar properties \citep{Roshan.et.al.2021, Fragkoudi.et.al.2021}, their abundance \citep{Rosas-Guevara.et.al.2022}, bar formation pathways \citep{Izquierdo-Villalba.et.al.2022}, bar evolution \citep{Rosas-Guevara.et.al.2022, Fragkoudi2024}. \citep{Irodotou2022} showed that bars grow stronger in the absence of AGN feedback due to formation of disky bulges that led to stronger bars in Auriga simulations \citep{Grand.et.al.2017}. Galaxies also contain a multi-phase, turbulent ISM, which many studies of bar dynamics have not considered.  Past studies have also shown that bars can direct material into the central supermassive black hole in a galaxy, coupling bar formation to black hole feedback \citep{Shlosman.et.al.1989, Hopkins.and.Quataert.2011, Angles-Alcazar.et.al.2021, Gonzalez-Alfonso.et.al.2021}. All the above processes can simultaneously affect bar formation in cosmological simulations, making it difficult to disentangle the contribution of each phenomenon. Even after decades of studies, bar formation in the cosmological context is a subject of debate.

The properties of bars have recently been studied in both large volume simulations like Illustris \citep{Vogelsberger.et.al.2014a, Vogelsberger.et.al.2014b}, IllustrisTNG  \citep{Weinberger.et.al.2017,Pillepich.et.al.2018} and EAGLE  \citep{Schaya.et.al.2015}, and in zoomed simulations like Auriga \citep{Grand.et.al.2017} and NIHAO \citep{Wang.et.al.2015}. While the large boxes can produce a substantial number of galaxies with different morphologies, they are constrained to balance mass and spatial resolution with computational time. Hence their ability to resolve the detailed dynamics of bars is somewhat limited. Cosmological zoom simulations like Auriga \citep{Grand.et.al.2017, Grand2024}, NIHAO \citep{Wang.et.al.2015}, and FIRE-2 (Feedback in Realistic Environments\footnote{\url{http://fire.northwestern.edu}})  \citep{Hopkins.et.al.2018} have moderately-sized samples of galaxies in a narrow galaxy mass range, but can resolve the disk scale height and some of these simulations have a multiphase ISM (FIRE-2 and NIHAO) \citep{Ma.et.al.2017, Garrison-Kimmel.et.al.2018, Gensior.et.al.2023}. Most importantly they contain enough particles in the bar-forming region to resolve bar dynamics \citep{Weinberg.1994, Dubinsky.et.al.2009, Weinberg.and.Katz.2007a, Weinberg.and.Katz.2007b} and the coupling to gas and dark matter. 

The zoomed cosmological simulations thus provide useful tests of how differences in bar properties are related to differences in the implementation of physical processes that affect bar formation, which differ widely between simulation suites. 

In this work, we look at the 13 high resolution Milky Way mass galaxies from the FIRE-2 suite \citep{Hopkins.et.al.2018}. We use the FIRE-2 simulations to determine how bars form, and when and how bar formation fails. We investigate whether the FIRE-2 bars are transient or long-lived and whether the disk properties affect the bar formation process. 

We first describe the FIRE-2 simulations and the sample of galaxies that we use in our study (Section \ref{sec:FIRE-2_sims}). We then present the different methods we use to analyse the sample galaxies in Section \ref{sec:methods}. We present the results in Section \ref{sec:results} on the different mechanisms of bar formation in FIRE-2 galaxies, how bar formation is affected by the host disk and halo properties. We discuss our results in Section \ref{sec:discussion} and in Section \ref{sec:summary} we summarize and discuss the major findings of this study.

\defcitealias{Samuel.et.al.2020}{S20}
\defcitealias{Wetzel.et.al.2016}{W16}
\defcitealias{Garrison-Kimmel.et.al.2019}{G19a}
\defcitealias{Hopkins.et.al.2018}{H18}
\defcitealias{Garrison-Kimmel.et.al.2017}{G17}
\defcitealias{Garrison-Kimmel.et.al.2019b}{G19b}

\begin{deluxetable*}{ccccccc}
\tablenum{1} \label{table:FIRE-2_galaxy_summary}
\tablecaption{Summary of $\redshift\sim 0$ properties of the FIRE-2 Milky Way-mass galaxies used in this study. }
\tablewidth{0pt}
\tablehead{
\colhead{Simulation} & \colhead{$\rm M_{\star, 90} (M_{\odot})$} & \colhead{$\rm M_{200} (M_{\odot})$} & \colhead{$\rm R_{90} (kpc)$} & \colhead{$\rm V_{max} (km s^{-1})$} & \colhead{$\rm R_{max} (kpc)$} & \colhead{Reference}
}
\decimalcolnumbers
\startdata
m12r &  $1.3\times10^{10}$  &  $1.10\times10^{12}$ &  10.4  &  147  &  9.8  &  \citepalias{Samuel.et.al.2020}  \\
Louise & $2.3\times10^{10}$ & $1.15\times10^{12}$ & 11.2 & 182 & 9.8 &  \citepalias{Garrison-Kimmel.et.al.2019}  \\
Juliet & $3.3\times10^{10}$  &  $1.10\times10^{12}$ &  8.1  & 209 & 1.4 & \citepalias{Garrison-Kimmel.et.al.2019} \\
Remus & $4.0\times10^{10}$ & $1.22\times10^{12}$  & 11.0 & 298 & 1.5  & \citepalias{Garrison-Kimmel.et.al.2019b} \\
m12w & $4.8\times10^{10}$ & $1.08\times10^{12}$ &  7.3  &  247  &  3.2  & \citepalias{Samuel.et.al.2020} \\
m12c & $5.1\times10^{10}$ &  $1.35\times10^{12}$ &  9.1  &  241 & 3.2 & \citepalias{Garrison-Kimmel.et.al.2019} \\
m12i &  $5.5\times10^{10} $  &  $1.18\times10^{12}$ & 8.5 &  265  &  2.0  &  \citepalias{Wetzel.et.al.2016}   \\
Romeo & $5.9\times10^{10}$ & $1.32\times10^{12}$ & 12.4 & 255 & 3.4 & \citepalias{Garrison-Kimmel.et.al.2019} \\
Thelma & $6.3\times10^{10}$ & $1.43\times10^{12}$ & 11.2 & 237 & 6.6 & \citepalias{Garrison-Kimmel.et.al.2019} \\
m12f & $6.9\times10^{10}$ & $1.71\times10^{12}$ & 11.8 &  276  & 2.2 & \citepalias{Garrison-Kimmel.et.al.2017}   \\
m12b & $7.3\times10^{10}$ & $1.43\times10^{12}$ &  9.0  &  316  & 1.8 & \citepalias{Garrison-Kimmel.et.al.2019} \\
Romulus & $8.0\times10^{10}$ &  $2.08\times10^{12}$  & 12.9 & 299 & 1.5  & \citepalias{Garrison-Kimmel.et.al.2019b} \\
m12m & $1.0\times10^{11} $ & $1.58\times10^{12}$ & 11.6 &  284  & 7.1 & \citepalias{Hopkins.et.al.2018}   \\
\enddata
\tablecomments{Columns: 1. Simulation name; 2. $\rm M_{\star,90}:$ Stellar mass within a cylindrical radius ($R=\sqrt{x^2 +y^2}$) that encloses 90\% of the stellar mass within $R=20$ kpc and $|z|<2$ kpc \citep{Santistevan.et.al.2020}; 3. $\rm M_{200}:$ total mass of the dark matter halo within the spherical radius $ r=\rm R_{200}$ ($r=\sqrt{x^2 +y^2 + z^2}$) inside which the mean density is 200 $\times$ the matter density of the Universe; 4. $\rm R_{90}:$ radius that encloses 90\% of the stellar mass within $R=20$ kpc and $|z|<2$ kpc \citep{Santistevan.et.al.2020}; 5. $\rm V_{max}:$ maximum circular velocity; 6. $\rm R_{max}:$ spherical radius corresponding to the maximum circular velocity; 7. References: 
\citealp[][(S20)]{Samuel.et.al.2020}; \citealp[][(W16)]{Wetzel.et.al.2016}; \citealp[][(G19a)]{Garrison-Kimmel.et.al.2019}; \citealp[][(H18)]{Hopkins.et.al.2018};\citealp[][(G19)]{Garrison-Kimmel.et.al.2017}; \citealp[][(G19b)]{Garrison-Kimmel.et.al.2019b}. For more details of the individual simulations, see \cite{Wetzel.et.al.2022}. }
\end{deluxetable*}

\section{The FIRE-2 Simulations} \label{sec:FIRE-2_sims}
We use the set of 13 Milky Way-mass galaxies in the FIRE-2 suite, all of which are run with the \textsf{GIZMO} code \citep{Hopkins.2015b, Hopkins.2017}. \textsf{GIZMO} implements a gravity solver (TREE+PM) with a Lagrangian meshless, finite-mass method for hydrodynamics with the FIRE-2 physics model \citep{Hopkins.et.al.2018}. In this prescription stellar feedback is determined and injected locally by following the evolution of the individual mono-age, mono-abundance stellar populations represented by each star particle using the stellar evolution models from \textsf{STARBURST99} \citep{Leitherer.et.al.1999} and the \cite{Kroupa.2001} initial mass function (IMF). The simulations evolve from a redshift $\redshift \sim 100$ to $\redshift = 0$. The galaxies form hierarchically, individually transitioning from an early phase of multiple interactions and large starburst events to well settled disks  \citep{Ma.et.al.2017, Yu.et.al.2021, Gurvich.et.al.2022,Hopkins.et.al.2023,Gensior.et.al.2023, McCluskey.et.al.2023} with a variety of star formation histories. 

FIRE-2 simulations account for multiple physical phenomena to implement stellar feedback: radiative heating and cooling with free-free photoionization/recombination, photoelectric, Compton effect, dust collision and cosmic rays). FIRE-2 considers molecular, metal-line and fine structure processes by tracking 11 different species separately. Star formation is implemented in self-gravitating, Jeans-unstable molecular gas that is self-shielding (following \citealt{Hopkins.et.al.2013}) and has number density n $> 1000$ $\rm cm^{-3}$. The FIRE simulations are able to produce disk galaxies with masses, scale radii, and scale heights that are comparable to observed Milky Way mass galaxies  \citep{Ma.et.al.2017,Garrison-Kimmel.et.al.2018, Sanderson.et.al.2020, Yu.et.al.2021, Gurvich.et.al.2022,Gensior.et.al.2023} and also realistic Giant Molecular Cloud populations \citep{Guszejnov.et.al.2020, Benincasa.et.al.2020}. Importantly for this study, the kinematic ``coldness'' of the stellar and gas disks of FIRE-2 galaxies have recently been shown by \citep{McCluskey.et.al.2023} to be consistent with observed galaxies, agreeing well with measurements of M31, M33 and galaxies from the PHANGS survey \citep{Sun2020, Pessa2023}, and that the MW is somewhat kinematically cold relative to this population.
The FIRE-2 model does not include feedback from black hole accretion.

The merger history of halos provides important information about the satellite interactions that can have a high tidal impact on the disk and lead to bar formation. The \textsf{Rockstar} code \citep{Behroozi.et.al.2013a} is used on each of the simulation runs to generate halo catalogues. To study the evolution of the identified halos in each of the simulation snapshots the code \textsf{Consistent-trees} \citep{Behroozi.et.al.2013b} was used to link the subhalos over time for each of the snapshots to from a merger tree. The stars inside the virial radius of each of the dark matter halos, having velocities less than twice the halo circular velocity in the simulations are identified and linked with the respective halos. 

In this work, we specifically study 13 galaxies in FIRE-2 (see Table \ref{table:FIRE-2_galaxy_summary}) having present day mass similar to the Milky Way, $\rm M_{200}= 1 \text{--} 2 \times 10^{12}$ $\rm M_{\odot}$, 7 of which are isolated hosts (no large neighbor within 10 Mpc) and 6 that are in systems analogous to the MW--M31 pair. The initial mass of gas particles is $7100$ $\rm M_{\odot}$ in the isolated systems and $3500$ $\rm M_{\odot}$ in the paired systems; star particles inherit the gas particle mass, which then decreases due to stellar evolution, leading to a typical star particle mass of 2000--5000 $\rm M_{\odot}$ at late times. The stellar and dark matter softening lengths are fixed based on the typical inter-particle spacing and the gas particles have an adaptive softening length (minimum of 1 pc) \citep{Hopkins.et.al.2018}. The simulation snapshots are saved at a frame rate of $\Delta t \sim 25$ Myrs between snapshots for most of the simulation and with $\Delta t = 2.2$ Myr for the last 22 Myr before $\redshift=0$.  We use \textsf{gizmo\_analysis} \citep{2020ascl.soft02014Wetzel} and \textsf{halo\_analysis} \citep{2020ascl.soft02015Wetzel} for post-processing of the simulation data.

We investigate whether the FIRE-2 galaxies form bars from about $\redshift\sim 2$ onward, after the galaxies have settled into a disk (i.e. when the majority of disk particles have more kinetic energy in circular motion compared to random motion; see \citealt{Garrison-Kimmel.et.al.2018}). One of the galaxies in this simulation suite (\textsf{m12m}) has been studied in detail by \cite{Debattista.et.al.2019}, where the authors have found a bar that transforms into an X-shaped bulge at the end of evolution (more detail in Section \ref{sec:fire_bar_strengths}).

In the early phase of evolution, galaxy interactions are frequent and the host galaxy is mostly dispersion dominated. Each of the disks becomes rotation dominated at a slightly different epoch, as studied in detail by \citet{McCluskey.et.al.2023,Hopkins.et.al.2023,Gruvich.et.al.2022}. In general, the FIRE-2 galaxies are slightly more dispersion-dominated than MW estimates at $\redshift=0$, but not more so than the general population of MW-mass galaxies \citep{Garrison-Kimmel.et.al.2018,McCluskey.et.al.2023}. 

\begin{figure}
	\includegraphics[width=\linewidth]{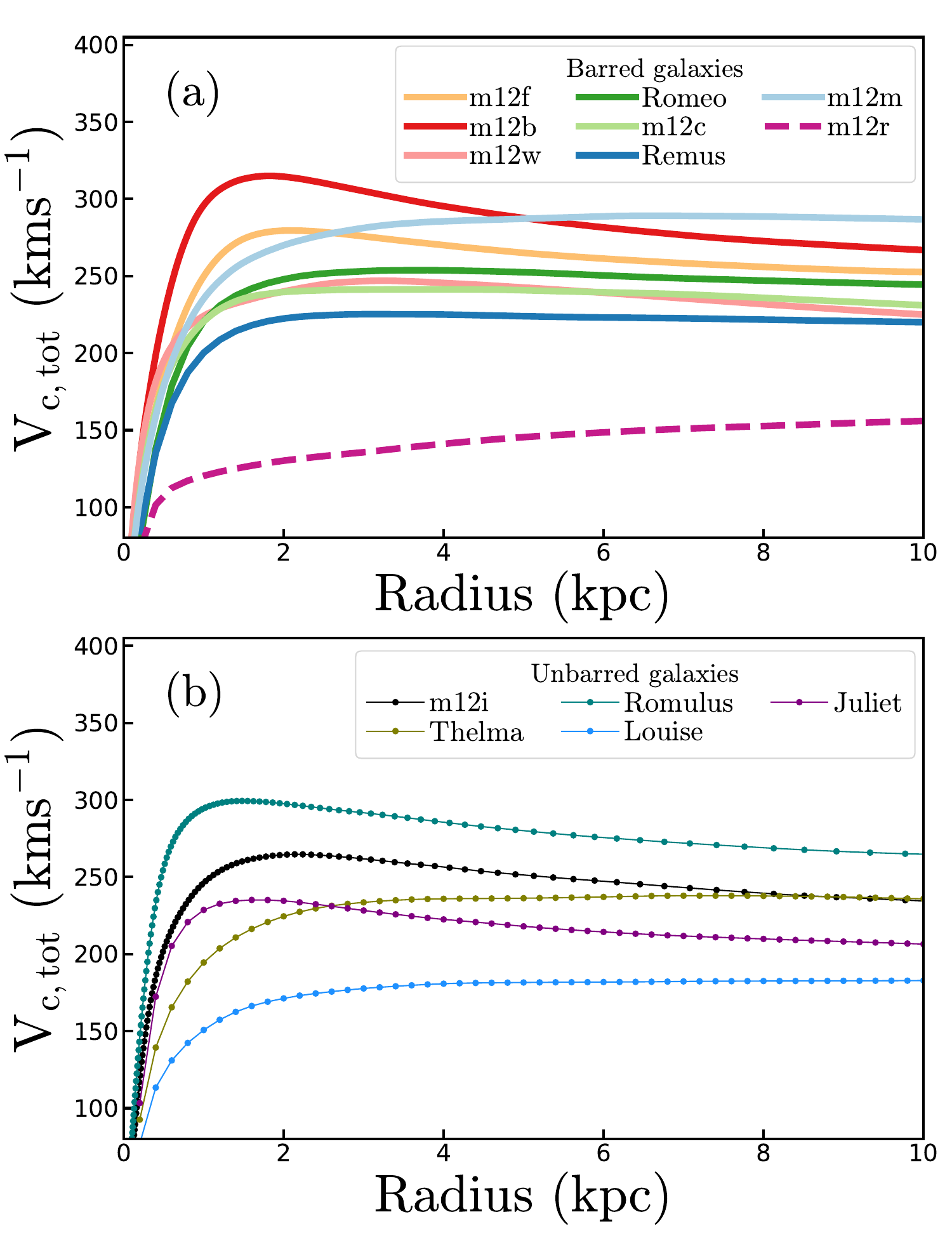}
    \caption{ {\bf Circular velocity of all the FIRE-2 galaxies in Table \ref{table:FIRE-2_galaxy_summary} at $\redshift=0$}. {\bf Panel a:} The coloured thick lines show the galaxies that form bars (some of them do not survive till $\redshift=0$), and the dashed thick line is for \textsf{m12r} which has low mass compared to the rest of the galaxies in this sample. {\bf Panel b:} The coloured thin lines with circular marks show the unbarred galaxies. We use the above choice of color and line style for the simulations throughout the article. 
    }
    \label{fig:all_rotation_curves}
\end{figure}

In Figure \ref{fig:all_rotation_curves} we present the total circular velocity curves of the galaxies in our sample within a radius of 10 kpc at $\redshift=0$. Most galaxies in the sample have similar mass distributions in the central region, except \textsf{m12r} and \textsf{Louise} which have significantly smaller total stellar mass than the others. The circular velocity $V_{c, i}(r)$ at radius $r$ is defined by taking into account the mass of the $i^{th}$ component (stars, gas or dark matter) in the galaxy:
\begin{equation}
\label{eq:circular_velocity}
    V_{c, i}(r)=\sqrt{\frac{G M_{i}(r)}{r} } \textbf{ .}
\end{equation} 
The total circular velocity curve is due to the combined mass of all components.  The maximum rotation velocity lies in the range $140<V_{max}/(\rm km s^{-1})<320$ and the corresponding radius range is $1.4<R_{max}/(\rm kpc)<9.8$ (see Table \ref{table:FIRE-2_galaxy_summary}).

\section{Detecting and Characterizing Bars in FIRE-2} \label{sec:methods}
In this Section, we first describe how we determine whether the galaxies in the FIRE-2 simulations from Table \ref{table:FIRE-2_galaxy_summary} develop bars during any time of evolution. We then estimate the different bar characteristics for all the galaxies and present the features of these bars.

\subsection{The strength of FIRE-2 bars} 
\label{sec:fire_bar_strengths}

We search each galaxy in Table \ref{table:FIRE-2_galaxy_summary} for bar-like features from $\redshift \sim 2$ to the present day, by calculating the \emph{mass-weighted} bar strength as a function of radius in the plane of the stellar disk. The bar strength is defined as the maximum amplitude of the $m=2$ Fourier mode of the 2D decomposition of the face-on stellar surface density distribution of the galaxy  \citep{Athanassoula.and.Misiriotis.2002,Athanassoula.et.al.2013MNRAS},
\begin{equation}
    \frac{A_{2}}{A_{0}}= \left( \frac{\sqrt{ a^{2}_{2} + b^{2}_{2} }}{\Sigma^{N}_{i=1} m_{\star i}}  \right) \textbf{,}
    \label{bar_strength}
\end{equation}
where, in general for the $m^{th}$ mode, $a_{m}= \Sigma^{N}_{i=1} m_{\star i} \cos(m \theta_{i}) $, $b_{m}= \Sigma^{N}_{i=1} m_{\star i} \sin(m \theta_{i})$, $\theta_{i}$ is the azimuthal coordinate and $m_{\star i}$ is the mass of i$^{th}$ stellar particle. 

We calculate the bar strength from the star particles as follows. First we define and rotate into a coordinate system in which the $z$ direction is perpendicular to the ``disk plane,'' defined by computing the principal-axis system from the youngest 25\% of star particles among the ones constituting the 90\% mass within a 10 kpc radius and using the $z$ direction as the eigenvector corresponding to the smallest eigenvalue. Then we select stars in the disk: within 10 kpc of the galaxy center in $R \equiv \sqrt{x^2+y^2}$ and within 2 kpc of the mid-plane in $z$. We then divide the disk into annuli of width $\Delta R=50$ pc\footnote{Even with a larger bin size of 100-200 pc, the barred structures in the m=2 component persist.} and calculate $A_{2}/A_{0}$ by applying Equation \ref{bar_strength} to the star particles in each annulus. We repeat this calculation in each snapshot of each simulation from $\redshift=2$ to the present day. 

This measurement of bar strength has two limitations. First, it captures not only the bar but also any other asymmetric features in the disk that have approximate $m=2$ symmetry, including spiral arms or remnants of galaxy mergers. Thus, we also require that the phase angle of the $m=2$ Fourier mode, 
\begin{equation}
\label{eq:PA}
    \phi_{2}=\tan^{-1}(b_{2}/a_{2})/2,
\end{equation}
be nearly constant over the length of the bar, which should rotate like a rigid object. Second, since our measure is mass-weighted, it is not directly comparable to the \emph{light-weighted} bar strength measure calculated similarly from images. However, in this work we are primarily interested here in the mechanisms governing bar formation, for which the mass-weighted version is appropriate. 

\begin{figure*}
\includegraphics[width=\textwidth]{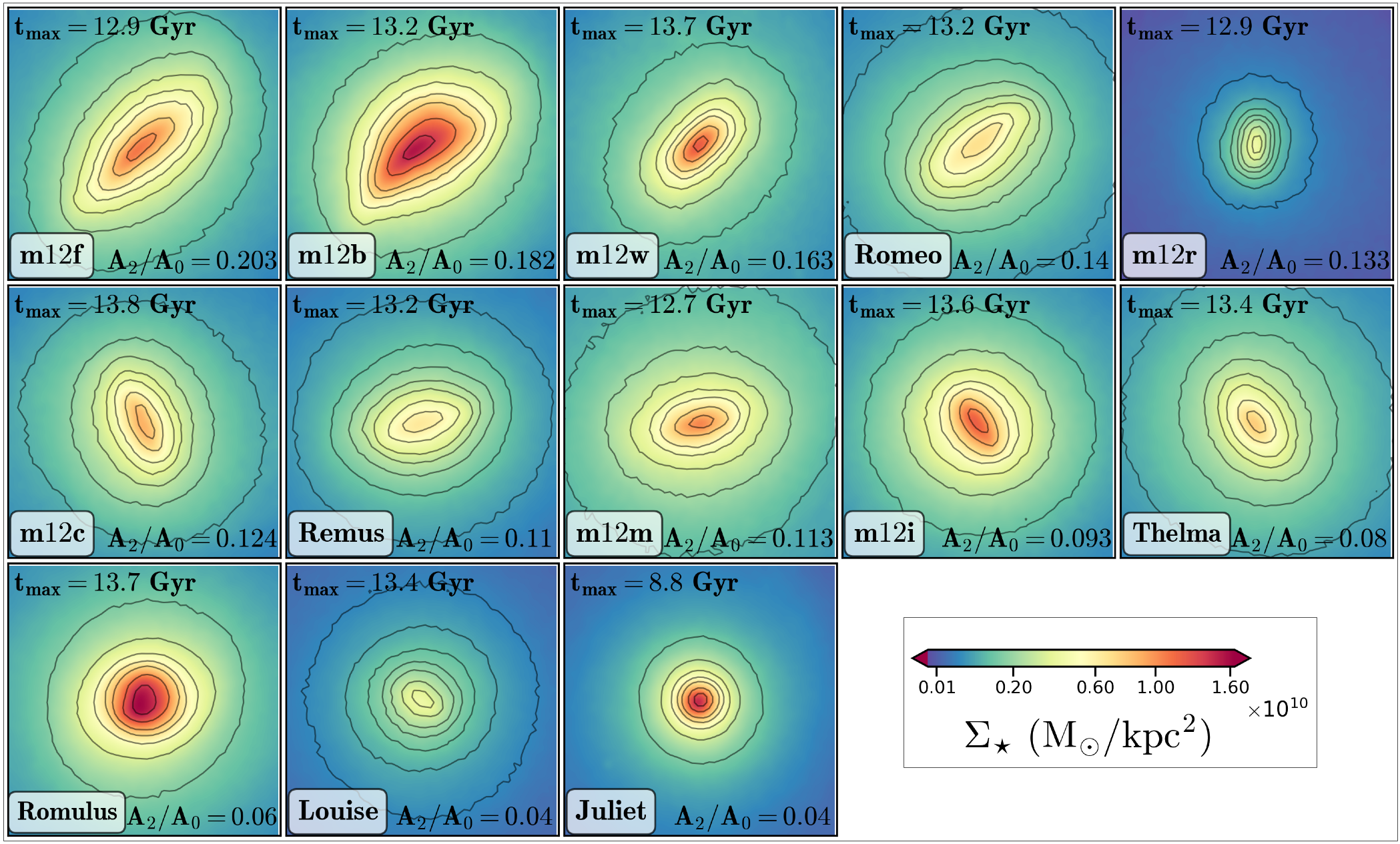}
    \caption{ {\bf Bars in Milky-Way-mass galaxies from the FIRE-2 suite.} Face-on projected stellar mass density for all star particles in the central (4 kpc)$^{3}$ volume is shown in each panel at the time of maximum bar strength $A_2/A_0$, with high-density regions in each case scaled to red, medium density region in yellow and low-density in blue following the color bar of stellar surface density $\Sigma_{\star}$ (M$_{\odot}$ kpc$^{-2}$). The contours are at 90\%, 70\%, 50\%, 40\%, 30\%, 20\% and 10\% of peak density in each galaxy. The value of the bar strength and the time in Gyr are provided in each panel. Bars and bar-like structures show a variety of shapes and sizes; for example, \textsf{m12b} (top row, second panel from right) is lopsided (see \citet{Lokas2021} for more examples of lopsided bars), and \textsf{m12w} (top row, middle panel) is distinctly rectangular. Some galaxies (eg, \textsf{Juliet}, \textsf{Louise} and \textsf{Romulus}) do not show bars according to our definition (Section \ref{sec:fire_bar_strengths}).
    }
    \label{fig1:all_bars}
\end{figure*}

\begin{figure}
	\includegraphics[width=\linewidth]{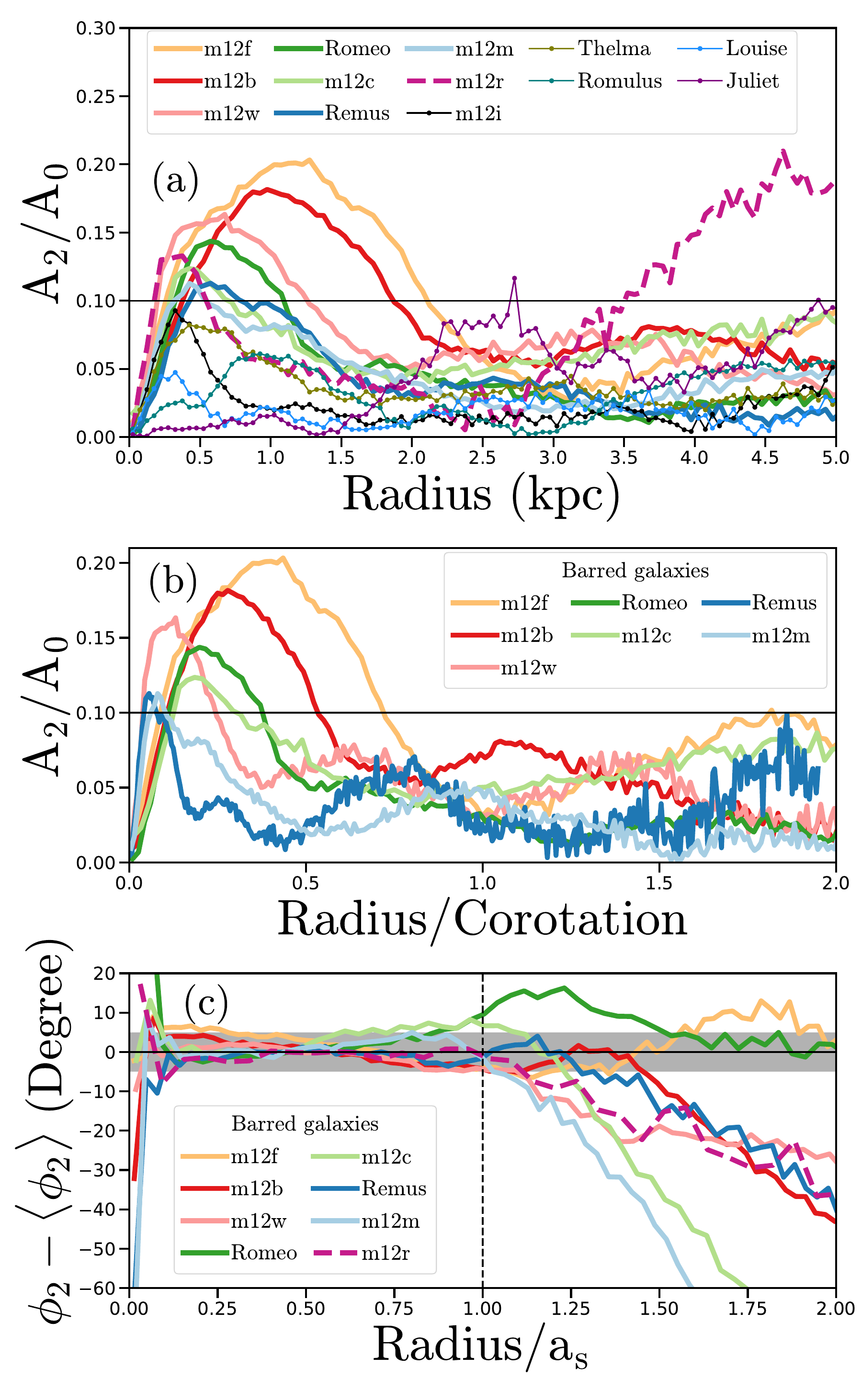}
    \caption{ {\bf Bar profiles.}  {\bf Panel (a)} shows $A_{2}/A_{0}$ profiles as a function of disk radius for all the galaxies, both barred (thick lines) and unbarred (dotted thin lines), at the time of peak bar strength (see Figure \ref{fig1:all_bars}). {\bf Panel (b)} shows only the barred galaxy profiles as a function of disk radius, scaled by the bar corotation radius. For all the barred galaxies the peak $A_{2}/A_{0}$ is well inside corotation. We do not add \textsf{m12r} in panel (b) as we cannot measure its pattern speed to determine its corotation radius (Section \ref{sec:bar_pattern_speed}). {\bf Panel (c)} shows the mean subtracted phase angle $\phi_{2} -\langle \phi_{2} \rangle$ as a function of disk radius, scaled by bar radius ($a_{s}$), where the mean is over the five data points around $\phi_{2}$ at the radius of peak $A_{2}/A_{0}$. 
    The bar phase $\phi_{2}$ must be within $0 \pm 5^{o}$ (shaded) to maintain the nearly constant PA characteristic of a bar. 
    Within the bar length (black dashed vertical line) the PA is nearly constant for all the barred galaxies, except in the very central region where the bar strength drops below $A_{2}/A_{0}<0.1$. }
    \label{fig2:all_bar_strength}
\end{figure}

In Figure \ref{fig1:all_bars} we present face-on `images' of the central 4$\times 4$ kpc region of each of the 13 FIRE-2 galaxies at the time of their maximum bar strength $A_{2}/A_{0}|_{max}$. Each panel also shows the time $t_{\mathrm{max}}$ of maximum bar strength for that galaxy in Gyrs. The bar strength profile of each galaxy at $t_{\mathrm{max}}^{\mathrm{lbt}}$ is shown in panel (a) of Figure \ref{fig2:all_bar_strength}. \textsf{m12f} has the highest bar strength of $A_{2}/A_{0}|_{max}=0.203$, while  \textsf{Juliet} shows the lowest peak bar strength $(A_{2}/A_{0})_{max}=0.038$. We find bar strengths $\gtrsim 0.1$ in 10 of the 13 galaxies. 

In a different simulation run of \textsf{m12m}, the bar has an X-shape in the younger population of stars as studied previously by \cite{Debattista.et.al.2019}, where the bar is of higher strength ($A_{2}/A_{0}|_{max}\sim 0.18$). However we find maximum bar strength of $(A_{2}/A_{0})_{max}\sim 0.11$ for the version of \textsf{m12m} in this article. These two runs include slightly different physics prescriptions: the run analyzed here includes an implementation of turbulent metal diffusion in the gas phase while the one analyzed in \cite{Debattista.et.al.2019} does not. We do not expect this difference to have a direct effect on bar formation; however, any variation of the physics model will affect the star formation history of the galaxy, which varies stochastically between runs due to the randomness involved in triggering supernovae \citep{Hopkins.et.al.2018}, and will therefore result in a slightly different evolution. We therefore attribute this difference to variations in the star formation and feedback history between the runs. This degree of variation underlines from a theoretical standpoint the relatively arbitrary use of a cutoff value of $A_{2}/A_{0}$, especially since we see continuous variations in this measure with both time (as we will discuss in Section \ref{sec:bar_duration}) and radius (Figure \ref{fig2:all_bar_strength}) in all our simulations.

One notable characteristic of the bars in Figure \ref{fig1:all_bars} is that their isodensity contours appear more rounded and less rectangular in shape, as compared to the more rectangular bars commonly seen in N-body simulations and some observations. As we will argue in Section \ref{sec:mass_profile}, this is because the circular velocity curves of the DM component in the FIRE-2 barred galaxies are slow-rising, similar to the model in \citet{Athanassoula.and.Misiriotis.2002}.

According to the most commonly used definition of a bar, $(A_{2}/A_{0})_{max} > 0.2$, none of the FIRE-2 systems would be considered to host a bar. However, this criterion was developed using light-weighted bar strengths calculated from images rather than the mass-weighted version that we calculate, making the value of this cutoff somewhat arbitrary. Furthermore, even for bar strengths below this nominal cutoff we observe long-lived features in our simulations with $m=2$ symmetry in the inner galaxy. Therefore, we instead use the following operational criteria to define a bar:
\begin{itemize}
    \item {For an asymmetry to be called a bar, the instantaneous peak bar strength at any radius should be $(A_{2}/A_{0})_{max} > 0.1$ for at least $1.5 T_{o}$, where $T_{o}$ is the orbital time $2\pi r /V_{c}(r)$ at the outer edge of the bar.} 
    \item {The bar position angle (PA) $\phi_{2}$ (Equation \ref{eq:PA}) should be constant within $\pm 5^{o}$ over the entire bar length (Figure \ref{fig2:all_bar_strength}, panel c). }
\end{itemize}

Figure \ref{fig2:all_bar_strength} panel (a) shows the bar strength as a function of cylindrical radius $R$ in the disk plane for all 13 FIRE-2 galaxies at the time of their peak bar strength. All the galaxies having bar strength $A_{2}/A_{0}>0.1$ in panel (a) are candidate bars, some of which we confirm in Section \ref{sec:bar_duration} after measuring their duration. In panel (b), we show the bar strength profiles of \emph{confirmed} bars (i.e. those lasting longer than $1.5T_o$) as a function of $R$ scaled by the bar corotation radius (Section \ref{sec:corotation}). The bar in \textsf{m12r} is missing from panel (b) as we cannot measure the pattern speed, and consequently the corotation of its extremely short bar (more details in Section \ref{sec:bar_pattern_speed}). In panel (c) we show the variation of mean-subtracted PA $\phi_{2} - \langle \phi_{2} \rangle$ as a function of $R$ scaled by the bar radius $a_{s}$ (Section \ref{sec:bar_length_measurement}). Here $\langle \phi_{2} \rangle$ is the mean of phase $\phi_{2}$ around the radius of the peak bar strength. The PA of each barred galaxy is constant within $\pm 5^{o}$ over the bar's length ($R/a_{s}<1$).

Figures \ref{fig1:all_bars} and \ref{fig2:all_bar_strength} show that by our criteria, eight of the FIRE-2 galaxies have bars: \textsf{m12f}, \textsf{m12b}, \textsf{m12m}, \textsf{m12w}, \textsf{Remus}, \textsf{Romeo}, \textsf{m12c} and \textsf{m12r}. Some other galaxies, such as \textsf{m12i} and \textsf{Thelma}, have asymmetries that maintain a constant PA but have $A_{2}/A_{0}<0.1$ throughout their evolution. The rest of the galaxies (\textsf{Louise}, \textsf{Juliet} and \textsf{Romulus}) do not satisfy either criterion; nor could we identify a bar-like structure in these galaxies through visual inspection of their stellar disks all throughout their evolution. Instead, we see a bulge-like stellar distribution at the centers of \textsf{Louise}, \textsf{Juliet} and \textsf{Romulus} (see Figure \ref{fig1:all_bars}).

\subsection{Bar duration} \label{sec:bar_duration}

\begin{figure}
\centering
	\includegraphics[width=\columnwidth]{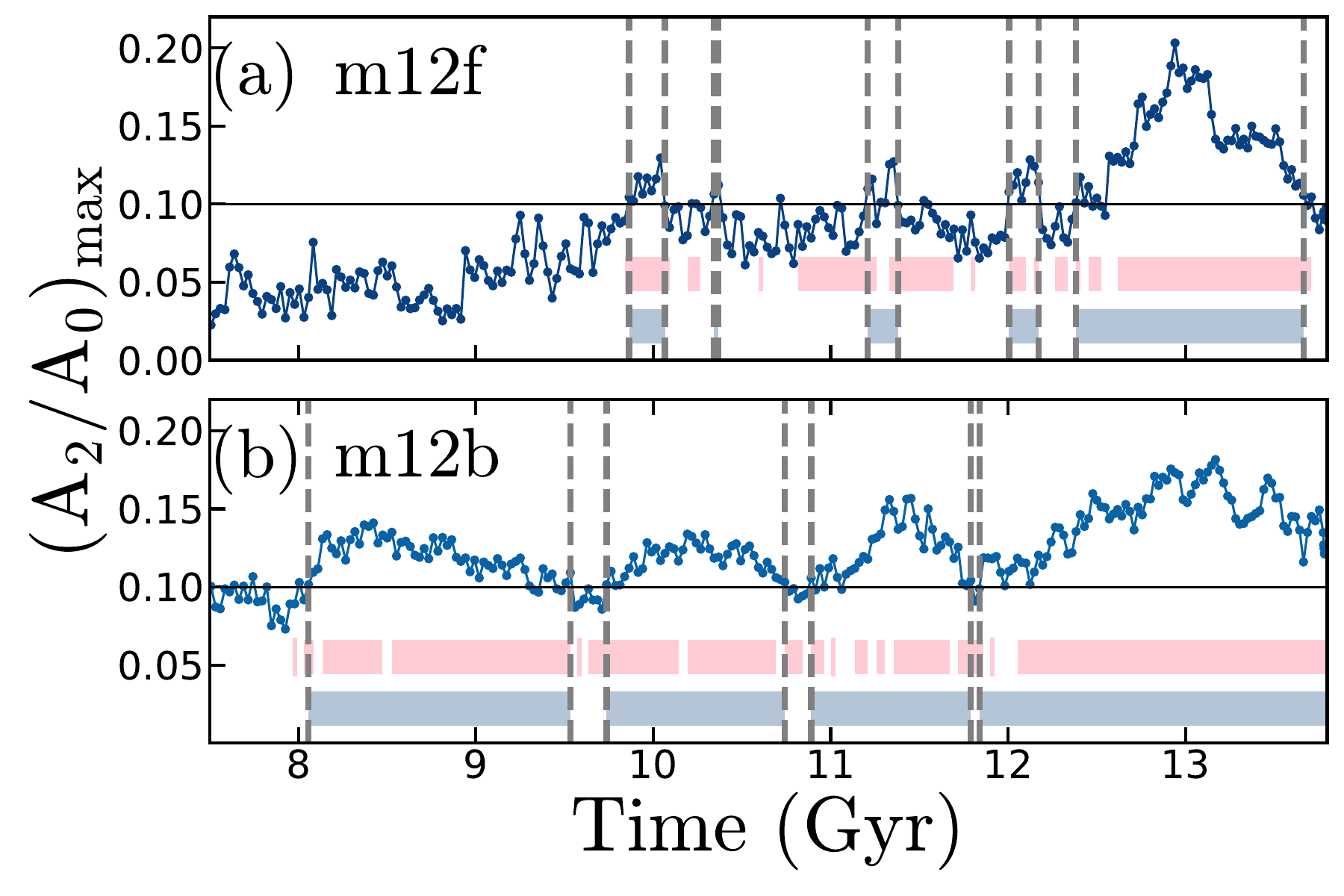}
    \caption{ {\bf Bar duration in \textsf{m12f} (panel a) and \textsf{m12b} (panel b):} This Figure shows the evolution of maximum bar strength within a radius of 2 kpc, $A_{2}/A_{0, (r<2 \rm kpc)}$ of two FIRE-2 galaxies \textsf{m12f} and \textsf{m12b} {\bf (having the strongest bars)} with time (in Gyrs). When bar strength $(A_{2}/A_{0})_{max}> 0.1$ (blue shaded regions bounded by dashed vertical lines) for a time interval $\Delta t >T_{o}$ (as per our bar definition), and also when the bar PA is nearly constant (pink shaded regions; for details on PA see Section \ref{sec:fire_bar_strengths}) we call it a bar episode (overlapping time interval of blue and pink shades). For \textsf{m12f} there are two major bar episodes: 1) $t=9.86-10.04$ Gyr and 2) $t>12.38$ Gyr. For \textsf{m12b} the bar episode is from $t=8.05-13.8$ Gyr ($\sim5.7$ Gyrs) as there is more uniform overlap between the high bar strength (blue shaded region) and constant bar phase $\phi_{2}$ (pink shaded region) for 5.7 Gyrs. For more details of bar episodes see Section \ref{sec:bar_duration}.}
    \label{fig:m12f_m12b_bar_duration}
\end{figure}

We next determine the duration over which each bar has $(A_{2}/A_{0})_{max}>0.1$ and constant phase $\phi_{2}$. Figure \ref{fig:m12f_m12b_bar_duration} gives an example of the evolution of these quantities with time for two of the strongest bars, in \textsf{m12f} (top panel) and \textsf{m12b} (bottom panel). Both bars undergo episodes of varying length where $A_{2}/A_{0}>0.1$ (blue shaded regions) and constant bar phase $\phi_{2}$ (pink shaded regions).

Generically, we define any time interval with bar strength $(A_{2}/A_{0})_{max}>0.1$ \emph{and} constant bar PA to be a ``bar episode.'' However, the full picture is slightly more complex, as illustrated by the examples shown in Figure \ref{fig:m12f_m12b_bar_duration}. In \textsf{m12f} (panel a), there is an early bar episode from $t=9.86 \text{--} 10.04$ Gyr, followed by a long interval where $(A_{2}/A_{0})_{max}>0.1$ occasionally but $\phi_{2}$ is not constant. Then the bar strength again rises to $(A_{2}/A_{0})_{max}>0.1$ for $t>12.38$ Gyr with constant $\phi_{2}$, and remains stable for a second bar episode of $\sim$ 1 Gyr. The bar-like asymmetry in \textsf{m12f} does not satisfy both bar criteria between 10.04-12.38 Gyr; hence we consider \textsf{m12f} to have two bar episodes of durations 0.18 and 1 Gyr separated by an unbarred period of 2.3 Gyr. On the other hand, in \textsf{m12b} (panel b), the bar strength dips below 0.1 for several very short intervals (0.20 Gyr, 0.148 Gyr and 0.048 Gyr), while the bar phase $\phi_{2}$ is almost continuously constant from $t=8.05-13.8$ Gyr. We consider this to be one continuous bar episode with a duration of nearly 5.7 Gyrs.

Our distinction between bar episodes is somewhat sensitive to the lower bound on $(A_{2}/A_{0})_{max}$. If we choose a minimum strength of 0.08 instead of 0.1 there would be more bar episodes for \textsf{m12f} according to our bar criteria. This is a limitation of our bar definition criteria and similar choices in the literature, as the choice of this lower limit to bar strength is largely empirical. Additionally, we note that the number of bar rotations is sensitive to the radius at which we estimate orbital time $T_{o}$. We evaluate $T_{o}$ at the edge of the bar that varies with evolution (see Appendix \ref{appendix:sec:bar_length_evolution}).

All other galaxies in our simulation suite (\textsf{m12m}, \textsf{Romeo}, \textsf{Remus}, \textsf{m12w}, \textsf{m12c} and \textsf{m12r}) undergo a single episode of high bar strength that ends when the bar strength either decreases below $A_2/A_0<0.1$ (\textsf{m12m}, \textsf{Romeo} and \textsf{m12r}) or remains high ($A_2/A_0\gtrsim 0.1$) until the end of the galaxy's evolution (until $\sim 13.8$ Gyrs for \textsf{m12w}, \textsf{Remus} and \textsf{m12c}). We list the durations of the bar episodes for all the galaxies in Table \ref{table_bar_duration}. 

Using the criteria for defining a bar in Section \ref{sec:fire_bar_strengths}, eight of the 13 FIRE-2 galaxies have bars at some point in their evolution: \textsf{m12f}, \textsf{m12b}, \textsf{Remus}, \textsf{Romeo}, \textsf{m12w}, \textsf{m12m}, \textsf{m12c} and \textsf{m12r}. The other five never reach the $A_{2}/A_{0}> 0.1$ limit during their evolution: \textsf{m12i}, \textsf{Thelma}, \textsf{Louise}, \textsf{Romulus} and \textsf{Juliet}.

\begin{deluxetable}{cccc}
\tablenum{2} \label{table_bar_duration}
\tablecaption{Bar Duration Estimates}
\tablewidth{0pt}
\tablehead{
\colhead{Simulation} & \colhead{Duration $T_{d}$} & \colhead{Orbital time $T_{o}$} & \colhead{$\frac{T_{d}}{T_{o}}$} \\
\colhead{} & \colhead{(Gyr)} & \colhead{$\frac{2\pi r}{V_{c}(r)}$ (Gyr)} & \colhead{} }
\decimalcolnumbers
\startdata
m12f-(episode-1) & 0.18 & 0.024 & 7.4  \\
m12f-(episode-2) & 1.28 & 0.041 & 31.2  \\
m12b & 5.74 & 0.035 & 163.5  \\
Romeo & 0.77 & 0.034 & 22.6 \\
Remus & 0.79 & 0.035 & 22.4 \\
m12w & 0.13 & 0.037 & 3.5 \\
m12m & 0.069 & 0.04 & 1.73  \\
m12c & 0.066 & 0.037 & 1.78 \\
m12r &  0.39 &  0.052 &  7.6   \\
\enddata
\tablecomments{Columns: 1. Simulation name; 2. Duration $T_{d}$ for a single bar episode according to the criteria in Section \ref{sec:fire_bar_strengths} (\textsf{m12f} has two bar episodes); 3. Orbital time $T_{o}=2\pi r/V_{c}(r)$ estimated at the edge of the bar, $r=a_{s}$ (see Section \ref{sec:fire_bar_strengths} and Appendix \ref{appendix:sec:bar_length_evolution}); 4. Number of times the bar rotates during its duration.  }
\end{deluxetable}

\subsection{Bar length}\label{sec:bar_length_measurement}
We estimate the bar length from the face-on stellar density distribution (as shown in Figure \ref{fig1:all_bars})  at the time of peak bar strength, using four different methods from the literature: (1) ellipse fitting of the bar region and bar PA measurement, (2) radial bar strength profile, (3) the difference between surface densities along the bar major and minor axes \citep{Athanassoula.and.Misiriotis.2002, Erwin.2018} and (4) bar and inter-bar density contrast \citep{Aguerri.et.al.2000}. Each method gives a slightly different bar length, as discussed in Appendix \ref{sec:appendix_bar_length_measurement}. See also \citet{Michel-Dansac.Wozniak2006, Ghosh.DiMatteo2024} for discussions of different methods to measure bar length. For the galaxies that have bars according to our definition, we report the semi-major axis length $a_{s}$ (bar radius) averaged across all four methods in Table \ref{table_bar_properties}. The lengths measured by each of the individual methods are given in Table \ref{table:appendix_bar_length} of Appendix \ref{sec:appendix_bar_length_measurement}.

\subsection{Bar pattern speed}\label{sec:bar_pattern_speed}
Except just before $\redshift=0$, the simulation snapshots are spaced too far apart to estimate the bar pattern speed $\Omega_{p}$ directly, by calculating the rate of change of the $m=2$ mode Fourier phase angle $\phi_{2}$ with time. Instead we apply the Tremaine-Weinberg (TW) method \citep{Tremaine.Weinberg.1984} to optimally projected 2D density maps of star particles older than 100 Myr\footnote{The TW method is based on the continuity equation and is therefore applicable to a population of stars of the same age; as the stars in bar orbits are mostly old, we select stars older than 100 Myr.}, following the method outlined in \cite{Merrifield.Kuijken.1995}. Complete details of how we apply the TW method to the simulated bars are in Appendix \ref{appendix:TW_method}.

To validate the TW method on our simulations, we compare the result with the pattern speed measured directly from the time derivative of the $m=2$ Fourier phase angle, $\Omega_{p}=d\phi_{2}/dt$, in the four galaxies hosting a bar during the last 22 Myrs of evolution, for which 10 snapshots are saved with a time interval of 2.2 Myrs. We can compare the pattern speed determined by both methods for \textsf{m12b}, \textsf{m12c} and \textsf{Remus}. We find that estimates from both methods are consistent for bars with $A_{2}/A_{0}|_{max} \gtrsim 0.1$ \emph{and} $2a_s \gtrsim 2.5$ kpc.\footnote{The TW method requires an inclined disk ($\sim 45^{o}$) that shortens the apparent bar length by a factor $1/\sqrt{2}$.} Among our sample, this allows us to measure the pattern speed using the TW method with confidence for \textsf{m12b}, \textsf{m12f}, \textsf{Romeo}, and \textsf{Remus}. We can additionally get a direct measurement of the pattern speed at $\redshift=0$ for the bars in \textsf{m12c}, \textsf{m12m}, and \textsf{m12w}, for which the bars are too weak and short to apply TW. These values are listed in Table \ref{table_bar_properties}. We discuss the consistency of the TW method with direct measurements of the pattern speed in more detail in Appendix \ref{Appendix:patterspeed}.

We also find a match between the measurements of pattern speed from the TW method and from the code \textsf{patternSpeed.py} developed by \citet{Dehnen2023} (see Appendix \ref{Appendix:patterspeed}).  In one of the cases in \textsf{m12b}, we find a steady decrease in bar pattern speed, which for a fixed bar length and mass would indicate a loss of angular momentum (see Appendix \ref{sec:appendix:m12b_patternspeed}). However, as the bar in \textsf{m12b} evolves, it lengthens and accumulates more stars, which could compensate for a decrease in pattern speed to maintain the angular momentum.
The change in bar angular momentum results from two opposing effects: the loss of angular momentum from the bar, which slows down its pattern speed, and the gain in angular momentum from adding stars in the bar orbits, which increases the bar's length. External torques from satellite galaxies can also impact bar angular momentum. It is challenging to separate the effects of angular momentum changes among the bar, disk, satellites, and DM halo in a cosmologically evolving system.

\begin{figure}
\centering
\includegraphics[width=0.95\columnwidth]{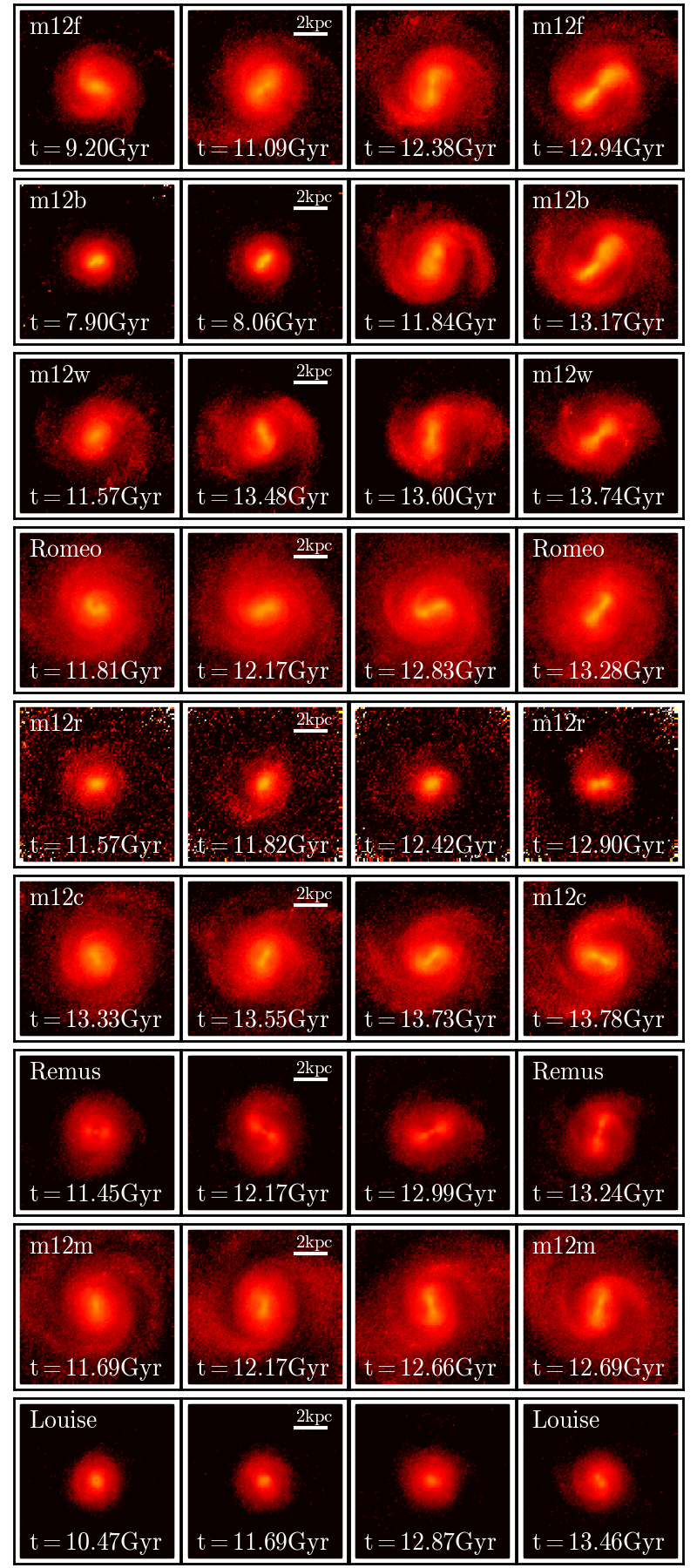}\\
    \vspace{-0.3cm}
\includegraphics[width=0.5\columnwidth]{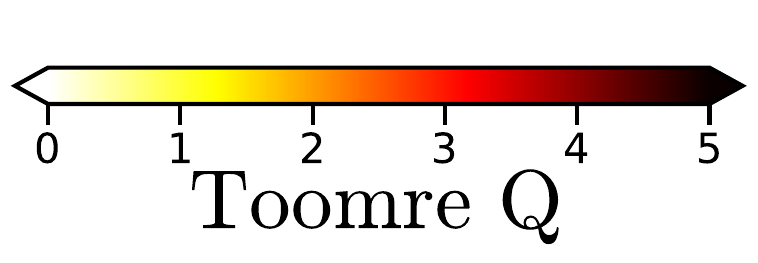}
    \caption{{\bf Evolution of $Q_\star$ during bar formation} for the 8 barred galaxies in descending order of peak bar strength (rows 1--8) and a typical unbarred galaxy (\textsf{Louise}; row 9). The first column shows each system prior to bar formation; the second shows the onset of bar instability, the third shows when $A_2/A_0$ first exceeds 0.1 and the fourth shows the snapshot of peak bar strength.}
    \label{fig:toomreq-evolution}
\end{figure}

\subsection{Corotation} \label{sec:corotation}
We calculate the corotation radius, \rco, at which the circular speed in the disk plane, $\Omega(R)$, is equal to the pattern speed of the bar obtained in Section \ref{sec:bar_pattern_speed}. To determine the circular speed $\Omega(R)$ as a function of cylindrical radius $R$, we use a composite potential model of each FIRE-2 galaxy as described in \cite{Arora.et.al.2022} for the appropriate snapshots of each simulation. In brief, the models are implemented using the Agama dynamics package \citep{Vasiliev.2019} with a spherical harmonic basis expansion for the halo (dark matter and hot gas) and a cylindrical spline basis expansion for the disk (cold gas and stars). Both components are constrained to be axisymmetric for consistency with the assumptions used to define \rco.
Using this smooth model we calculate the derivative of the axisymmetric potential $\Phi(R,z)$ at the disk midplane to determine the circular speed: 
\begin{equation}
    \Omega(R) =  \frac{1}{R} \left.\frac{\partial \Phi(R,z)}{\partial R}\right|_{z=0}.
\end{equation}
The radius at which $\Omega(r)=\Omega_{p}$ then gives the bar corotation radius $\rco$ as presented in Table \ref{table_bar_properties}. All the bars are well within the corotation radius. 

Using the bar length $2a_{s}$ and the corotation radius, we estimate the $\mathcal{R}$-parameter $\mathcal{R} \equiv \rco/a_{s}$, also provided in Table \ref{table_bar_properties}. $\mathcal{R}$ is greater than 1.4 for all the bars. 
However, the pattern speeds we measure are not appreciably different from those measured for real systems: they lie between 36--97 km s$^{-1} \rm kpc^{-1}$ (Table \ref{table_bar_properties}). All the bars in FIRE-2 are shorter than their corotation radius, with $a_s < \rco$ (see Table \ref{table_bar_properties}).

\subsection{Disk stability}
\label{sec:toomreq_methods}

Global disk instabilities like bars can form via swing amplification \citep{Julian.and.Toomre.1966}, a process that depends sensitively on the properties of the central regions of the disk. The tendency to swing-amplification is usually quantified using one of several metrics, including the Toomre $Q$ parameter that compares the radial velocity dispersion of the disk stars to the mass surface density of a disk. Even though the Toomre $Q$ parameter denotes instability against the m=0 axisymmetric perturbation and not the m=2 perturbations, a value of $Q$ lower than one indicates that the disk is unstable to gravitational instabilities. While there are other methods for probing disk instability \citep{Ostriker.and.Peebles.1973, Efstathiou.et.al.1982, Fujii2018, Marioni.et.al.2022}, we find that they all generally yield results similar to those obtained using the Toomre $Q$ parameter for weak bars. Therefore, we continue to use the well-established Toomre $Q$ parameter to assess disk instability.

One of the necessary conditions for swing amplification is that $Q \gtrsim 1$, so that local overdensities have a chance to grow by swing-amplification rather than gravitational instability, but not by so much that the random motion measured by $\sigma_{z,\star}$ excludes bar-supporting radial orbits. \cite{Toomre1981} showed that $1<Q<2$ is necessary to form a bar in a Mestel disk.

To investigate whether this condition is still reflective of our cosmological bars, and to understand whether gas flows in the central galaxy can contribute to bar formation, we calculate $Q$ for both the stellar disk and the gas disk of each simulated galaxy in our sample. 
We calculate $Q_{\star}$ for the stellar disk by modifying the expression in \cite{Toomre1964} to include the gravity of the gas and dark matter,
\begin{equation} \label{eq:toomreQ}
    Q_{\star}=\frac{\sigma_{R} \kappa}{3.36 G \left(\Sigma_{\star}+ \Sigma_{gas}+\Sigma_{DM}\right) }
\end{equation}
where $\sigma_{R}$ is the stellar dispersion in the plane of the disk, $\kappa$ is the epicyclic frequency ($\kappa^{2}(R)= \left. \partial^{2} \Phi(R,z)/\partial^{2} R \right|_{z=0} + \left(3/R\right) \left. \partial \Phi(R,z)/\partial R\right|_{z=0}$; estimated using the potential model of each FIRE-2 galaxy presented in \citealt{Arora.et.al.2022, Arora2024}; see Section \ref{sec:corotation}), and $\Sigma_{\star}+ \Sigma_{gas}+\Sigma_{DM}$ are the surface densities of stars, gas, and dark matter respectively within $\pm 0.5$ kpc of the disk plane, as we intend to measure the local instability close to the mid-plane.
The gas disk \Qgas is determined using an analogous expression from \citet{Orr.et.al.2020},
\begin{equation}
    \Qgas =\frac{\sqrt{2} \sigma_{z} \Omega}{\pi G (\Sigma_{gas} +\gamma \Sigma_{\star})},
\end{equation}
where $\sigma_{z}$ is the gas dispersion perpendicular to the disk plane, $\Omega$ is the circular speed, and $\gamma$ is the fraction of the stellar component that lies within the gas disk scale height and therefore contributes to the self-gravity of the gas disk. 

Although the disk instability criterion is based on the minimum value of Q, the Q-maps in Figure \ref{fig:toomreq-evolution} show the regions that fall under this criterion at different times of evolution \citep{Orr.et.al.2020}. We calculate $Q_{\star}$ and \Qgas\ for all barred galaxies in our sample in the central $10\times10$ kpc$^2$ region of the disk with $|z|\leq 0.5$ kpc. We evaluate the surface densities using face-on bins of $0.1\times 0.1$ kpc$^2$ area. The mean \Qgas\ is in the range of $0.36<\langle \Qgas \rangle<1.1$ in the inner 5 kpc of most of the galaxies' centers. We do not find a correlation between bar strength and $Q_{gas}$, so we focus on $Q_\star$ in the rest of this Section.

\begin{deluxetable*}{cccccccccccc}
\tablenum{3} \label{table_bar_properties}
\tablecaption{Bar Characteristics}
\tablewidth{0pt}
\tablehead{
\colhead{Name} & \colhead{$\left(\frac{A_2}{A_0}\right)_{\rm max}$ } & \colhead{$\rm T_{peak}$ } &  \colhead{\redshift$_{,\mathrm{peak}}$ } & \colhead{$\Omega_{\rm p, D}$} & \colhead{ $\langle\Omega_{\rm p, TW}\rangle$  } & \colhead{ $\left. \Delta \redshift \right|_{\rm TW}$ } & \colhead{ $\Omega_{\rm p, peak}$  }   & \colhead{$\rco$ } & \colhead{$a_{s}$} & \colhead{$\mathcal{R}$} & \colhead{ Q$_{\star, min}$}  \\
\colhead{} & \colhead{} & \colhead{(Gyr) } &  \colhead{ }  & \colhead{ $\rm (km s^{-1} kpc^{-1})$ } & \colhead{ $\rm (km s^{-1} kpc^{-1})$} & \colhead{ } & \colhead{ $\rm (km s^{-1} kpc^{-1})$} & \colhead{(kpc)} & \colhead{(kpc)} & \colhead{} & \colhead{}
}
\decimalcolnumbers
\startdata
\multicolumn{3}{c}{Disks with bars: $A_2/A_0>0.1$} \\
\hline
\textsf{m12f} & 0.203 & 12.94  & 0.064 & 73.9 & 70.50 & 0.07--0.05   & 91.07  &  2.93  & 1.86 & 1.67 & 0.94 \\
\textsf{m12b} & 0.182 & 13.17  & 0.046 & 89.5 & 85.37 & 0.0016--0.0  & 81.6   &  3.5   & 1.79 & 1.95 & 0.87 \\
\textsf{m12b} & 0.182 & 13.17  & 0.046 & 89.5 & 84.23 & 0.055--0.036 & 81.6   &  3.5   & 1.79 & 1.95 & 0.87 \\
\textsf{m12w} & 0.163 &  13.79 & 0.0   & 46.8 & -     & -            & -      &  5.17  & 1.35 & 3.82 & 1.08 \\
\textsf{Romeo} & 0.144 & 13.28 & 0.038 & -    & 69.37 & 0.055--0.023 & 82.9   &  2.9   & 1.32 & 2.19 & 1.07 \\
\textsf{m12r} & 0.133 & 12.98  & 0.061  & -   & -     & -            &  -     &  -     & 0.81 & -    & 1.06 \\
\textsf{m12c} & 0.123 & 13.77  & 0.001 & 67.0 & -     & -            & 97.4   &  2.3   & 1.27 & 1.81 & 0.97 \\
\textsf{Remus} & 0.113 & 13.21 & 0.041 & 93.8 & 42.81 & 0.048--0.034 & 36.09  &  10.25 & 1.58 & 6.48 & 1.72 \\
\textsf{m12m} & 0.112 & 12.68  & 0.084 & -    & -     &  -           & 49.5   &  5.3   & 1.53 & 3.59 & 1.14 \\
\hline
\multicolumn{2}{c}{Disks with $A_{2}/A_{0}<0.1$} \\
\hline
\textsf{m12i} & 0.092 & 13.62 & 0.013 & 90.19 & -     & -            & -      & -      & 0.7   & -   & 0.99 \\
\textsf{Thelma} & 0.082 & 13.46 & 0.024 & 57.50 & -   & -            &  -     & -      & 1.0   & -   & 1.18 \\
\enddata
\tablecomments{Columns: 1. Simulation name; 2. Peak bar strength overall times; 3. Time of peak bar strength; 4. redshift of peak bar strength; 5. Bar pattern speed measured directly from high-cadence snapshots ($0.0016<\redshift<0.0$); 6. Median bar pattern speed measured using the TW method; 7. range of redshifts used for the TW method;  8. TW pattern speed at peak bar strength (for more details see Appendix \ref{Appendix:patterspeed});  9. \rco, corotation radius at peak bar strength; 10. $a_{s}$, bar semi-major axis length (bar radius) at peak bar strength; 11. $\mathcal{R}=\rco/a_{s} $ at peak bar strength; 12. mimimum $Q_\star$ within 5 kpc (see Section \ref{sec:toomreq_methods}) at peak bar strength. The missing values of $\Omega_{\rm p, D}$ (column 5) indicate the bar does not exist at the end of the simulation and there is no constant pattern speed in the bar region. The missing value in columns 6, 7, 8, 9 and 11 means that we cannot measure these quantities as the bars are weak and short in length. We could measure the TW pattern speed for two episodes in \textsf{m12b} (see columns 6 and 7 for \textsf{m12b}) and both episodes have all other quantities same for the rest of the columns (1--5, 8--12). }
\end{deluxetable*}

In Figure \ref{fig:toomreq-evolution}, we show the evolution of $Q_\star$ in the central region ($10\times10$ kpc$^{2}$, $|z|<0.5$ kpc) of the disk of each barred galaxy during bar formation. The first eight rows show the barred galaxies in descending order of bar strength, and the last row shows a typical example of an unbarred galaxy: \textsf{Louise}, which has $(A_{2}/A_{0})_{max} = 0.06$. The first column shows each disk in an unbarred phase ($A_{2}/A_{0}<< 0.1$). The second column shows the disk when the bar instability has begun to grow, but still $A_{2}/A_{0}< 0.1$. The third column is when $A_{2}/A_{0}\sim 0.1$ in each barred system, and the fourth column shows $Q_{\star}$ at the peak bar strength. At peak bar strength, we can easily trace the $1<Q_\star<2$ region (yellow and orange colors) associated with each bar. Extended spiral structures with nearly as low Q extend from each end of the bars in many cases, and the bars often show a two-lobed structure. These features all trace the combined high stellar and gas surface density ($\Sigma_{\star}+\Sigma_{gas}$) from the denominator of Equation \ref{eq:toomreQ}, showing that this factor plays the leading role in setting $Q_\star$. The other quantities in the equation that vary with position ($\sigma_{R}$, $\kappa$ and $\Sigma_{DM}$) do not show any prominent features in their maps. \textsf{Louise}, the unbarred galaxy in the last row of Figure \ref{fig:toomreq-evolution} looks very different from the barred galaxy maps, showing no features with low $Q_{\star}$.

Every galaxy in Figure \ref{fig:toomreq-evolution} reaches its lowest $Q_\star$, close to 1.0, in the center of the galaxy at the time of peak bar strength. Table \ref{table_bar_properties} lists these values. At larger $R$, $Q_\star$ gradually increases, creating a fairly large region where $Q_\star$ is high but not too high, appropriate for the growth of the bar by swing amplification. The typical $Q_\star$ in this region is between 2 and 3 rather than 1 and 2 as for the Mestel disk example, which is strikingly similar given the significant differences between the idealised case and our simulations.

We carried out several other tests to examine and quantify disk instability \citep{Ostriker.and.Peebles.1973, Efstathiou.et.al.1982, Marioni.et.al.2022} (also see Section \ref{sec:kinematic_coldness}), that lead to a similar conclusion. The FIRE-2 bars are weak in strength and have properties that lie in the boundary of strong and unbarred galaxies.

\section{The Origins of FIRE-2 Bars} 
\label{sec:results}
Bars can form during satellite interactions with galaxies and during internal evolution \citep{Gadotti.2011, Sellwood.2014} in the disk under favourable conditions as the disk undergoes gravitational instabilities \citep{Hohl.1971, Ostriker.and.Peebles.1973}. A cold disk \citep{Ostriker.and.Peebles.1973} with high surface density and low gas content, which is centrally baryon dominated, favours bar formation. On the other hand, the presence of a centrally concentrated bulge or DM halo, that does not participate in transfer of angular momentum from the disk to the DM halo, may make the disk stable against bar formation  \citep{Ostriker.and.Peebles.1973}. However, previous studies have also shown that the distribution of dark matter (in live halos) in the central region of the disk affects the bar morphology by transfer of angular momentum from the disk to the DM halo \citep{Combes.and.Sanders.1981,Athanassoula.2002,Athanassoula.and.Misiriotis.2002}, making the bar stronger and more rectangular for a disk dominated by dark matter.

In this Section, we investigate how many of these phenomena affect bar formation and bar morphology in the FIRE-2 galaxies: satellite interactions (Section \ref{sec:sat_interaction}), internal evolution of the disk (Section \ref{sec:secular_evolution}), the relative concentration of different components in the central galaxy (Section \ref{sec:mass_profile}), kinematic coldness of the disk (Section \ref{sec:kinematic_coldness}) and disk gas fraction (Section \ref{sec:gas_fraction_stellar_feedback}).

\subsection{Bar Formation from Satellite Interactions}
\label{sec:sat_interaction}
In this subsection, we investigate which FIRE-2 simulations likely had bar formation triggered by interactions with satellites. 

\citet{Karachentsev.and.Makarov.1999} defined a tidal index,  $\Theta = \log_{10} \left(\frac{\Msat}{D_{H}^{3}} \right)$, and used it to quantify the effects of satellites and minor mergers on the disks of host galaxies \citep[see also][]{Weisz.et.al.2011,Pearson.et.al.2016}, where  \Msat\ is the mass of the satellite (in units of $\rm M_{\odot}$) and \dhost\  is the distance of the satellite from the host galaxy's center (in units of kpc). 

However, disks with different masses respond differently to satellite interactions. To compare the impact of satellites on the various galaxies in FIRE-2, we need to consider how the properties of the host disks vary across the simulation suite. We therefore define a scaled tidal index: 
$\Gamma = \log_{10} \left(\frac{\Msat/\Mhost}{\left(\dhost/\rhost\right)^{3}} \right)$,
where \rhost is the effective radius of the host and \Mhost is the host galaxy's mass. The quantity \Mhost/\rhost$^3$ can also be expressed in terms of the circular velocity $V_c$ at a given radius $R$ in the host galaxy, \Mhost/\rhost$^3 = V_c^2/G R^2$,
and measures the concentration of the host galaxy. We are interested in interactions close enough to the galaxy center to affect orbits on the rising part of the rotation curve (see Section \ref{sec:toomreq_methods}) so we evaluate the quantity  $\Mhost/\rhost^3$  at the peak of the circular velocity curve:
\begin{equation}
    \Gamma = \log_{10} \left(\frac{\Msat/\dhost^3}{V^2_{c, \mathrm{max}}/G R^{2}_{\mathrm{max}}} \right).
    \label{eq:tidal_index}
\end{equation} 
where $V_{c, \mathrm{max}}$ is the maximum circular velocity and $R_{\mathrm{max}}$ is the corresponding radius. Both these quantities are given in Table \ref{table:FIRE-2_galaxy_summary}). In this expression, we take masses in units of $\rm M_{\odot}$, distances in kpc, and speed in km s${}^{-1}$.

We search for the satellites that interact with the host disks in FIRE-2 simulations at different times during evolution by investigating the merger trees of dark matter subhalos in each system. Merger trees are constructed using the codes ROCKSTAR \citep{Behroozi.et.al.2013a} and Consistent Trees \citep{Behroozi.et.al.2013b}, that identify dark matter subhalos and trace their evolutionary history. We present the details of the steps we use for finding the satellites in Appendix \ref{Appendix:tidal-index}.

To compare the impact of different satellites on bar formation in each of the FIRE-2 simulations, 
we first calculate $\Gamma$ for each satellite in each of the FIRE-2 simulations as function of time. Since \dhost\ varies as a function of time, so will $\Gamma$, which reaches its maximal value when a satellite is at pericenter or during merger. 

We select satellites for which  $\Gamma  > -2.45$ during at least one pericentric passage. This value for $\Gamma$ is based on work by
 \cite{Purcell2011}, who showed that a Sagittarius-like dwarf galaxy ($M_{sat} = 10^{10.5}$ M$_{\odot}$) at a pericenter of 30 kpc produced a pronounced bar in their isolated N-body simulations of a MW-like galaxy with $M_{host} = 3.59\times 10^{10} $ M$_{\odot}$. We take $R_{host} = R_{eff}= 4.77 $ kpc and use this to calculate $\Gamma$ since a rotation curve is not provided in the paper from which we could determine the peak velocity and radius. 
Whether a satellite induces bar formation also depends on other factors such as the kinematic temperature of the inner disk and the orbit of the satellite \citep[see e.g.,][]{Cavanagh2020}. Thus, the choice of $\Gamma$ in this Section is not meant as an ultimate lower limit, but is instead used to select satellites of interest in each of the FIRE-2 simulations. If we set a lower limit to the tidal index $\Gamma$ we will find more satellites but none of those seems to lead to bar formation within 10 dynamical times.  We explore the effects of varying this limit in $\Gamma$ in Appendix \ref{Appendix:tidal-index}. 

We consider a satellite with sufficiently high $\Gamma$ to be the trigger of bar formation if the bar begins forming within 10 dynamical times of its pericentric passage. We also calculate the angle between the satellite orbital angular momentum and the spin of the disk, $\cos\theta_{orbit}=\hat{j}_{orbit}\cdot\hat{j}_{\star}$ where $\cos\theta_{orbit}=+1$ for a perfectly prograde orbit, -1 for a perfect retrograde orbit, and 0 for a perfect polar orbit. Prograde satellite encounters are considered more likely to trigger bar formation \citep{Lokas.2018}, but we include all satellite orbits in our analysis for now. In Appendix \ref{Appendix:tidal-index}, we describe the detailed steps we follow to determine which satellites meet the criteria outlined above using the scaled tidal index and in Table \ref{table:satellite_properties} we summarise the properties of the satellites in FIRE-2 galaxies.

\begin{figure*}
\centering
\includegraphics[width=0.485\textwidth]{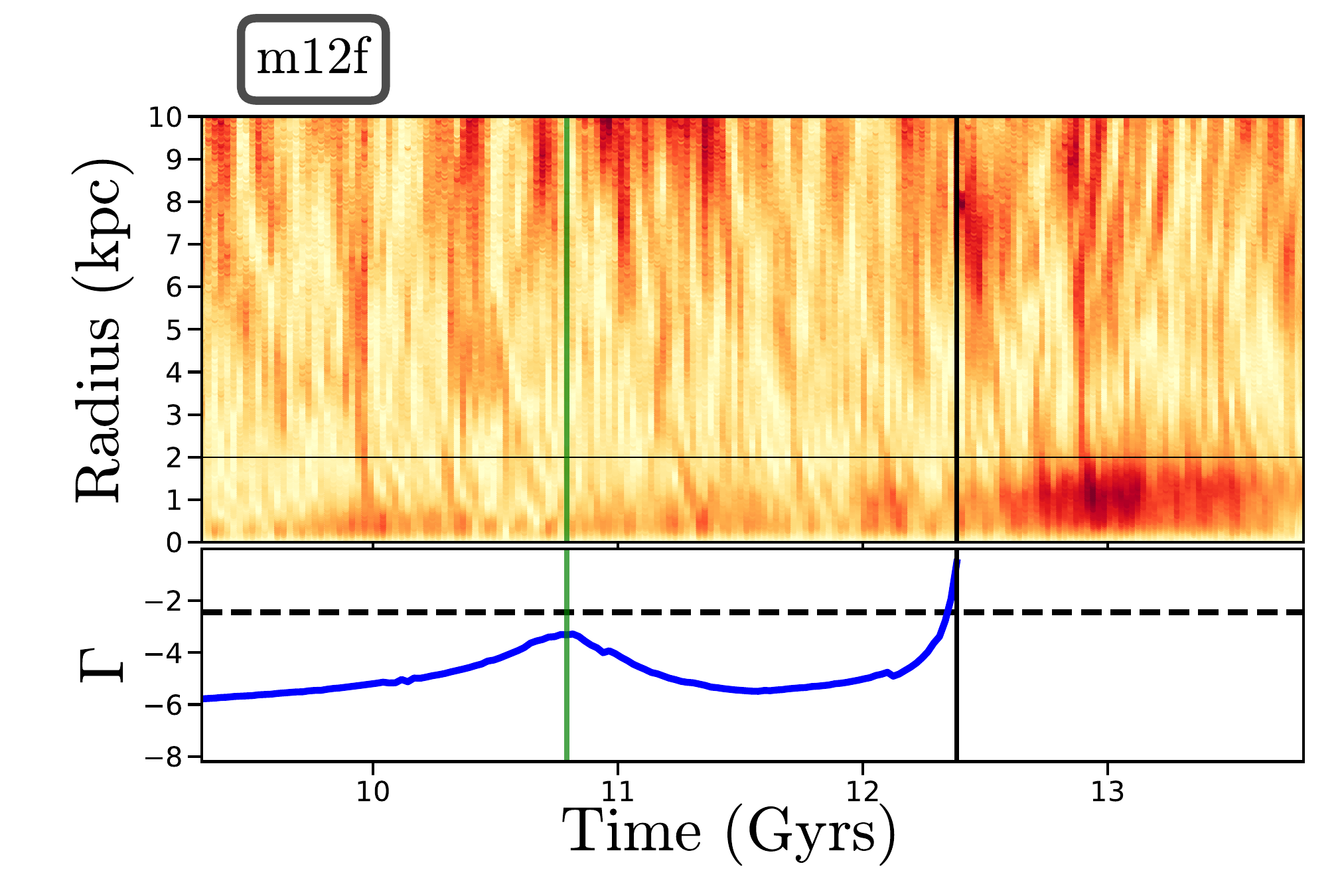}
\includegraphics[width=0.485\textwidth]{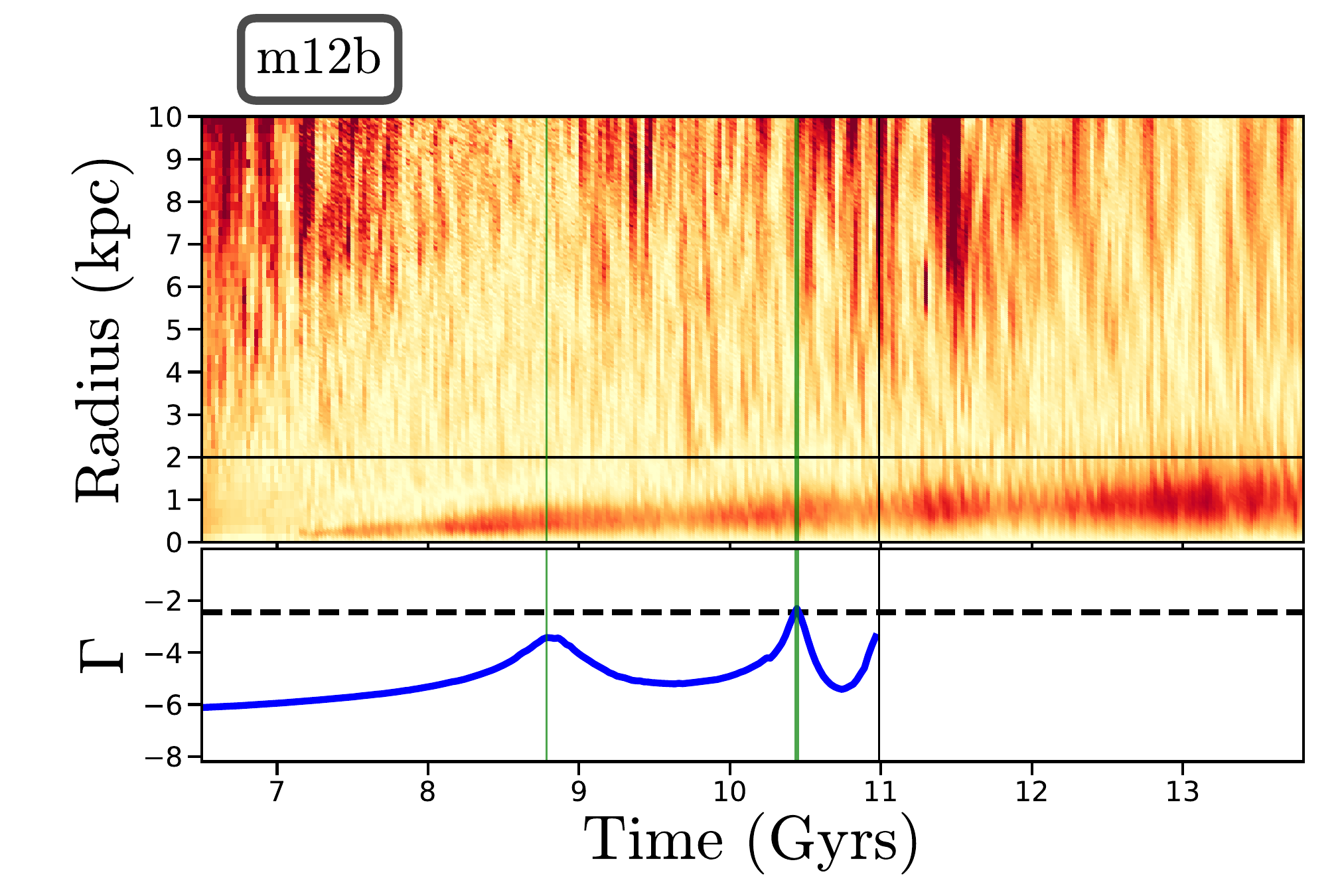} \\
\includegraphics[width=0.485\textwidth]{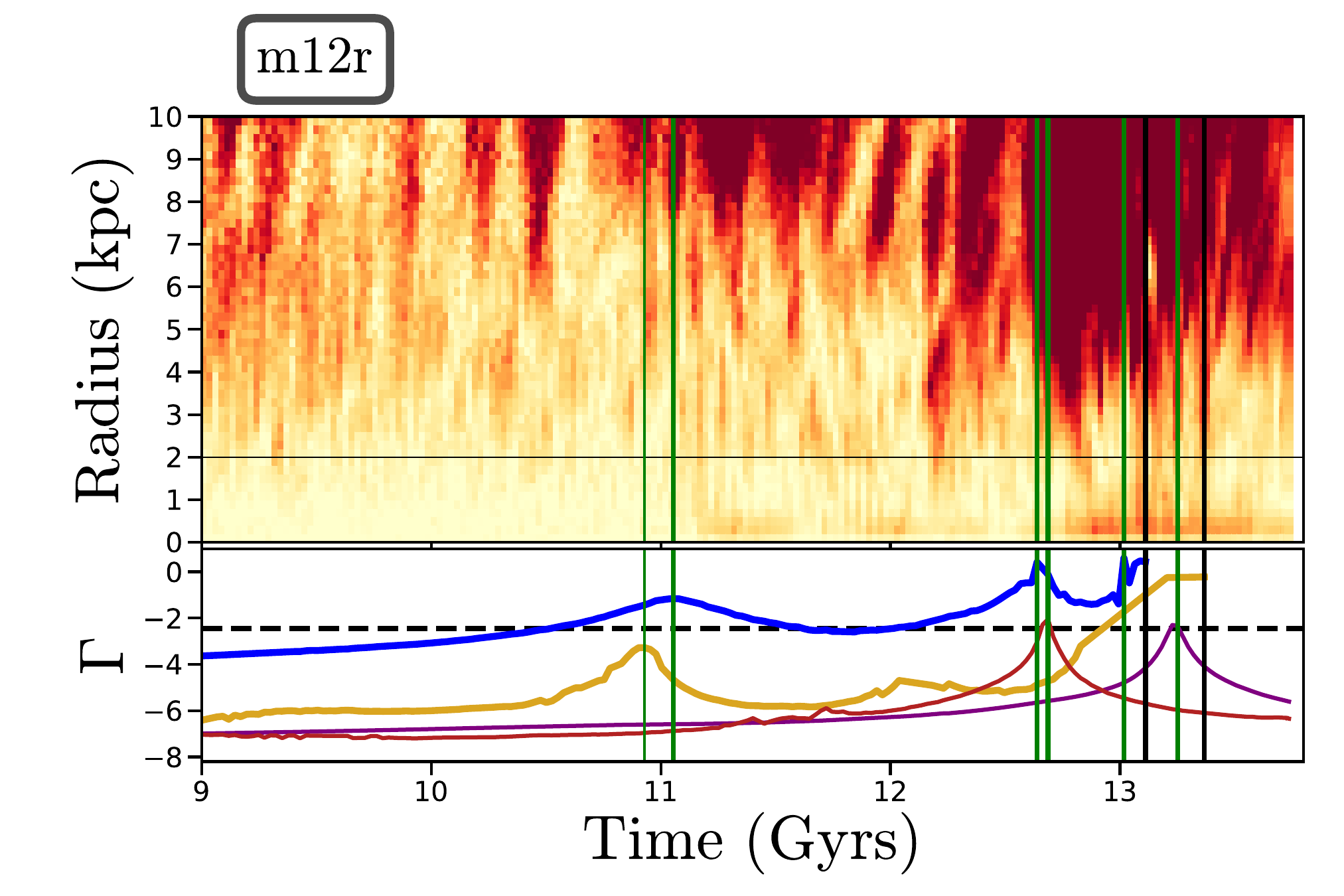}
\hspace{0.5cm} 
\includegraphics[width=0.14\textwidth]{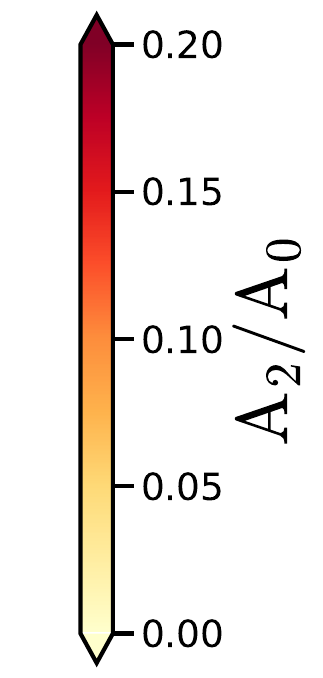}
    \caption{{\bf Bar formation due to satellite interaction in \textsf{m12f}, \textsf{m12b} and \textsf{m12r}.}  {\bf The upper half of all panels} show the evolution of the bar strength ($A_{2}/A_{0}$) with time (in Gyrs), at different radii (in kpc) of the stellar disk. The color bar shows the range of bar strength, $0.0<A_{2}/A_{0}<0.2$, where yellow is for the lower limit and dark red is for the upper limit. The dark red patch in the central region (radius $<2$ kpc) shows the presence of the bar. The dark red stripes at the outer part of the disk may be perturbations due to the interaction of different satellites (black lines for mergers and green lines for pericenter passage) or due to transient spiral arms in the disk. {\bf The lower half of each panel} shows the evolution of the tidal indices $\Gamma$ (y-axis; Equation \ref{eq:tidal_index}) of the satellites with $\Gamma>-2.45$ (dashed horizontal line; see Section \ref{sec:sat_interaction}) at least once during evolution. The x-axis is time (in Gyrs).  The blue thick curve of $\Gamma$ evolution shows the strongest satellite interaction in each system. In \textsf{m12r}, a satellite with a similar maximum tidal index as the blue one is shown in golden color. The thick vertical lines (black and green) represent the strongest tidal impact interactions, while the thin lines are for smaller tidal index satellites. The dashed vertical lines represent the retrograde motion of the satellite with respect to the stellar disk (see details in Table \ref{table:satellite_properties}). For more details on each system see Section \ref{sec:sat_interaction}.
    }
    \label{fig:satellite_interaction_examples}
\end{figure*}

In Figure \ref{fig:satellite_interaction_examples}, we show the three FIRE-2 simulations (m12f, m12b, and m12r) that had satellite encounters which met these criteria. In the top panels, we show the evolution of bar strength as a function of disk radius ($R < 10$ kpc) in time. The color bar represents the bar strength, $A_2/A_0$, for each simulation.
In the bottom panels we show the evolution of the scaled tidal index $\Gamma$ for each satellite that meets the criteria. The different colored lines indicate different satellites within each simulation. 
Consistent with prior work, all satellites that meet our criteria are on prograde or partially prograde orbits ($0<\cos\theta_{orbit}<1$).

All simulations in Figure  \ref{fig:satellite_interaction_examples} have bars in the central regions of the disks, which appear as dark red patches within $r<2$ kpc in the top panels. Similar dark red patches in the outer parts of the disks are either spiral arms or other deformations of the stellar density from an azimuthally symmetric distribution during the satellite interaction. 

Below, we discuss each of the cases in Figure \ref{fig:satellite_interaction_examples} in more detail. 

\textsf{m12f}: In Figure \ref{fig:satellite_interaction_examples}, one of the massive satellites in \textsf{m12f} (blue line in bottom panel) undergoes close pericentric passage with the disk and interacts for more than a Gyr before attaining a maximum tidal index of $\Gamma=-0.53$ at the time of merger at 12.4 Gyrs (the time when $A_2/A_0$ crosses 0.1).  This event forms the strongest bar among all the FIRE-2 galaxies. During the merger, the mass ratio of the stellar disk of the host to the satellite is $M_{\star, \rm host}/M_{sat}=0.5$ and this event causes a large perturbation in the outer disk (note the dark red patches in the outer disk that coincide with the black vertical line tracing the merger). These perturbations propagate to the inner disk. The bar grows stronger (dark patch within radius $r<2$ kpc, around 13 Gyrs ) after the merger. After 13.6 Gyr the bar weakens significantly and the bar strength dips below $0.1$ in the last 0.11 Gyr of evolution (see Figure \ref{fig:m12f_m12b_bar_duration}). 

\textsf{m12b}: The bar in \textsf{m12b} is lopsided (Figure \ref{fig1:all_bars}) and grows inside-out, maintaining its lopsidedness over a period of more than 5.7 Gyrs (Figure \ref{fig:satellite_interaction_examples}) even as it increases in size. See \citet{Lokas2021}, where the authors study lopsided bars in cosmological simulations. One of the massive satellites in \textsf{m12b}, which has $\rm M_{sat}=6.1\times10^{10}$ M$_{\odot}$ at pericenter with $\Gamma=-2.3$ (blue curve in Figure \ref{fig:satellite_interaction_examples}), interacts with the disk for more than 2 Gyrs, during which it undergoes two pericentric passages before merging with the disk at 10.9 Gyrs.  The rise and fall of the bar strength is in phase with the two pericentric passages. By the time the satellite ultimately merges with the disk, its mass has decreased to ${1/8}^{\mathrm{th}}$ the disk mass at $\rm D_{H}=14.8$ kpc, yielding $\Gamma=-3.39$ (see Table \ref{table:satellite_properties}), and is no longer massive enough to affect the bar.

The bar gains strength again via internal evolution and remains in the disk till the end of the simulation. Other than the above satellite, \textsf{m12b} undergoes multiple satellite interactions of lower $\Gamma$ between 7--12.5 Gyrs (not shown in Figure \ref{fig:satellite_interaction_examples}). As a result, large perturbations appear in the outer disk throughout its evolution.

\textsf{m12r}: Not all strong satellite interactions lead to the formation of a strong bar. Galaxy \textsf{m12r} undergoes multiple satellite interactions, two of which are on highly prograde orbits ($cos\theta_{orbit}\sim 0.9$) and have high tidal indices that disturb and distort the outer disk (thick blue line with $\Gamma_{max} =0.61$ and golden line with $\Gamma_{max}= -0.22$ in Figure \ref{fig:satellite_interaction_examples}; see also Table \ref{table:satellite_properties}).   
The pericentric passage of the satellite close to 12.6 Gyrs (in blue) raises $A_2/A_0$ in the outer disk ($r>3$ kpc) and increases the bar strength in the inner disk ($r<2$ kpc). This satellite's final pericentric passage at 13.11 Gyr has $\Gamma=0.43$, one of the highest values of any interaction in the simulation suite. Another satellite in a highly prograde orbit ($\cos \theta_{\mathrm{orbit}}=0.9$) with high tidal index ($\Gamma=-0.22$; shown in gold) merges with the disk at nearly the same time (13.37 Gyr). These strong tidal interactions with multiple satellites create large asymmetries in stellar density that are also observed in $A_{2}/A_{0}$. A weak, short bar is seen to form momentarily in the inner disk of \textsf{m12r} but is quickly dissipated under the strong tidal forces, consistent with results from N-body simulations that show that mass ratios above $\Msat/\Mhost \sim 0.1$ are not conducive to bar formation \citep[e.g.][]{Cavanagh2020}. In Section \ref{sec:kinematic_coldness} we explore further the transient nature of the bar in \textsf{m12r}.

\begin{deluxetable*}{ccccccccc}
\tablenum{4} \label{table:satellite_properties}
\tablecaption{Satellites in FIRE-2 galaxies}
\tablewidth{0pt}
\tablehead{
\colhead{Name} & \colhead{$\frac{A_{2}}{A_{0}}_{\rm max}$} & \colhead{$\log_{10}(\rm \frac{M_{sat}}{M_{\odot}})$} & \colhead{D$_{\rm H}$} & \colhead{$\rm \frac{M_{\star, host}(r< D_{H})}{M_{sat}}$}  & \colhead{Tidal index} & \colhead{ $\cos\theta_{orbit}$  } & \colhead{Time of merger (M)} & \colhead{Time of} \\
\colhead{} & \colhead{} & \colhead{}  & \colhead{(kpc)} & \colhead{}  & \colhead{$\Gamma$} & \colhead{ } & \colhead{or pericenter (P) (Gyr)} & \colhead{$\frac{A_{2}}{A_{0}}=0.1$ (Gyr)} }
\decimalcolnumbers
\startdata
\multicolumn{3}{c}{Simulations with bars affected by satellites} \\
\hline
m12f & 0.203 & 10.97  & 4.44  &  0.50  &  -0.53  &  0.99   & 12.38 (M)    &   12.38   \\
m12b & 0.182 &  11.21  & 39.1 & 0.36 & -3.42 & 0.07  & 8.78 (P) & 8.06  \\
    & &  10.79  &  12.01 &  1.08   &  -2.3  & 0.92  & 10.44 (P)  &   \\
    & &  9.96  &  14.81 &  7.6   &  -3.39  & 0.84  & 10.96 (M)  &   \\
    & &  9.29  &  12.16 &   40.5   &  -3.81  & -0.94  & 12.43 (P)  &  \\
m12r & 0.133 &  9.63  & 5.17  &  2.29   &  -0.22  & 0.94  & 13.37 (M) &  12.92  \\
           & &  10.69 & 7.13  &  0.18   &  0.42  & 0.99 & 12.64 (P)   &      \\
           & &  10.47 & 5.21  &  0.29   &  0.61  & 0.88 & 13.01 (P)   &      \\
           & &  10.78 & 7.54  &  0.19   &  0.43  & 0.94 & 13.11 (M)   &      \\
\hline
\multicolumn{5}{c}{Simulations with bars not significantly affected by satellites} \\
\hline
Romeo & 0.144 &  8.42  &  8.7 &  187  & -3.52   &   -0.60  & 10.24 (P)  &   12.64   \\
      & &  9.1  &  17.08 &  50.11  & -3.71   &   -0.04  & 10.92 (P)  &    \\
Remus & 0.113 & 10.23 & 26.34 & 6.36  & -3.99  &  0.19  & 13.64 (P)  & 12.22 \\
      &       & 9.56 & 20.55 & 22.4   & -4.34  &  0.22  & 10.65 (P)  & \\
m12m & 0.112 & 8.9 & 18.54 & 129 & -3.45 & -0.42 & 11.64 (P) & 12.96 \\
     &       &  9.2  &  56.26 &  76.6 &  -4.61  & 0.33  & 12.77 (P) &       \\
m12w & 0.163 &  7.14   & 6.66  & 3206  & -4.47   &  0.39  & 12.73 (P)  &  13.66     \\
m12c & 0.123 &  11.02  & 18.12  &  0.58 &  -1.87   &  0.79   & 12.89 (P)  &  13.73   \\
      &      &  7.81  &   5.85  &  699 &  -3.61  &  -0.22   & 13.78 (M)  &      \\
\hline
\multicolumn{3}{c}{Simulations with weak bar-like perturbations} \\
\hline
m12i & 0.092  & 9.58   & 29.99  &  9.1  & -4.46  &   -0.73  & 9.86 (P) & --  \\
     &        & 9.08   & 10.43  &  25.48  & -3.57  &   -0.53  & 7.79 (P) &      \\
Thelma &  0.082 &  10.53  & 7.26  & 0.49  &  0.79  &   0.89  & 8.01 &  --    \\
       &        &  9.80  & 3.38  & 1.58  & 1.06   &   0.06  & 8.11 &     \\
\hline
\multicolumn{2}{c}{Simulations without bars} \\
\hline
Romulus & 0.07 &  10.18  & 41.41  & 4.27   &  -5.26  &    & 7.74 &  -- \\
        &      &  9.14  & 8.87  &  25.65  &  -4.29 &    & 6.61  &    \\
Louise  & 0.06 &  10.29  & 5.12  &  0.93  &   2.33 &    & 13.48 &  --  \\
Juliet  & 0.05 & 10.23  & 35.45  & 2.46  &  -4.43 &    & 13.10  &  -- \\
        &      & 8.22  &  6.99 &  139.3  &   -4.32 &    & 8.32 &      \\
\enddata
\tablecomments{Summary of the massive ($M_{halo}>10^7 M_{\odot}$) satellite interactions with varied tidal index $\Gamma$ in the FIRE-2 galaxies that induce bar formation in some galaxies during merger or pericenter passage and in some galaxies do not. The table has 4 parts with the galaxies separated by horizontal lines: (1) galaxies that undergo interactions with satellites with $\Gamma>-2.45$ at pericenter and form bars (\textsf{m12f}, \textsf{m12b}, and \textsf{m12r}), (2) galaxies that form bar but not due to satellite interactions (\textsf{Romeo}, \textsf{Remus}, \textsf{m12m}, \textsf{m12w} and \textsf{m12c}), (3) galaxies that undergo strong interactions (not during growth of $A_{2}/A_{0}$) but only form weak bar-like perturbations (\textsf{m12i} and \textsf{Thelma}), and (4) galaxies that do not form bars, but have strong bulges that hinder bar formation even with strong satellite interactions. Columns: 1. name of the simulation; 2. peak bar strength; 3. M$_{200}$ mass of satellite in log scale; 4. distance of the satellite from host galaxy center (at pericenter or merger); 5. mass ratio of the satellite and the host galaxy within a radius of $\rm D_{H}$ from the host center; 6. the tidal index $\Gamma$ at pericenter or merger (Equation \ref{eq:tidal_index}; see Section \ref{sec:sat_interaction}); 7. cosine of the angle between angular momentum vector of the satellite orbit and the disk angular momentum, 8. time of the merger (M) or pericenter passage (P) of the satellite; 9. time when the bar strength crosses $A_{2}/A_{0}=0.1$, i.e., when the bar forms. If the time interval between the time of the interaction (merger or pericenter passage) and the time of bar formation is greater than 10 dynamical times, we do not associate bar formation with satellite interaction. 
}
\end{deluxetable*}

\begin{figure*}
\centering
	\includegraphics[width=0.45\textwidth]{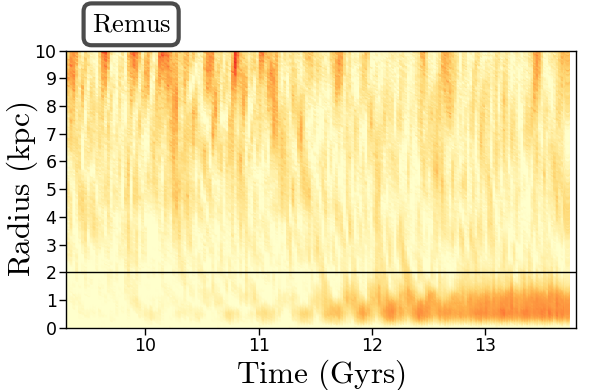}
	\hspace{1cm}
	\includegraphics[width=0.45\textwidth]{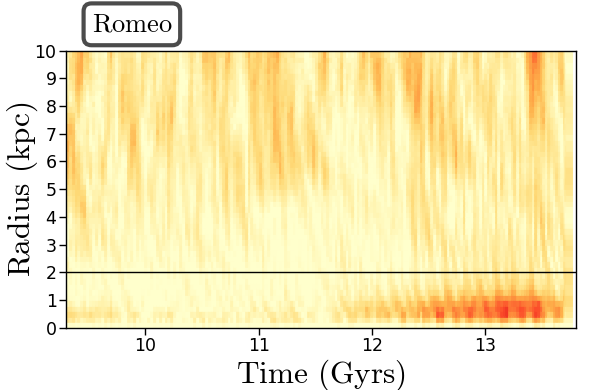} 
	\vspace{0.0cm}\\
 	\includegraphics[width=0.45\textwidth]{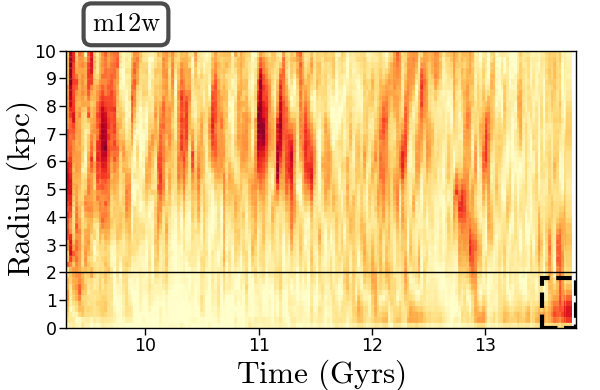}
	\hspace{1cm}
	\includegraphics[width=0.45\textwidth]{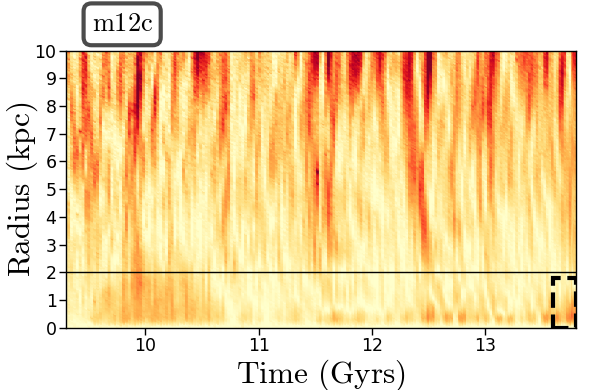} 
	\vspace{0.0cm}\\
	\includegraphics[width=0.55\textwidth]{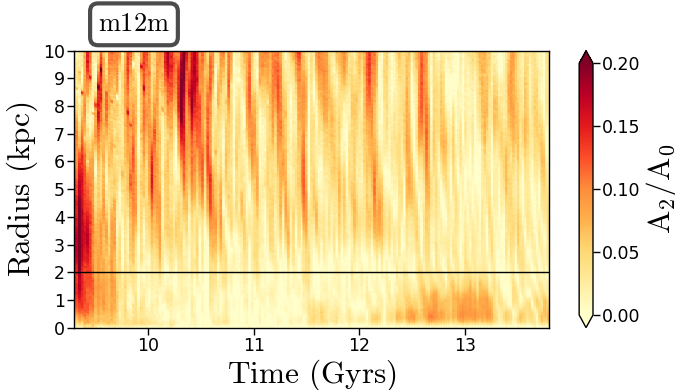}
    \caption{{\bf Bar formation during internal evolution of the disk in \textsf{Remus}, \textsf{Romeo}, \textsf{m12w}, \textsf{m12c} and \textsf{m12m}}. Similar to Figure (\ref{fig:satellite_interaction_examples}), the dark patch in the central region (radius $<2$ kpc) shows the presence of the bar. We highlight the bars in \textsf{m12w} and \textsf{m12c} with dashed rectangles as their time duration is short. The dark red stripes at the outer part of the disk are perturbations. In these five cases either the satellite interactions are less impactful (seen from the very calm outer disk for \textsf{Remus} and \textsf{Romeo}), or time between the satellite interaction and rise of the bar strength above 0.1 is more than 10 dynamical times (\textsf{m12w}, \textsf{m12c} and \textsf{m12m}). However, bar instabilities arise in the disk.  In Appendix \ref{sec:appendix:m12m-satellite-interactions} we present the satellite interactions that have failed to trigger bar instability in \textsf{Remus}, \textsf{Romeo}, \textsf{m12w}, \textsf{m12c} and \textsf{m12m}.}
    \label{fig:secular_evolution_examples}
\end{figure*}

\subsection{Bar Formation from disk internal Evolution}
\label{sec:secular_evolution}
Five of the FIRE-2 galaxies host bars that do not form through any kind of satellite interaction (merger/pericenter), but during internal evolution of the disk (Table \ref{table:satellite_properties}). In addition, only some of the bars in other galaxies are instantiated by a satellite passage. In Figure \ref{fig:secular_evolution_examples} we present the evolution of the bar strength $A_2/A_0$ with radius and time for the galaxies \textsf{m12m}, \textsf{Remus}, \textsf{m12w}, \textsf{m12c} and \textsf{Romeo}. The bar length and strength (reddish color) in the central regions of the disks ($r<2$ kpc) appears to oscillate at several frequencies, and the outer regions of the disks in \textsf{Remus} and \textsf{Romeo} are much less perturbed compared to the disks hosting bars that form through satellite interactions in Figure \ref{fig:satellite_interaction_examples}. This is unsurprising since these galaxies have far fewer satellite interactions than those hosting bars triggered by satellites, and of those even fewer have high tidal index $\Gamma$ (Table \ref{table:satellite_properties}). None of the satellites in these systems meet the two criteria for bar formation outlined in Section \ref{sec:sat_interaction} simultaneously. For example, one of the satellites in \textsf{m12c} has a pericenter passage with tidal index $\Gamma=-1.87$, which fits the first selection criterion, but a bar does not form for more than 20 dynamical times after this interaction so we consider it uncorrelated with bar formation.

In the absence of satellite interactions, the bar strength grows from the center outward, showing oscillations in $A_{2}/A_{0}$ at multiple frequencies. Once formed, the bar persists until the end of the simulation for \textsf{Remus}, \textsf{m12w} and \textsf{m12c} but dissolves for \textsf{Romeo} and \textsf{m12m}. The bar in \textsf{m12w} appears to form in somewhat unique circumstances, as gas is expelled from the central region during an epoch of high star formation rate. As the gas content decreases, a rectangular bar arises in the disk at 13.6 Gyr and stays till the end of the simulation.

All five bars in Figure \ref{fig:secular_evolution_examples} have lower $A_{2}/A_{0,\mathrm{max}}$ than the bars in Figure \ref{fig:satellite_interaction_examples} formed through satellite interactions. It is complicated to directly predict the $\mathcal{R}$ ratio of a bar that is satellite-induced, compared to a bar that forms during internal evolution \citep{Miwa.and.Noguchi.1998,Lokas.et.al.2016, Martinez-Valpuesta.et.al.2017}, however here we find that bars formed through satellite interaction have lower $\mathcal{R}$ ratios ($\rco/a_{s}$) than bars formed through internal evolution (see Table \ref{table_bar_properties}).

Given that FIRE-2 galaxies can clearly form bars through internal processes, we now consider which properties of the central region create favourable conditions for bar formation to occur via internal evolution. Are the intrinsic properties of the galactic disk and dark matter halo playing any role that makes it more likely for bars to form in some disks rather than others, or in the eventual length and strength of the bar? To address these questions, we looked at a variety of disk properties for the simulations that form bars through internal processes: the central density and concentration of stars, gas and DM (\S \ref{sec:mass_profile}); the kinematic coldness of the disk (\S \ref{sec:kinematic_coldness}); and the role of gas fraction and stellar feedback (\S \ref{sec:gas_fraction_stellar_feedback}).

\subsubsection{The central density profile} \label{sec:mass_profile}
\begin{figure*}
\centering
\includegraphics[width=\textwidth]{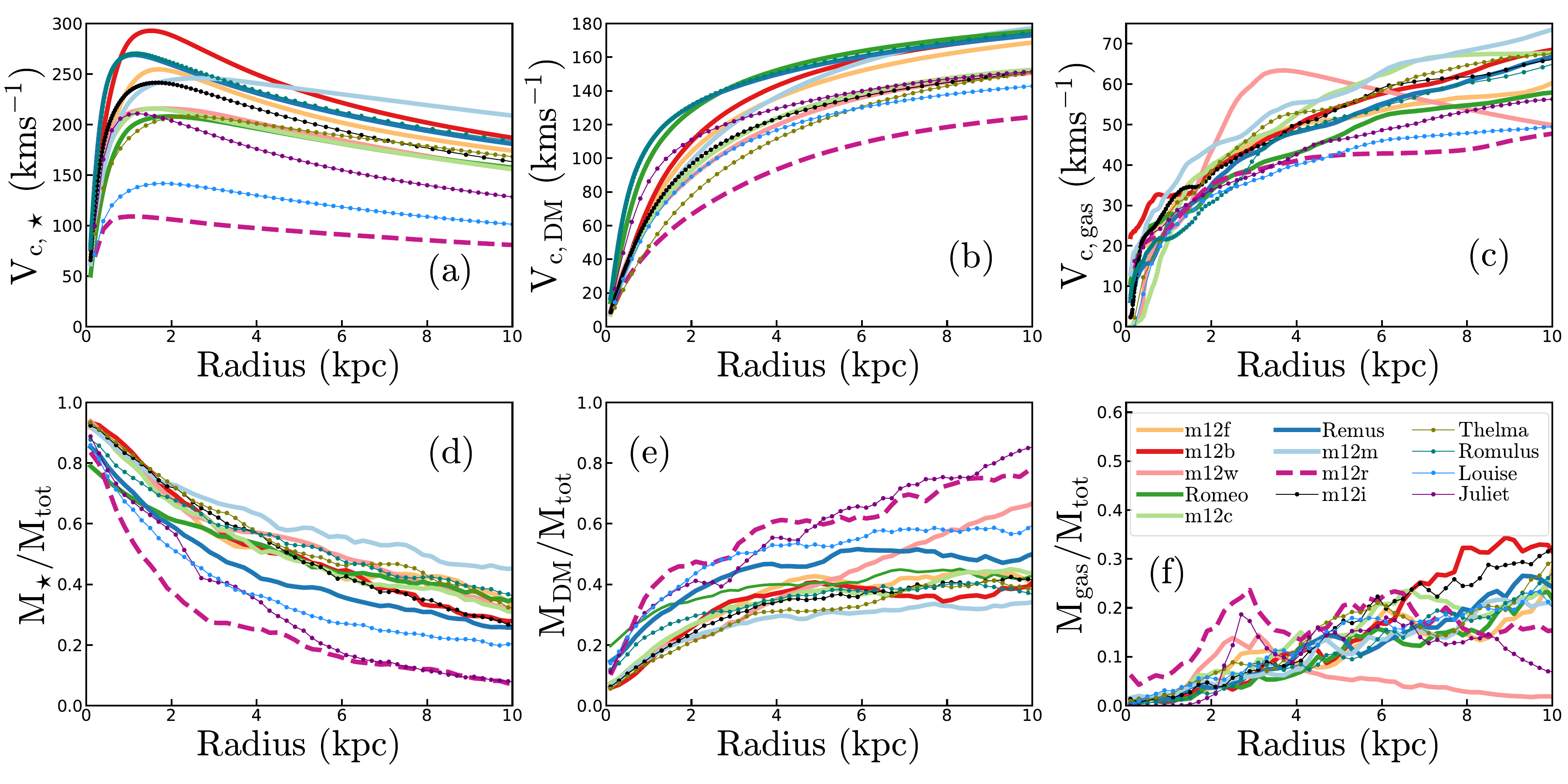}
    \caption{{\bf Rotation curves and mass fractions.} The top row (panels a--c) shows rotation curves (Equation \ref{eq:circular_velocity}) of all the FIRE-2 galaxies at their peak bar strength for stars (a), gas (b) and DM (c); the bottom row (panels d--f) shows relative mass fractions of different species within $|z| \leq 2$ kpc for the same species. All quantities are plotted as a function of spherical radius $r$ (not cylindrical $R$). Thick lines represent the barred galaxies, thin lines with circular markers represent unbarred galaxies. The thick dashed line is for \textsf{m12r}, which is significantly less massive than all other galaxies. Note that the vertical axis range is different for each component.}
    \label{fig:Vc_stars_dm_gas}
\end{figure*}

We examine the central mass profiles of stars, gas and DM in the FIRE-2 disks to investigate their effect on bar formation, since the relative distribution of stars, gas and DM in the central galaxy has a significant effect on bar properties \citep{Athanassoula.2002}. \citet{Athanassoula.and.Misiriotis.2002} show that stronger, more rectangular bars form in more dark matter-dominated disks, provided there is interaction between the DM halo and the disk \citep{Athanassoula.2002}. Furthermore, a standard condition for swing amplification of a bar instability is a rising central rotation curve that is dominated by the stellar circular velocity at the peak and falls to the circular velocity of the DM halo at higher radii \citep{Efstathiou.et.al.1982,Debattista.and.Sellwood.1998, Marioni.et.al.2022}. 

In Section \ref{sec:FIRE-2_sims} we presented the total circular velocity profiles of all the FIRE-2 galaxies (see Figure \ref{fig:all_rotation_curves}). In Figure \ref{fig:Vc_stars_dm_gas}, we present the rotation curves and mass profiles in the inner 10 kpc of the FIRE-2 galaxies separated by species (stars, gas, and DM). 
The top panel of Figure \ref{fig:Vc_stars_dm_gas}, which shows the rotation curves, shows that the barred and unbarred galaxies in FIRE-2 have sharply rising stellar circular velocity curves (panel a) that dominate over the DM (panel b) and gas (panel c) components, in agreement with bar instability theory. For example, in \textsf{m12f} the circular velocity of the DM component rises slowly and matches the circular velocity of the stellar component at a radius of $\sim 10$ kpc. \textsf{m12f} has $\rm R_{90}=11.8$ kpc (Table \ref{table:FIRE-2_galaxy_summary}) and within nearly this entire region the mass distribution is dominated by the stellar component. The other FIRE-2 galaxies are similarly baryon-dominated at the center. 

In the bottom row of Figure \ref{fig:Vc_stars_dm_gas}, we show the mass fractions of the different components with radius. Similar to the results from the circular velocity curves (top row), we find that FIRE-2 galaxies have more stars than DM in their centers (panels d and e). However, we do not see any prominent difference in mass fractions between the barred and unbarred galaxies in FIRE-2. Hence the barred and unbarred galaxies have similar proportions of their mass in stars, DM and gas. Interestingly, the central gas mass fraction contributes significantly to the rotation curve (panels c and f), which is thought to suppress bar formation (\citealt{Masters.et.al.2012}; see \S \ref{sec:gas_fraction_stellar_feedback}).

We also compare bar length to the radius corresponding to the peak of the circular velocity curves ($\rm R_{max}$). The peak circular velocity for the barred galaxies is in the range $V_{max}\sim 147-316$ kms$^{-1}$ and the corresponding radius $R_{max}\sim1.5-9.8$ kpc, with \textsf{m12r}, the lowest stellar mass galaxy, having the lowest peak velocity $V_{max}=147$ kms$^{-1}$ and largest radius $R_{max}=9.8$ kpc in our sample. Nearly all the bars have $R_{max}> a_{s}$; in other words, they lie within the rising part of the total circular velocity curve (except \textsf{Remus}; see Table \ref{table:FIRE-2_galaxy_summary} and Table \ref{table_bar_properties}) Furthermore, the peak circular velocity ($\rm V_{max}$) is greater than the maximum circular velocity of the DM halo in all systems except for \textsf{m12r}, another classic criterion for bar formation \citep{Efstathiou.et.al.1982, Debattista.and.Sellwood.1998}, although not a sufficient criterion \citep{Efstathiou.et.al.1982, Marioni.et.al.2022}.

\begin{figure}
\centering
\includegraphics[width=\columnwidth]{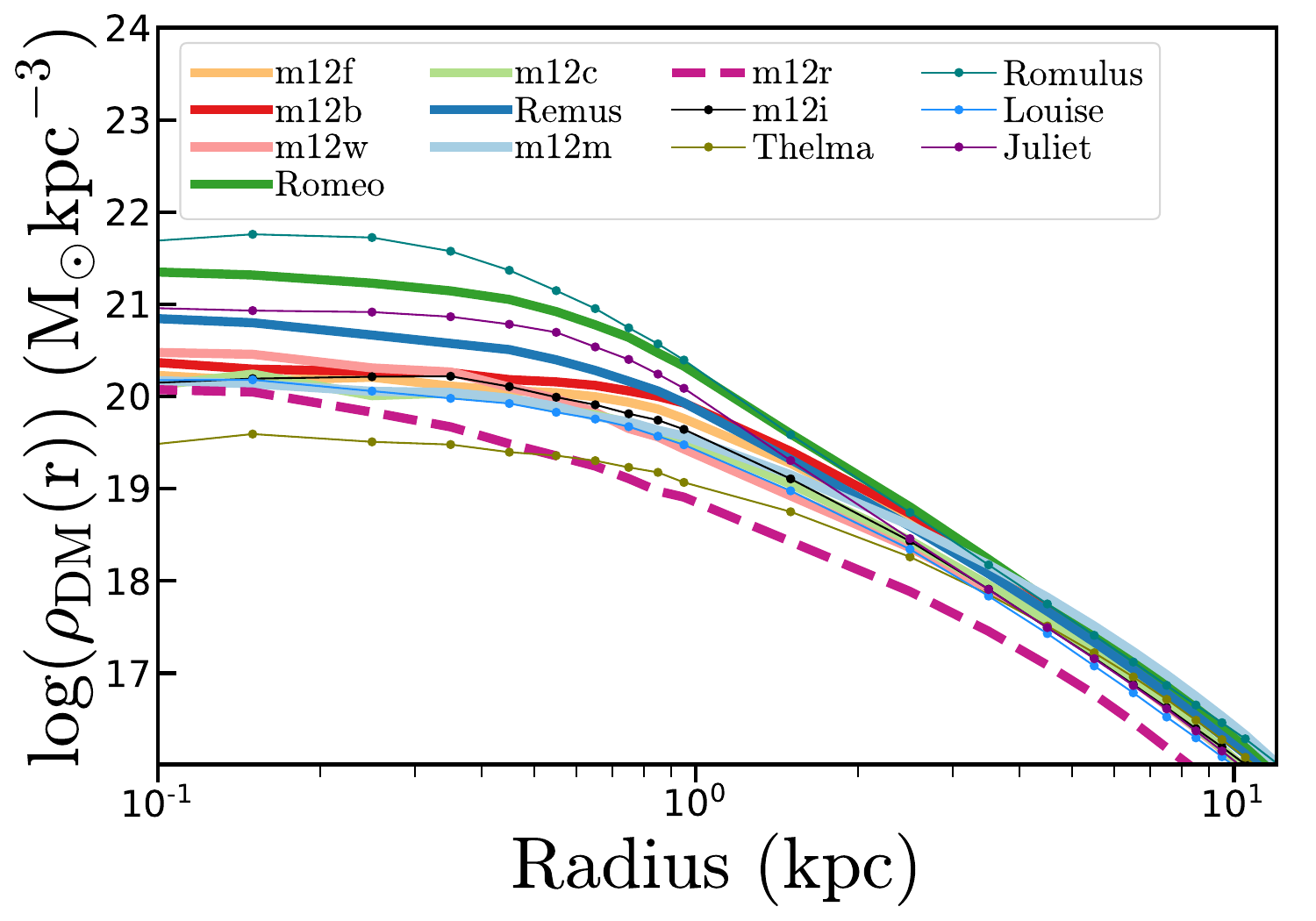} 
    \caption{{\bf Cores in the DM density profile at peak bar strength.} Thick lines represent barred galaxies, thin lines show unbarred galaxies and the thick dashed line shows \textsf{m12r} (one of the lowest DM densities). The central density profiles are similar for barred and unbarred systems. }
    \label{fig:dm_density}
\end{figure}

Additionally, we investigate the density profile of the DM component of the barred and unbarred galaxies in FIRE-2 to examine the role of the DM density in the central region close to the disk (also see \citealt{Lazar.et.al.2020}). Figure \ref{fig:dm_density} shows the DM density profile $\rho_{\rm DM}(r)$ (in $\rm M_{\odot} kpc^{-3}$, in log scale) as a function of radius (in kpc) of the FIRE-2 galaxies, using the same color scheme and line-style as Figure \ref{fig:all_rotation_curves}. Figure \ref{fig:dm_density} shows that the barred and unbarred galaxies both have flat or ``cored'' central density profiles with a range of maximum densities. There is no distinct difference between the barred and unbarred galaxies, which means that the DM density profile is not an indicator of barred/unbarred disk morphology in this sample. However, preliminary studies using N-body simulations show that cored DM profiles are less likely to form strong bars compared to cuspy DM profiles (Ansar et al. in prep).

The slow-rising DM component of the rotation curve in most of the FIRE2 galaxies can explain why bars formed in FIRE-2 galaxies are more rounded and less rectangular, similar to the idealized N-body model MD in \citet{Athanassoula.and.Misiriotis.2002} which forms a rounded bar in a stellar disk with similar mass and a slow-rising DM component. The more rounded bar morphology also results in lower bar strength $A_{2}/A_{0}$, which shows a higher amplitude for rectangular bars. However, \emph{all} the FIRE systems, both barred and unbarred, fit many of the classic criteria for bar formation, with few obvious differences between the two groups, and all have similarly cored DM profiles that have previously been associated to weaker bars. Thus the relative distribution and concentration of the stars, gas, and DM can explain the shape, and perhaps the weakness, but not the origin of bars in FIRE.

Round-shaped bars have been found in the 3.6 micron Spitzer images of barred galaxies \citet{Erwin2023}. Even though we do not conduct synthetic observations, the old stellar population observed in 3.6 micron images can be compared to the bars in the simulations. The bar major-axis stellar surface density profiles of the FIRE-2 galaxies are similar to the “two-slope” and “flat-top” type of bar profile in \citet{Erwin2023} (see Figure \ref{appendix:fig:stellar_surfacedensity} in Appendix \ref{sec:appendix:stellar_surfacedensity}).

\begin{figure*}
\centering
    \includegraphics[width=\textwidth]{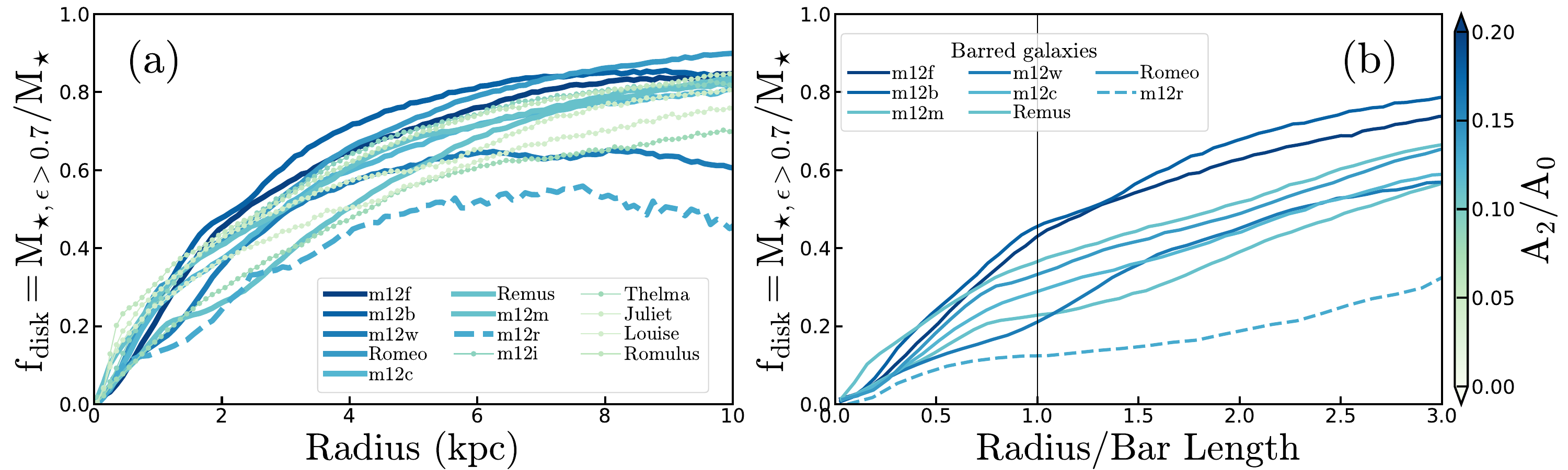} 
    \caption{{\bf Kinematically colder disks host strongly barred galaxies.} Higher $f_{disk}$ shows the large fraction of star particles in rotational orbits in the disk plane, which is a proxy for a kinematically cold disk that is prone to bar instability. {\bf Panel (a):} $f_{disk}(r)$ profile of all FIRE-2 galaxies at their peak bar strength, with the annular bin size of $\Delta R=0.1$ kpc. {\bf Panel (b):} The same $f_{disk}(r)$ profile for the barred galaxies at their peak bar strength, with the x-axis scaled by the bar length. The black vertical line denotes the edge of the bar for all the galaxies. In panel a, thick lines are for barred galaxies and thin lines with solid circles for unbarred galaxies. In both panels, the galaxies are color coded with the value of their peak bar strength from the color bar of $A_{2}/A_{0}$. Strongly barred galaxies (for example, \textsf{m12f} and \textsf{m12b}) are kinematically colder compared to weakly barred ones (for example, \textsf{m12m} and \textsf{Romeo}). The dashed line in both panels is for \textsf{m12r}, which has very different properties than the rest of the galaxies.  }
    \label{fig:fdisk_all_gal_barred_gal}
\end{figure*}

\subsubsection{The kinematic coldness of the disk} \label{sec:kinematic_coldness}

In idealized disk galaxy simulations, ``kinematically cold'' disks with a high fraction of stars in rotational motion in a common plane are the most susceptible to the bar instability \citep{Ostriker.and.Peebles.1973}. To determine whether this is so in our simulations we measure the fraction of stellar mass that contributes to rotationally supported orbits in near circular motion in the disk plane, $f_{disk}$ as a proxy for kinematic coldness. To estimate $f_{disk}$, we calculate the circularity $\epsilon=j_{z,i}/j_{circ,i}(E_{i})$ (defined in \citealt{Abadi.et.al.2003}) for each star particle $i$ in the disk with cylindrical radius $R<10$ kpc and $|z| \leq 2$ kpc. Here, $j_{z, i}=\textbf{j}_{\textbf{z,i}} \cdot \textbf{j}_{\textbf{net}}/|\textbf{j}_{\textbf{net}}|$ is the component of the specific angular momentum of the $i^{th}$ star particle towards the direction of the total specific angular momentum $\textbf{j}_{\textbf{net}}$ of the disk (also see \citealt{El-Badry.et.al.2018}), and $j_{circ,i}(E_{i})$ is the specific angular momentum of the circular orbit with the same energy $E_{i}$ as the star particle. The maximum specific angular momentum an orbit can have is for a circular orbit, so $\epsilon$ lies between $\pm 1$, with $+1$ indicating a prograde circular orbit and $-1$ for a retrograde circular orbit in the galaxy disk's plane. A value of $\epsilon\sim0$ can represent either a circular orbit nearly perpendicular to the disk plane or a radial orbit. 

The energy $E_{i}$ can be written in terms of the radius $R_{c,i}$ of the circular orbit in the plane of the disk with the same energy as star particle $i$: 
\begin{equation}
    E_{i}= \frac{G M(r< R_{c,i})}{2 R_{c,i}} + \Phi(R_{c,i}, 0);
\end{equation}
the corresponding specific angular momentum is 
\begin{equation}
    j_{circ}(E_{i})= R_{c,i} V_{c,i}(R_{c,i})= \sqrt{G M(r< R_{c,i}) R_{c,i}}.
\end{equation}
Here $\Phi(R_{c,i},0)$ is the total potential due to all components evaluated at radius $R_{c,i}$ in the disk plane. For each star particle we calculate $E_{i}$, $R_{c,i}$, and $j_{circ,i}(E_{i})$ to obtain $\epsilon_{i}$ for each star and calculate the mass fraction of stars with circularity greater than 0.7; that is, $f_{disk}(r)=M_{\star,\epsilon>0.7}/M_{\star}$, in annular radial bins of width $\Delta R=0.1$ kpc.

Figure \ref{fig:fdisk_all_gal_barred_gal} shows $f_{disk}$ as a function of disk radius $R$, for all the galaxies in our FIRE-2 sample at $z_r=0$ (panel a) and separately for the barred galaxies only (panel b) at their peak bar strength, which is at a different time for each galaxy. The line color in panel b is proportional to maximum bar strength. 
Overall, kinematically colder disks with high $f_{disk}$ and galaxies with strong satellite interactions (for example, \textsf{m12f} and \textsf{m12b}) form stronger bars than the cold disks without any strong satellite interactions (for example, \textsf{m12m}, \textsf{Remus} and \textsf{Romeo}) in the FIRE-2 galaxy sample.
The central regions of the disks are also clearly kinematically hotter than the outer disks in all the galaxies.

\begin{figure*}
\centering
\includegraphics[width=0.9\textwidth]{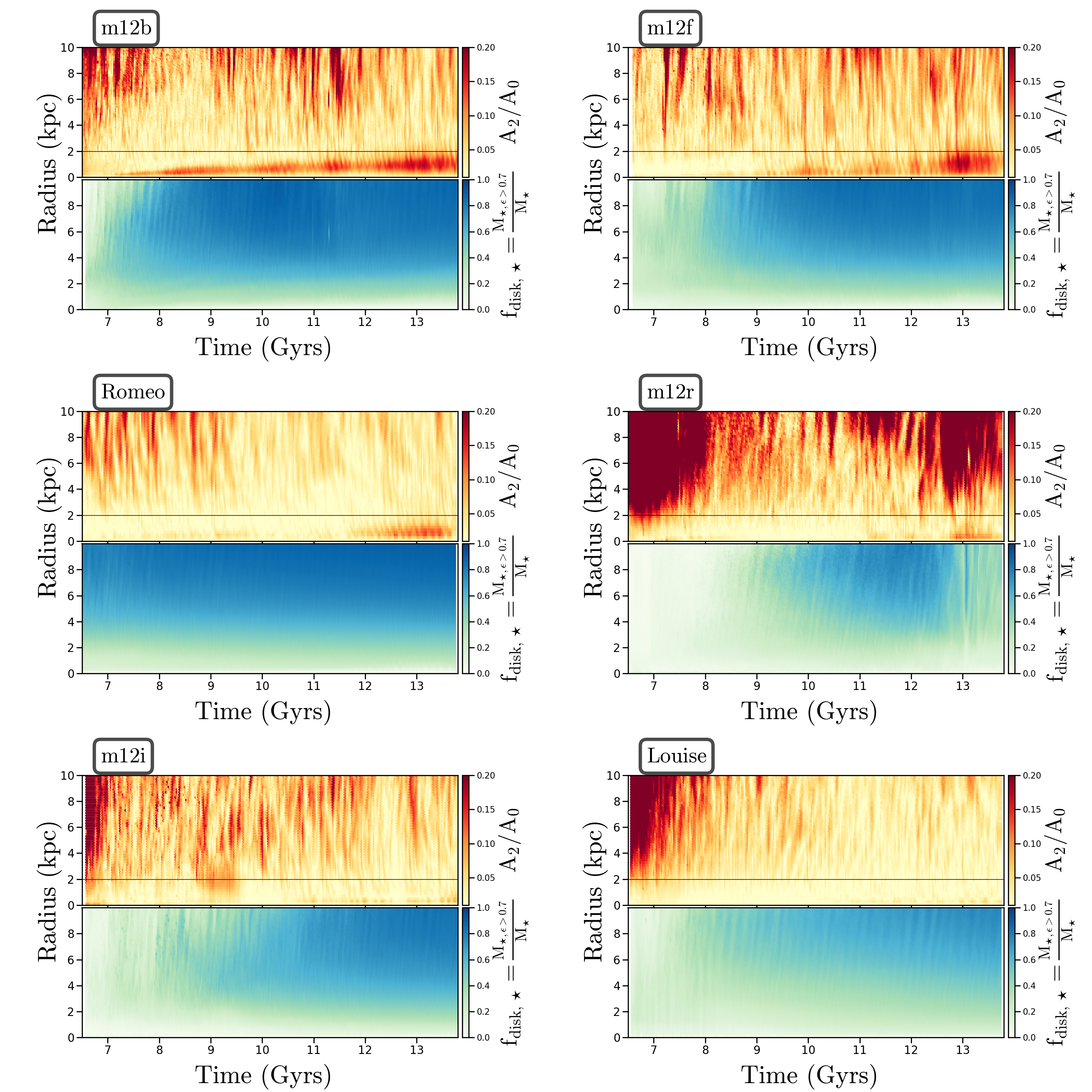}
\caption{ {\bf Stronger bars in kinematically colder disks}. The history of bar strength ($A_{2}/A_{0}$) evolution (top panels for each simulation with color bar for bar strength: $0.0<A_{2}/A_{0}<0.2$) along with kinematic coldness parameter $f_{disk}(r)=M_{\star,\epsilon>0.7}/M_{\star}$ of the stellar disk (bottom panels for each simulation with color bar: $0.0<f_{disk}<1.0$) for all the barred galaxies in FIRE-2. In the lower half of all panels, the higher the $f_{disk}$ value, the colder is the disk. Lighter shades of grey show the high $f_{disk}$, darker shaded of blue show very low $f_{disk}$, while different shades of orange, green and purple have intermediate values of $f_{disk}$. This Figure shows examples of bars forming during satellite interactions (\textsf{m12f} and \textsf{m12b}) and internal evolution (\textsf{Romeo}) in kinematically cold disks, a weak, short and transient bar in kinematically hot disk (\textsf{m12r}), a failed bar (\textsf{m12i}) and an unbarred galaxy with a bulge and kinematically hot disk (\textsf{Louise}).
}
\label{fig:kinematic_coldness}
\end{figure*}

To establish a causal connection between the kinematic coldness of the stellar disk and bar strength, we consider the evolution in time of the bar strength and kinematic coldness at different radii for all 13 FIRE-2 galaxies. We expect the bar strength $A_2/A_0$ to be anticorrelated with the coldness parameter $f_{disk}$, with the bar growing in strength after the disk becomes cold. 
We find that the disks in the FIRE-2 galaxies have a broad range of kinematic temperatures. 
In Figure \ref{fig:kinematic_coldness}, we present some instructive examples across this range; similar plots for all 13 galaxies are in Figures \ref{appendix:fig:kinematic_coldness} and \ref{appendix:fig:kinematic_coldness_unbarred} of Appendix \ref{sec:appendix:kinematic_coldness_history}. The color scheme for the bar strength is the same as in Figures \ref{fig:satellite_interaction_examples} and \ref{fig:secular_evolution_examples}. For $f_{disk}$ we use darker shades of blue for kinematically cold regions of the disk, while lighter shades of blue and green to show the kinematically hot regions (see Figure \ref{fig:kinematic_coldness}).

\textsf{m12b} and \textsf{m12f} (top row of Figure \ref{fig:kinematic_coldness}) are the strongest barred galaxies in the sample, and both have very kinematically cold stellar disks. The bar in \textsf{m12b} begins to form around 7.5 Gyr, just as the disk has become significantly cooler. When \textsf{m12f} reaches a similar coldness, around 9.5 Gyr, a bar also begins to grow, although it does not get very strong until a satellite interaction around 12.7 Gyr. 
Both \textsf{m12f} and \textsf{Romeo} (middle row, left panel) show that even if the disk is very kinematically cold, the bar subsequently formed by internal evolution is not very strong (other examples are \textsf{m12m} and \textsf{Remus}, both shown in Figure \ref{appendix:fig:kinematic_coldness}). 
\textsf{m12r} (middle row, right panel) forms a weak, short bar in a kinematically hot disk only after a strong satellite interaction, around 12.7 Gyr. This bar quickly fades out as it is dissolved by the high fraction of random motion. 
In the absence of a strong external perturbation, kinematically hot disks in our sample such as \textsf{m12r}, \textsf{m12i}, and \textsf{Louise} (bottom row, left panel) do not form bars (see also  \textsf{Juliet} and \textsf{Romulus} in Figure \ref{appendix:fig:kinematic_coldness_unbarred}). Thus, we find that in the cosmological environment of galaxy evolution the kinematic coldness of the disk is still an important factor bar formation. Although a cold disk by itself does not appear to be a sufficient condition for bar formation, it is clearly a necessary one for bars that form via internal evolution. The lifetime of a bar after it forms, whether due to internal evolution or satellite interaction, also depends on the state of the kinematic coldness of the disk.

Additionally, we note that \citet{McCluskey.et.al.2023} showed that $v_{\phi, \star}$, $\sigma_{\star}$ and $v_{\phi, \star}/\sigma_{\star}$ for ``young stars" in FIRE-2 agree well with M31, M33, for galaxies in the PHANGS survey and in the cold gas. There are no evident signs that gas or young stars in the FIRE-2 galaxies are too hot dynamically. However, this may have some tension with the MW and the old stars that the authors did not compare with.

\subsubsection{Gas fraction and stellar feedback}
\label{sec:gas_fraction_stellar_feedback}

\begin{figure*}
\centering
 \includegraphics[width=\textwidth]{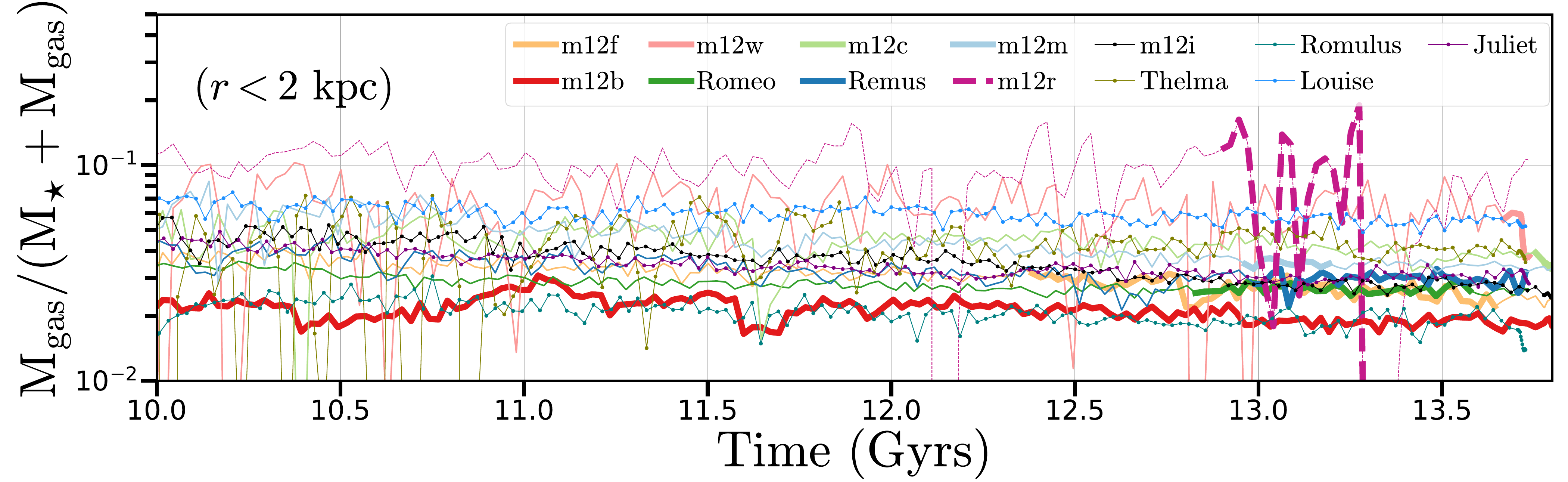}
\caption{ {\bf Evolution of central gas fraction for barred and unbarred galaxies.} Each line shows $M_{\rm gas}/(M_{\rm gas}+ M_{\star})$ in the central region of the disk ($R<2$ kpc and $|z|<1$ kpc). As in Figure \ref{fig:all_rotation_curves}, thin lines with solid circles are for unbarred galaxies and the dashed magenta line is for \textsf{m12r}. Thick lines are for the ``bar episodes" in the barred galaxies. No trend in barred vs unbarred systems is at first apparent, but \textsf{m12w} (pink) and \textsf{m12r} (dashed magenta), the two weak bars, have high gas fractions that vary substantially with time, while the two strongest bars, \textsf{m12f} (orange) and \textsf{m12b} (red), have low gas fractions that are relatively steady in time. }
    \label{fig:gas_fraction}
\end{figure*}

Observational studies have shown that bars are less likely to form in gas-rich star-forming galaxies, and are instead mainly found in gas-depleted red galaxies \citep{Barazza.et.al.2008, Masters.et.al.2012}. The gas fraction in the disk is expected to affect the morphological properties of bars, and bar formation in a gas-rich disk is thought to proceed more slowly \citep{Athanassoula.et.al.2013MNRAS}. As shown in Figure \ref{fig:gas_fraction}, the FIRE-2 galaxies mostly have a mass fraction of nearly ten percent in gas in the central 2 kpc, and would likely not be considered gas-depleted. There is also a significant amount of gas infall and outflow in the central regions of the FIRE-2 galaxies, driven by feedback from star formation.  Consequently, the central gas fraction $M_{\rm gas}/(M_{\rm gas}+M_{\star})$ fluctuates rapidly, modifying the disk potential on free-fall timescales. 

In Figure \ref{fig:gas_fraction}, we show the evolution of gas fraction $M_{gas}/(M_{gas}+M_{\star})$ in the central region ($r<2$ kpc) of the disk for each of the FIRE-2 galaxies. The color scheme and line thickness are same as in Figure \ref{fig:all_rotation_curves}. We use thin lines with solid circles for unbarred galaxies, thick lines for ``bar episodes" and thin lines for the unbarred phases in the barred galaxies, and a thick dashed line for \textsf{m12r}.

In Figure \ref{fig:gas_fraction}, we find that \textsf{m12w} (in pink) and \textsf{m12r} (dashed magenta line) have the highest gas fractions that also vary the most compared to the other galaxies. Our investigation shows that \textsf{m12w} and \textsf{m12r} undergo repeated episodes with large fluctuations in star formation rate over a short time interval.  

\begin{figure*}
    \centering
    \includegraphics[width=0.9\textwidth]{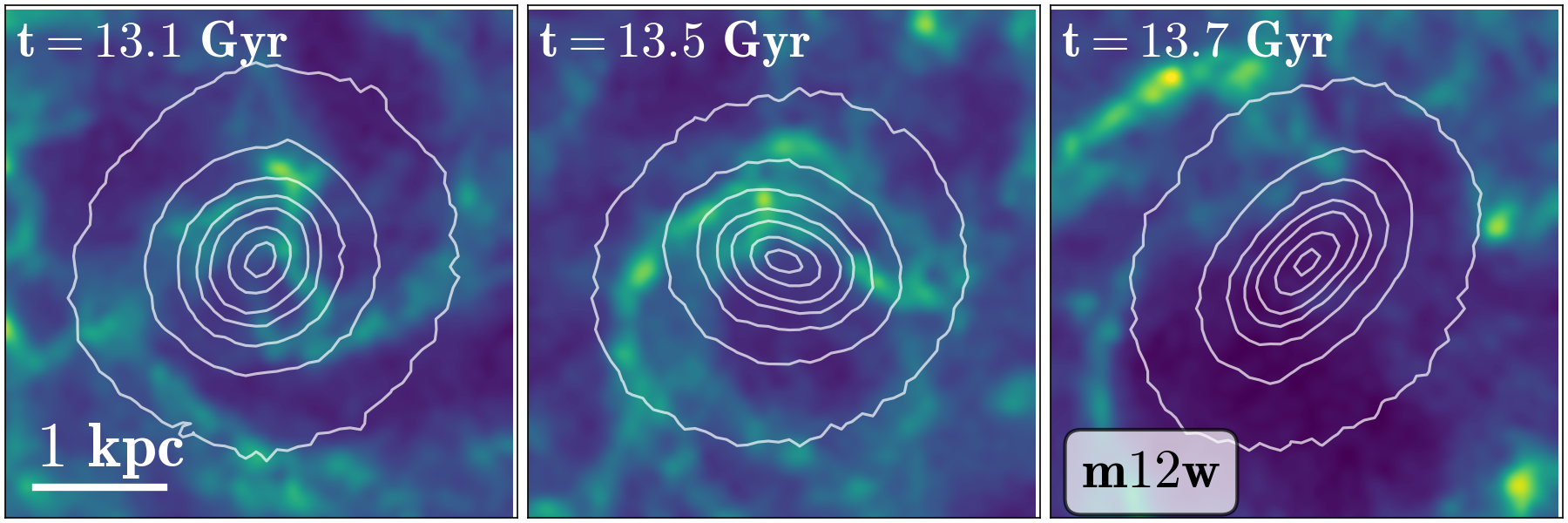}\\
     \includegraphics[width=0.3\textwidth]{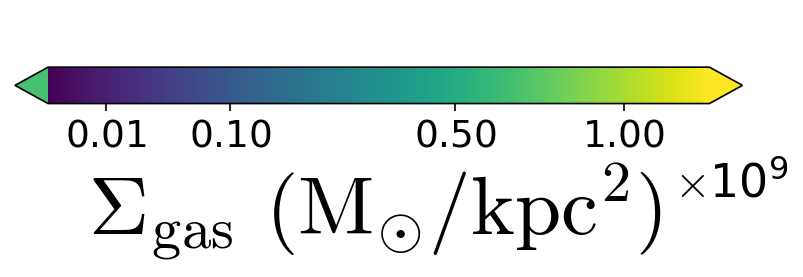}
    \caption{{\bf Bar formation following removal of gas in the inner galaxy.} In all three panels, white contours show face-on stellar distribution in the inner $4\times 4$ kpc$^2$ region of \textsf{m12w} and the colormap corresponds to gas surface density (blue=low, yellow=high). {\bf Left:} large gas fraction in center and no bar structure in the stellar component. {\bf Center:} as gas begins to flow out of the center, a bar structure begins to emerge. {\bf Right:} gas evacuated from the center and a strengthening bar in the stellar component.}
    \label{fig:m12w-gas}
\end{figure*}

This rapid change in the star formation rate (within $R<2$ kpc) leads to rapid fluctuations in the feedback strength and hence in the gas fraction \citep{El-Badry.et.al.2016, Orr.et.al.2018,Yu.et.al.2021,Gurvich.et.al.2022}. These changes in gas fraction alter the ability of the galaxy to form a bar. For example, in \textsf{m12w} a bar forms immediately following a large outflow of gas
(Figure \ref{fig:m12w-gas}), which itself is induced by a strong interaction with a satellite. Similarly, \textsf{m12r} undergoes a strong tidal interaction with a satellite galaxy that forms a bar, but has a high gas fraction that should suppress bar formation, and indeed the bar dissolves rapidly. Conversely, the barred galaxies with the highest strengths, \textsf{m12f} (in orange) and \textsf{m12b} (in red) show the lowest gas fractions by the end of their evolution. 

Rapid, quasiperiodic fluctuations in the gas fraction near the galaxy center, driven by feedback processes, can destabilize bar orbits by modifying the central potential, as a result of the mass redistribution of gas. For more discussion on fluctuations in potential affecting bar formation, see Section \ref{sec:stellar_feedback}.

\section{So why are FIRE bars weak and short?}\label{sec:discussion}

A superficial comparison with the bars arising in the zoomed cosmological simulations of MW-mass galaxies from Auriga \citep{Fragkoudi.et.al.2021} reveals that the bars that arise in the FIRE-2 galaxies appear much weaker ($A_{2}/A_{0}\lesssim 0.2$) and more ellipsoidal.  We caution that although the two suites share similar particle mass resolution, many other choices in the numerical models are quite different, especially the representation of the multi-phase ISM and clustered nature of star formation (we do; they don't), the locality of stellar feedback (scales of 3 vs 100 pc), and whether AGN feedback is included (we don't; they do). Taken together, these differing choices lead to both small- and large-scale differences in the environment of the inner galaxy that can explain the differences in the types of bars that are formed. Broadly, the stellar feedback model in Auriga leads to smoother ISM, smoother star formation, and less core formation, which may lead to the formation of stronger bars.

\subsection{Do the bars buckle during evolution?}
Buckling is a dynamic process in barred galaxies where a bar initially forms an asymmetric structure that extends out of the disk plane, eventually evolving into a thickened boxy/peanut bulge, clearly visible from an edge-on view \citep{Debattista2006, Quillen2014, Sellwood.Gerhard2020}.  When applying standard probes of bar buckling \citep{Debattista2006, Xiang2021} on the FIRE-2 bars, we observe some of the measures showing signatures of buckling in some cases, while some other probes do not detect any buckling event in those cases. Even though one of the \textsf{m12m} simulation runs studied in \citet{Debattista.et.al.2019} shows the presence of the X-shaped bar, a buckling event has not been confirmed as the origin of the X-shape. All X/peanut-shapes may not originate from a buckling event \citep{Li2023}. Furthermore, buckling events have been studied thoroughly for strong bars but not for weak bars, as in the case of FIRE-2. The evidence for bucking in these bars is ambiguous at best.

\subsection{Influence of the inner rotation curve}  \label{sec:discussion_inner_rotation_curve}
The FIRE-2 galaxies are consistently baryon-dominated in their centers, with a slow-rising circular velocity in the inner dark matter component compared with a very sharply rising baryonic component. They hence produce relatively short bars, since the region where the overall rotation curve is rising, permitting swing amplification to operate, is relatively small.   The subdominant DM component also leads to the formation of ellipsoidal bars \citep{Athanassoula.2002, Athanassoula.and.Misiriotis.2002, Berentzen.Shlosman.Jogee.2006}. Conversely, a larger DM fraction in the center at fixed stellar surface density can permit the formation of the longer, more rectangular bars apparent in simulations that form this type of galaxy in their halos \citep{Athanassoula.and.Misiriotis.2002}.

Another test for the relative influence of the central mass distribution comes from additional simulations in the FIRE-2 suite that include cosmic ray (CR) physics. These ``CR-run'' simulations use the same initial conditions as those in our study but form less massive disks in the same DM halos, since the cosmic rays help moderate star formation but are otherwise not expected to influence bar formation \citep{Hopkins.et.al.2023}. A bar does form in the CR version of \textsf{m12b}, \textsf{m12b-CR}, which has a similar inner circular velocity profile as the suite of galaxies without CRs: the stellar circular velocity peaks at the center ($V_{c, max}=248$ km s$^{-1}$ at $R_{max}=1.0$ kpc) and dominates over the slow-rising DM circular velocity up to a radius of $r\leq6$ kpc. The stellar disk in \textsf{m12b-CR} is also kinematically colder than the other CR-run galaxies, which favours bar formation (Section \ref{sec:kinematic_coldness}). The less massive \textsf{m12f-CR} fails to satisfy our bar criteria as it does not maintain $A_2/A_0>0.1$ for long enough. The other CR-run galaxies (\textsf{m12f-CR}, \textsf{m12c-CR}, \textsf{m12w-CR} and \textsf{m12m-CR}) are all less massive and have slow-rising DM rotation curves that peak at larger radii ($150<V_{c,max}/(\rm km s^{-1})<200$ and $10<R_{max}/(\rm kpc)< 22$); none of them form bars. Therefore, differences in rotation curves alone cannot fully explain our results.

\subsection{The impact of stellar feedback on bar formation}\label{sec:stellar_feedback}
The bars in FIRE-2 galaxies exhibit periodic oscillations in the bar strength, $A_{2}/A_{0}$, at a wide range of frequencies within the same system. The overall bar strength varies on long timescales of more than a Gyr in the longest-lived bars, sometimes leading to intervals spent below the cutoff of 0.1 that we use to define a bar (Figure \ref{fig:m12f_m12b_bar_duration}). At shorter timescales of about 0.1 Gyr the bar strength and length (in the sense of the region in which $A_{2}/A_{0}>0.1$) also oscillates. This timescale is consistent with the timescale of ``breathing modes'' initiated by cycles of star formation and feedback, similar to the process that can create cores in dwarf galaxies \citep{2006Natur.442..539M, 2012MNRAS.421.3464P}. Just as these relatively small but repetitive fluctuations in the potential can induce large scale changes in the inner DM distribution, so they can also suppress bar formation since the timescale of the breathing modes is comparable to the orbital period of the radial orbits needed to sustain the bar (both are essentially the dynamical time). Recent numerical experiments by \citet{Weinberg.2024} have shown that stochastic variations in the potential of the inner galaxy do not have to be very large to significantly change the eccentricity of stellar orbits caught at apocenter, thereby repeatedly disrupting the bar as it forms. \cite{Weinberg.2024} further shows that the characteristic frequency of the stochasticity needs to be only broadly consistent with the dynamical time to be effective at disrupting the bar. We believe this mechanism leads to the short-period oscillations seen in the FIRE-2 bars, and suppresses strong bar formation. This explanation is tentative, since we have not yet confirmed that their model of the stochastic fluctuations is consistent with the process in our simulations. We aim to demonstrate this quantitatively in a future paper.

The significant role of stellar feedback in suppressing bar formation leads to several important implications. First, it implies that bar length and strength should be anti-correlated to the star formation rate or gas fraction in the inner galaxy, as shown for our simulated bars in Figure \ref{fig:gas_fraction} and \ref{fig:m12w-gas}. In other words, we do not expect bars to coincide with periods in which the star formation rate fluctuates significantly, consistent with the results of \citet{Neumann.2020}, or be common in gas-rich galaxies, consistent with \citet{Masters.et.al.2012} and \citet{fraser-mckelvie2020}. Second, it implies that simulations that do not account for this process, either by ignoring stellar feedback entirely (as in collisionless N-body simulations, but also those that include gas but not star formation, such as \citealt{Athanassoula.et.al.2013MNRAS}) or implementing it in a way that does not produce these small-scale fluctuations in the potential, as in Auriga \citep{Fragkoudi.et.al.2021}, EAGLE \citep{Cavanagh.et.al.2022}, and Illustris/IllustrisTNG \citep{Zhao.et.al.2020}, are likely probing the upper limit of the strength of the bar instability in MW-mass galaxies. 

An important caveat to this discussion is that the FIRE-2 galaxies do not include a model for feedback from a central supermassive black hole, which can in principle regulate star formation in the inner galaxy by lowering the gas fraction  \citep{Wellons.et.al.2023, Mercedes-Feliz.et.al.2023}. Indeed, the FIRE-2 galaxies appear to have somewhat higher SFR at $z_r \sim 0$ than the Milky Way, although this conclusion depends somewhat on which measurements of the MW's SFR are used for the comparison \citep{Gandhi.et.al.2022}. \emph{Simulations} in which a subgrid AGN model does have this effect should therefore form longer and stronger bars; however, the efficiency of AGN to remove gas from galaxy centers at this mass scale is \emph{observationally} difficult to constrain thanks to the complexity of the multiphase gas \citep[e.g.][]{2012ARA&A..50..455F,2017FrASS...4...42M,2017NatAs...1E.165H,2021MNRAS.503.1568R, Wellons.et.al.2023}. The FIRE-3 suite does include a model for BH formation and feedback; however, it also includes multiple other changes to other physical prescriptions that prevent apples-to-apples comparisons, especially since the net effect is to lower the total stellar masses of the galaxies by a factor of a few relative to FIRE-2 \citep{Hopkins.et.al.2023}, which in turn changes the predisposition of the galaxies to form bars by lowering their surface density (as we found with the CR suite). Bars are also thought to funnel material toward a central BH \citep[e.g.][]{Hopkins.and.Quataert.2011}, altering the mass accretion rate and making this a deeply nonlinear relationship. We thus defer a detailed study of the impact of introducing a BH feedback model on simulated bars to future work.

Another noteworthy aspect of bar formation is whether there exists a connection between the bar strength and the disk formation time scale \citep{Khoperskov.et.al.2023}. We do not observe such a correlation between bar strengths at $\redshift=0$ and the lookback time of the Early-Disk Era (see \citealt{McCluskey.et.al.2023}) for the 13 galaxies in the FIRE-2 sample in the bar strength range $0.0<A_2/A_0<0.2$. Two of the earliest settling disks in the FIRE-2 sample (\textsf{m12m} and \textsf{Romeo}) form bars that are transient and do not last till $\redshift=0$.

\section{Summary}\label{sec:summary}

In this work we use the 13 MW-mass galaxies from the FIRE-2 zoom-in cosmological-hydrodynamical simulations \citep{Hopkins.et.al.2018} to study bar formation and destruction in a complex cosmological environment with realistic star formation histories and feedback. Here we summarize our main findings.\\

\noindent\textit{How bars form in FIRE-2 galaxies}: We define an asymmetric instability to be a bar if bar strength $A_2/A_0>0.1$ and bar phase $\phi_{2}$ is constant within $\pm 5$ degree within the bar length for at least $\sim 2$ bar rotations. We find that \textbf{in the FIRE-2 simulations, 8 out of 13 galaxies form bars} at some point in their evolution. Not all are barred at $\redshift=0$. We consider 3 of these bars (m12f, m12b, m12r) to result primarily from an interaction with a massive satellite, 4 (m12m, m12c, Remus, Romeo) to arise from internal evolution of the disk in the absence of any strong tidal forces, and 1 (m12w) to be due to the rapid evacuation of gas from the central few kpc. While the bars triggered by tidal interactions can form in slightly more dispersion-supported systems, bars arising from internal evolution preferentially form in the most strongly rotation-supported, and therefore most kinematically cold, systems in our sample, and bars induced by tides last longer in colder systems (Section \ref{sec:kinematic_coldness}). \\

\noindent\textit{How bars do NOT form in FIRE-2 galaxies}: 
Disk instabilities like bars and spiral arms arise in galaxies in the physical Universe either due to gravitational instabilities in the disk or due to satellite interactions or flyby events, or a combination of both. Their origin in our simulations is likewise physical and is not due to Poisson noise, as is sometimes observed for bars forming in constrained galaxy simulations at lower resolution \citep{Dubinsky.et.al.2009}. At FIRE-2 resolution, the N-body relaxation timescale for the region of interest is longer than a Hubble time. \\

\noindent\textit{Galaxies without bars}: 5 of the 13 galaxies in the FIRE-2 sample (\textsf{m12i}, \textsf{Thelma}, \textsf{Louise}, \textsf{Romulus} and \textsf{Juliet}) either do not form a bar at all or show only small increases in $A_2/A_0$ for short intervals. The underlying reason in most cases is that the galaxy is too kinematically hot in the inner region, either because it hosts a large dispersion-supported central bulge (\textsf{Louise}, \textsf{Juliet} and \textsf{Romulus}) or because it has a large number of satellite interactions that disturb and heat its disk (\textsf{Thelma}, \textsf{m12i}). \\

\noindent\textit{Bar characteristics}: The bars that form in FIRE-2 MW-mass galaxies have peak dipolar amplitudes $A_2/A_0$ in the range 0.1--0.2, lengths 1.4--3.7 kpc, and pattern speeds 36--97 km s$^{-1}$ kpc$^{-1}$ (Table \ref{table_bar_properties}). The corotation radius tends to be much larger than the bar length in most cases, so by the traditional $\mathcal{R}$ measure the bars in FIRE-2 would be considered ``slow'' ($0.15<a_{s}/\rco<0.59$) although their pattern speeds are not intrinsically low. However, directly comparing the $\mathcal{R}$ measure of the bars in FIRE-2 with that of bars from observations (where the fast bars have $\mathcal{R}<1.4$) is likely problematic, as most of the bars studied in observations are stronger and very different from the bars that we study here \citep{Corsini.et.al.2011, Aguerri.et.al.2015}. In contrast, in a recent study, \citet{Geron2023} measured $\mathcal{R}$ for a galaxy sample of weak and strong bars to find that 62\% of their bars are slow. Ideally, we need to compare $\mathcal{R}$ for similar types of bars in observations and in simulations. The lifetimes of the FIRE-2 bars range very widely: there are galaxies that host a bar nearly their entire existence (m12b) once the disk settles, those that form a bar only in the last $\sim500$ Myr (m12w), and nearly everything in between (Table \ref{table_bar_duration}). Our bar properties are measured directly from the simulation particles, not from synthetic observations, and therefore \textbf{we caution against their direct comparison to observational measurements} since the effects of light-weighting, extinction, projection, and observational selection have not been accounted for. We plan to consider the \emph{observed} properties of the simulated bars in future work. \\

Future observational surveys will play an imperative role in the study of formation of stellar bars and their properties across different redshifts. For a better understanding of bar formation and bar dissolution in a cosmological context, further study on the dynamical interactions between the bar, the stellar and gaseous disks, the surrounding DM halo and the external satellite interactions is crucial. \\

\section*{Data Availability}
FIRE-2 simulations are publicly available \citep{Wetzel.et.al.2023} at \url{http://flathub.flatironinstitute.org/fire}.
Additional FIRE simulation data is available at \url{https://fire.northwestern.edu/data}.
A public version of the \textsc{Gizmo} code is available at \url{http://www.tapir.caltech.edu/~phopkins/Site/GIZMO.html}.

\begin{acknowledgments}
This work was facilitated by the Pre-Doctoral Program of the Center for Computational Astrophysics at the Flatiron Institute; analysis was carried out on resources maintained by the Scientific Computing Core. The Flatiron Institute is supported by the Simons Foundation. 

SA is grateful to the Simons Foundation for the support during the Pre-Doctoral Program from February 2021 -- June 2021.

Support for SP was provided by NASA through the NASA Hubble Fellowship grant \#HST-HF2-51466.001-A awarded by the Space Telescope Science Institute, which is operated by the Association of Universities for Research in Astronomy, Incorporated, under NASA contract NAS5-26555. This work was supported by a research grant (VIL53081) from VILLUM FONDEN. 

RES gratefully acknowledges support from the Simons Foundation as well as from NSF grant AST-2007232 and NASA grant 19-ATP19-0068. 

Support for PFH was provided by NSF Research Grants 1911233, 20009234, 2108318, NSF CAREER grant 1455342, NASA grants 80NSSC18K0562, HST-AR-15800. 

AW received support from: NSF via CAREER award AST-2045928 and grant AST-2107772; NASA ATP grant 80NSSC20K0513; HST grants AR-15809, GO-15902, GO-16273 from STScI.

ECC acknowledges support for this work provided by NASA through the NASA Hubble Fellowship Program grant HST-HF2-51502 awarded by the Space Telescope Science Institute, which is operated by the Association of Universities for Research in Astronomy, Inc., for NASA, under contract NAS5-26555.

Numerical calculations were run on allocations AST21010 and AST20016 supported by the NSF and TACC, and NASA HEC SMD-16-7592.

\end{acknowledgments}

\vspace{5mm}

\software{astropy \citep{2022ApJ...935..167A,2018AJ....156..123A, 2013A&A...558A..33A}, 
          gizmo analysis \citep{2020ascl.soft02015Wetzel}, 
          halo analysis \citep{2020ascl.soft02014Wetzel}, 
          AGAMA \citep{Vasiliev.2019},
          photutils \citep{larry_bradley_2020_4049061}
          }

\appendix

\section{Measurement of bar length} \label{sec:appendix_bar_length_measurement}
\begin{figure*}
\centering
    \includegraphics[width=\textwidth]{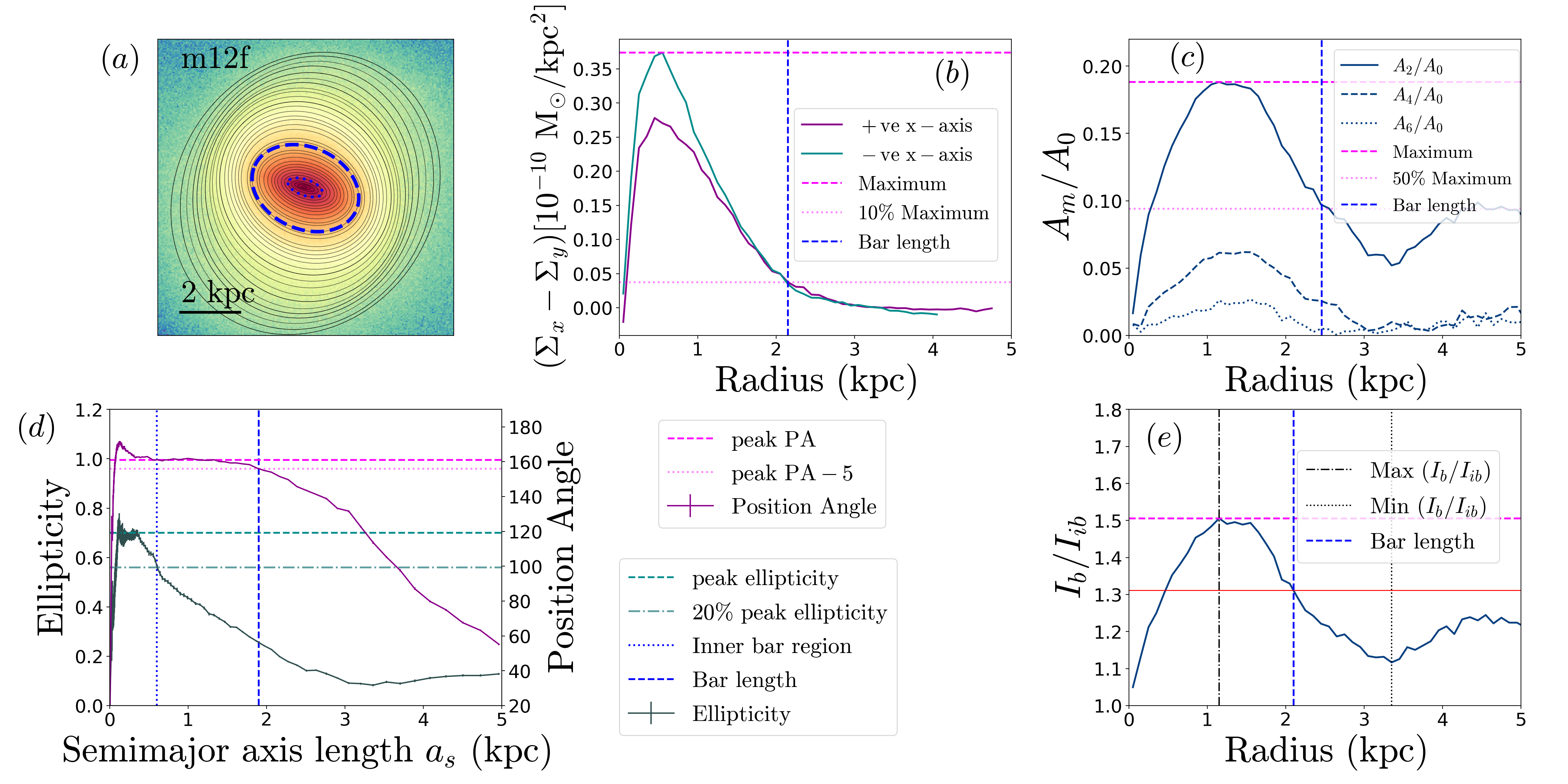}
    \caption{{\bf Estimation of bar lengths in the FIRE-2 galaxies using four methods.} Method 1: panel $(a)$ and $(d)$ show the ellipse fitting method. The central bar region is chosen within the radius where there is a $20\%$ decrease in peak ellipticity value (marked with a blue dotted vertical line) and the corresponding PA is marked with a dashed magenta horizontal line. A deviation of $\pm 5$ degrees in PA marks the edge of the bar/bar length (blue dashed vertical line). Method 2: in panel $(b)$, the difference in surface density $\Sigma_{x}$ along the bar axis (x-axis) and $\Sigma_{y}$ perpendicular to the bar axis (y-axis) is used to estimate bar length at $10\%$ of the peak value of $(\Sigma_{x}-\Sigma_{y})$ (marked with blue dashed line). Dissimilarity in $(\Sigma_{x}-\Sigma_{y})$ along the positive and negative x-axis shows a slight asymmetry in the distribution, although the bar length estimate is the same for both. Method 3: Bar length is measured from $50\%$ decrease in bar strength in $(c)$ (also marked with a blue dashed line). Method 4: in panel $(e)$, the bar length is measured (marked with a blue dashed vertical line) from the ratio of stellar density in the bar and inter-bar region $I_{b}/I_{ib}$ (from \cite{Aguerri.et.al.2000}). The radius of maximum $I_{b}/I_{ib}$ is marked with a black dashed vertical line and the minimum is marked with a black dotted vertical line. }
    \label{fig:appendix:m12f_bar_length}
\end{figure*}

\begin{deluxetable*}{ccccccc}
\tablenum{5} \label{table:appendix_bar_length}
\tablecaption{Bar semi-major axis length $a_{s}$ estimates using different methods}
\tablewidth{0pt}
\tablehead{
\colhead{Simulation} & \colhead{\redshift$_{,\mathrm{peak}}$} & \colhead{$a_{s,1}$} & \colhead{$a_{s,2}$} & \colhead{$a_{s,3}$} & \colhead{$a_{s,4}$} & \colhead{Average $a_s$} \\
\colhead{} & \colhead{} &  \colhead{ (kpc)} & \colhead{(kpc)} & \colhead{(kpc)} & \colhead{(kpc)} & \colhead{(kpc)} }
\decimalcolnumbers
\startdata
m12f & 0.064 & 1.4 & 2.1  & 2.0 & 1.97 & 1.86 \\
m12b & 0.046 & 1.6 & 1.96 & 1.85 & 1.77 & 1.79  \\
m12m & 0.084 & 1.62 & 1.5  & 1.68 & 1.32 & 1.53 \\
Romeo & 0.038 & 1.26 & 1.21 & 1.45 & 1.16 & 1.32 \\
Remus & 0.041 & 2.1 &  1.42 & 1.55 & 1.25 & 1.58\\
m12w & 0.0 & 1.35 & 1.41 & 1.45 & 1.18 & 1.35  \\
m12c & 0.0016 & 1.5 &  1.27  & 1.4 & 0.90 & 1.27  \\
m12r & 0.061 & 0.95 & 0.77 & 0.85 & 0.66 & 0.81 \\
\enddata
\tablecomments{Columns: 1. Simulation name; 2. Redshift of maximum bar strength; 3. bar semi-major axis length $a_{s}$ measured using Ellipticity/PA (Method 1); 4. $a_{s}$ measured using bar strength profile (Method 2); 5. $a_{s}$ measured using the difference in surface densities between the major and minor axis of the bar (Method 3); 6. $a_{s}$ measured using the bar--inter-bar density contrast method of \cite{Aguerri.et.al.2000} as discussed in Section \ref{sec:bar_length_measurement} (Method 4); 7. average of all $a_s$ measurements in columns (3)-(6). }
\end{deluxetable*}

We estimate the bar lengths in the FIRE-2 barred galaxies using four different methods that are frequently used in the literature \citep{Athanassoula.and.Misiriotis.2002,Aguerri.et.al.2000, Erwin.2018}. We also discuss how some of the methods are more suitable for certain types of bars than others and how the following methods perform for bar length estimation in the FIRE-2 galaxies. 
\begin{itemize}
    \item {Method 1: Fitting elliptical isophotes in the bar region.}
    \item {Method 2: Using the bar strength distribution ($A_{2}/A_{0}$) as a function of disk radius.}
    \item {Method 3: From the difference in surface densities between the bar major axis and the minor axis in the disk plane.}
    \item {Method 4: From decomposition of the stellar surface density into Fourier modes from \citep{Aguerri.et.al.2000}.}
\end{itemize}

In Figure \ref{fig:appendix:m12f_bar_length} we present examples of the above four methods to estimate bar length in \textsf{m12f} close to its peak bar strength. We show the fitted elliptical isophotes on the face-on density distribution (panel a) that is used to determine the PA (magenta curve in panel d) and ellipticity (dark green curve in pnel d) as a function of disk radius. Panel b shows bar length estimation using surface densities along the bar major axis and the minor axis; panel c shows how we estimate bar length using bar strength profile $A_2/A_0$ as a function of disk radius $R$ and panel e shows the method using bar and inter-bar intensities (method 4). The blue dashed vertical line in all panels mark the bar length of \textsf{m12f} using different methods.

In the first method, we fit the bar in \textsf{m12f} with elliptical isophotes (see panel a in Figure \ref{fig:appendix:m12f_bar_length}) using the python package {\it photutils} \citep{larry_bradley_2020_4049061} to find the PA and the ellipticity profile as a function of the disk radius (see panel d in Figure \ref{fig:appendix:m12f_bar_length}). We define a central bar region with an upper bound of radius set by the 20\% decrease in the peak ellipticity (vertical blue dotted line in panel d in Figure \ref{fig:appendix:m12f_bar_length}). We define the bar length (blue dashed vertical line in panel d) to be the radius at which the PA (magenta curve) changes by $\pm 5$ degrees from its value at the edge of the central bar region (blue dotted vertical line in panel d). The bar in \textsf{m12f} and other FIRE-2 galaxies are more round and do not show a sharp change in PA. The bars have the peak ellipticity close to the center ($R<1$ kpc), with a gradual decrease of ellipticity away from the peak (see dark green line of ellipticity in Figure \ref{fig:appendix:m12f_bar_length}). We, therefore, do not expect to find sharp changes in the PA and ellipticity at the edge of the bars as typically observed for rectangular/boxy-edge bars in studies with N-body simulations or in observations.

In the second method (panel c in Figure \ref{fig:appendix:m12f_bar_length}), we estimate the bar length at the radius where the bar strength $A_2/A_0$ drops to $50\%$ of its peak value (see blue vertical line in panel c).  

In the third method (panel b in Figure \ref{fig:appendix:m12f_bar_length}), we estimate the bar length from the difference in surface density along the bar major-axis (x-axis) and along the bar minor-axis (y-axis) $( \Sigma_{x} -\Sigma_{y} )$.  The bar length (blue vertical dashed line in panel b) is defined to be the radius at which $( \Sigma_{x} -\Sigma_{y} )$ decreases by 80\% of the peak value (magenta dotted horizontal line).

In the fourth method (panel e in Figure \ref{fig:appendix:m12f_bar_length}), we use the ratio of the density contrast in the bar ($I_{b}$) and inter-bar ($I_{ib}$) region, first introduced by \cite{Ohta.et.al.1990}) and redefined in \cite{Aguerri.et.al.2000}, where the authors defined the bar region as follows:
\begin{equation}
\label{eq:bar_interbar_fourier_modes_bar_length}
    I_{b}/I_{ib} > \frac{ \left( I_{b}/I_{ib} \right)_{max} - \left( I_{b}/I_{ib} \right)_{min}  }{2} + \left( I_{b}/I_{ib} \right)_{min} 
\end{equation}
where, $I_{b}= A_{0}+A_{2}+A_{4}+A_{6}$ and $I_{ib}= A_{0}-A_{2}+A_{4}-A_{6}$ and $A_{m}$ is the amplitude of the $m^{th}$ Fourier mode. Here we use the condition from \cite{Aguerri.et.al.2000} to determine the bar length (blue dashed vertical line in panel e). 

We present the semi-major axis length $a_s$ (or half bar length) of the bars estimated using the four methods, as well as the average $a_s$ from the four methods in Table \ref{table:appendix_bar_length}.

The ellipse fitting/PA method yields the shortest bar lengths for the strongest bars in \textsf{m12f} and \textsf{m12b}, while for other galaxies the method using the bar and inter-bar density contrast yield the shortest bar lengths. The largest estimate of bar length for \textsf{m12f} and \textsf{m12b} comes from the method using bar strength profile (method 2). While for other bars, the methods using surface density and using ellipticity/PA estimate provide the largest bar length. We note the similarity between the nature of ellipticity and PA profiles in the FIRE-2 galaxies and the model MD in \cite{Athanassoula.and.Misiriotis.2002}. This is expected given the structural resemblance of the FIRE-2 bars to the bars having a round shape in model MD in \citet{Athanassoula.and.Misiriotis.2002}.

\begin{figure*}
\centering
 \includegraphics[width=\textwidth]{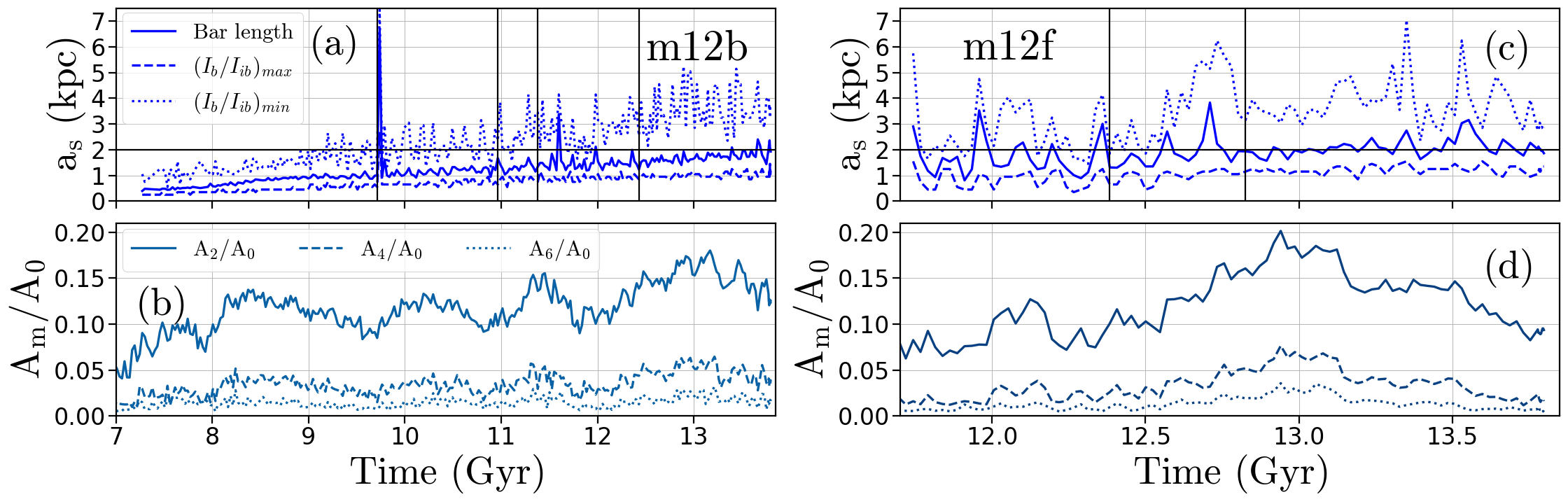}
    \caption{{\bf The evolution of bar length (top panels) along with the evolution of bar strength (bottom panels) in \textsf{m12b} (panel a and b) and in \textsf{m12f} (panel c and d).} {\bf Top panels:} bar length (blue lines with dots), Max($ I_{b}/I_{ib}$) (dashed blue line) and Min($ I_{b}/I_{ib}$) (dotted blue line) is shown for \textsf{m12b} (panel a) and \textsf{m12f} (panel c) \citep{Aguerri.et.al.2000}. The black vertical lines indicate the satellite mergers. {\bf Bottom panels:} The bar strength $A_{2}/A_{0}$ (dark blue line), $A_{4}/A_0$ (dashed blue line) and $A_{6}/A_0$ (dotted blue line) is shown for \textsf{m12b} (panel b) and \textsf{m12f} (panel d). The maximum and minimum values of $ I_{b}/I_{ib}$ are for comparison. The large fluctuations in the bar length measurement are due to oscillations in the bar strength (see bottom panels) and occasionally due to satellite mergers. The bar length in \textsf{m12b} grows to $\sim 2$ kpc over a period of more than 5.7 Gyrs and in \textsf{m12f} the bar length is most stable at 2 kpc when bar strength $A_{2}/A_{0}>0.1$.  }
    \label{appendix:fig:bar_length_evolution}
\end{figure*}

\section{Bar length evolution}\label{appendix:sec:bar_length_evolution}
Estimating bar length at different time intervals is essential for determining the orbital time $T_{o}=2\pi r/V_{c}(r)$ at the edge of the bars. We estimate the bar lengths of the FIRE-2 galaxies at multiple time intervals to estimate the orbital time $T_{o}$ (see Section \ref{sec:fire_bar_strengths} and Table \ref{table_bar_duration}). We show two examples of bar length estimation in the FIRE-2 galaxies \textsf{m12b} and \textsf{m12f} in Figure \ref{appendix:fig:bar_length_evolution}.

In Figure \ref{appendix:fig:bar_length_evolution} we present the evolution of bar length for \textsf{m12b} (left panel a and b) and \textsf{m12f} (right panel c and d), estimated with the bar and inter-bar intensity ratio $\rm I_{b}/I_{ib}$ from Section \ref{sec:appendix_bar_length_measurement}. The bar in \textsf{m12b} grows steadily in length inside-out for a long duration of 5.7 Gyrs, while, the bar in \textsf{m12f} maintains almost a constant length when bar strength $A_{2}/A_{0}>0.1$ (see panel c in Figure \ref{appendix:fig:bar_length_evolution}). Panels b and c show the evolution of the bar strength ($\rm m=2$) and other even Fourier modes ($\rm m=4$ and $\rm m=6$), that are in phase with the evolution of the higher amplitude of $m=2$ mode. Some of the large fluctuations in the bar length are due to large perturbations in the disk as a result of mergers (indicated by the black vertical lines), however, most of the mergers do not affect the bar length estimates.

\section{Bar pattern speed estimation with the TW method} \label{appendix:TW_method}
\begin{figure*}
\centering
	\includegraphics[width=\textwidth]{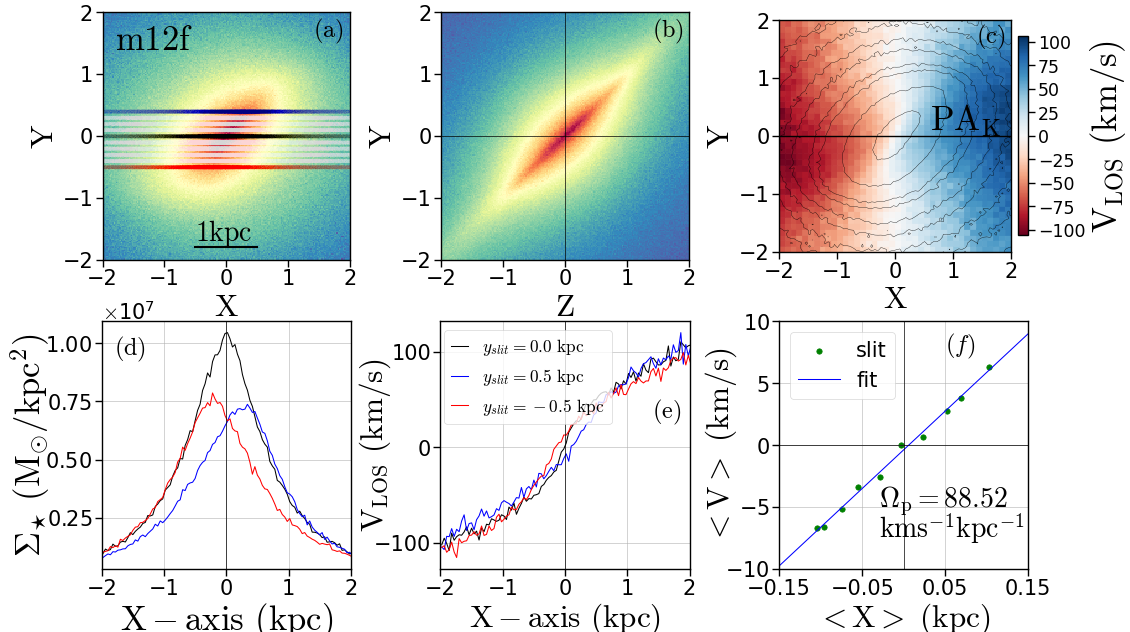}
    \caption{ {\bf Application of the TW method to measure bar pattern speed in \textsf{m12f} at its peak bar strength.} In the TW method slits are placed on the projected 2D density $(a)$ and velocity map $(c)$ of the galaxy \textsf{m12f} (at maximum bar strength) along the kinematic PA ($\rm PA_{K}$) when the disk is oriented at a most suitable inclination of $45^{o}$ from the line of sight $(b)$ and the bar PA is $\sim 41^{o}$ from the ($\rm PA_{K}$) of the disk. $\rm PA_{K}$ is kept along the x-axis (black horizontal line in $(c)$). $(d)$ shows the surface density distribution and $(e)$ shows line of sight velocity $V_{\rm LOS}$ distribution for the middle slit ($Y=0$ kpc black curve) and the extreme slits ($Y=0.5$ kpc (blue) and $Y=-0.5$ kpc (red)). The slope of $\langle V \rangle$ vs $\langle X \rangle$ gives the bar pattern speed $\Omega_{p}=88.5$ km s$^{-1}$ kpc$^{-1}$ for \textsf{m12f} at peak bar strength. }
    \label{fig:appendix_m12_TW_orientation}
\end{figure*}

In this Section, we present the TW method of pattern speed estimation of the bars in FIRE-2 in detail. The TW pattern speed is given by \citep{Merrifield.Kuijken.1995},
\begin{equation} \label{TW_expression}
\Omega_{p}\sin(i)=\frac{\int^{\infty}_{-\infty}\Sigma(X,Y) V_{\rm LOS}(X, Y) {\rm d}X   }{ \int^{\infty}_{-\infty}\Sigma(X,Y) X {\rm d}X } = \frac{\langle V \rangle}{\langle X \rangle}
\end{equation}
where, $X$ \& $Y$ are the sky coordinates, $\Sigma(X, Y)$ and $V_{\rm LOS}(X, Y)$ are the projected surface density and line of sight velocity on the sky plane. The numerator $\langle V \rangle$ and denominator $\langle X \rangle$ are weighted with surface density $\Sigma(X,Y)$ along slits that are kept parallel to the kinematic PA of the disk, here the x-axis (see panels a and c in Figure \ref{fig:appendix_m12_TW_orientation}).

The TW method works best when the galaxy disk is inclined by $\sim 45^{o}$ to the line of sight of the observer and the angular separation between the bar PA and the kinematic PA (x-axis marked with $\rm PA_{K}$ in panel c Figure \ref{fig:appendix_m12_TW_orientation}) of the disk is $\sim 41^{o}$. With the freedom to orient the simulation volume towards any line of sight, we fix the galaxy disk at the described orientation as shown in panels a and b in Figure \ref{fig:appendix_m12_TW_orientation}. The TW method is based on the asymmetry about $Y=0$ line in the surface density and line of sight velocities (red and blue curves in panels d and e) for slits that are off the $X-axis$ (red and blue slits in panel a). For each slit, we get a point in the $\langle V \rangle$ and $\langle X \rangle$ plot in panel f. Ideally, all the points in panel f should lie on a straight line passing through the origin. Practically, all the points do not lie exactly on a straight line as in panel f, and we estimate the bar pattern speed from the slope of the fitted line. At peak bar strength, the bar pattern speed in \textsf{m12f} is $\Omega_{p}=88 {\rm  km s}^{-1} {\rm kpc}^{-1}$. 

In the TW method, the disk is inclined by $45^o$ to the line of slight and that decreases the bar length by $1/\sqrt{2}$ on disk re-orientation. Hence the TW method fails for short and weak bars in FIRE-2, \textsf{m12w}, \textsf{m12c} and \textsf{m12r} even though their bar strength $A_{2}/A_{0}|_{max}>0.1$. For short and weak bars we find that the surface density $\Sigma(X, Y)$ and line-of-sight velocity $V_{\rm LOS}(X, Y)$ profile along the two extreme slits ($\pm 0.5$ kpc) overlap and the asymmetry about $Y=0$ in $\Sigma(X, Y)$ and $V_{\rm LOS}(X, Y)$ is lost. The similarity in the $\Sigma(X, Y)$ and $V_{\rm LOS}(X, Y)$ for different slits leads to large errors in the estimation of $\langle X \rangle$ and $\langle V \rangle$ (Equation \ref{TW_expression}), which results into wrong estimates of the \emph{slope} that is directly related to the pattern speed $\Omega_{p}$.

\section{Direct estimation of bar pattern speed from Fourier phase angle}
\label{Appendix:patterspeed}
We present the direct measurement of bar pattern speed in the FIRE-2 galaxies \textsf{m12b}, \textsf{m12f}, \textsf{m12w}, \textsf{m12c} and \textsf{Remus}, which have nearly constant bar PA (within $\pm 5^o$; similar to panel c in Figure \ref{fig2:all_bar_strength}) in the last 22 Myrs of their evolution. We have 10 high-cadence snapshots ($0.0016<\redshift<0$) with a time interval of 2.2 Myr between subsequent snapshots, for each of which we estimate the bar phase angle $\phi_{2}$ (Equation \ref{eq:PA}) in an annular ring of width 200 pc at their bar length $R=a_{s}$. We measure the bar length $a_{s}$ using the bar and inter-bar density contrast (see Section \ref{sec:appendix_bar_length_measurement} and \ref{appendix:sec:bar_length_evolution}). The slope of $\phi_{2}|_{R=a_s}$ vs $t$ is the bar pattern speed $\Omega_{p}=d\phi/dt$. 

\begin{figure*}
\centering
\includegraphics[width=0.32\textwidth]{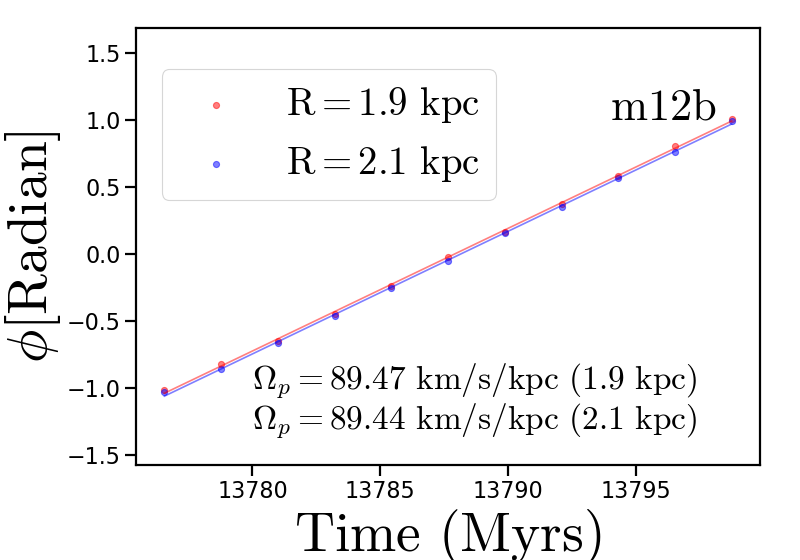} 
\includegraphics[width=0.32\textwidth]{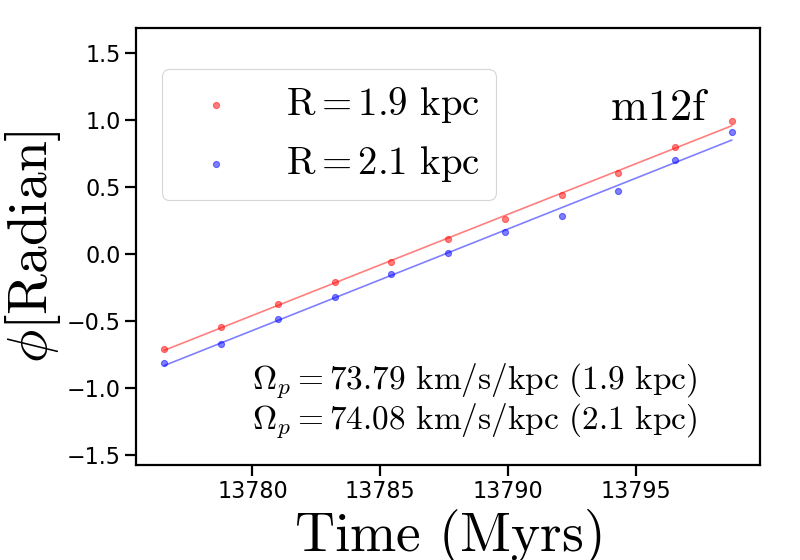}
\includegraphics[width=0.32\textwidth]{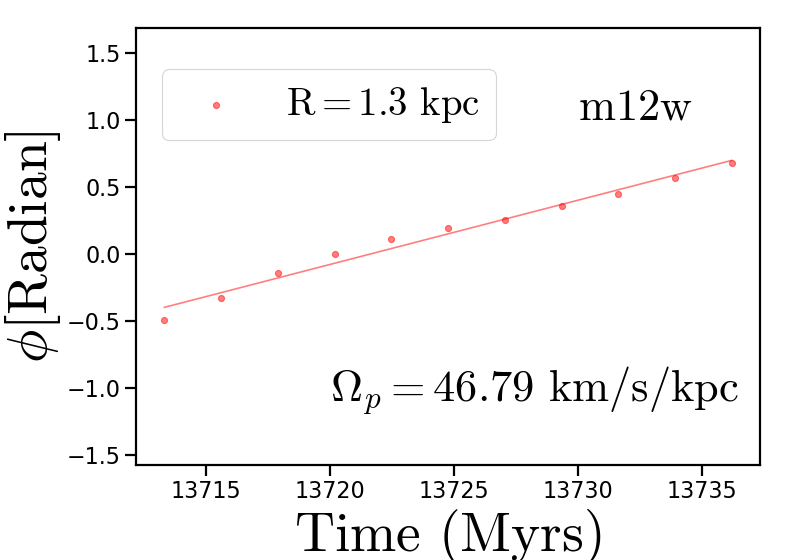} \\
\includegraphics[width=0.32\textwidth]{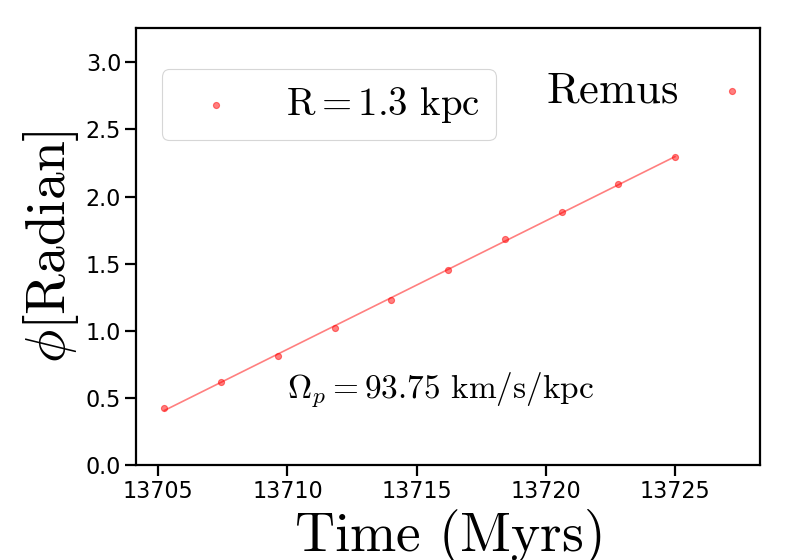}
\includegraphics[width=0.32\textwidth]{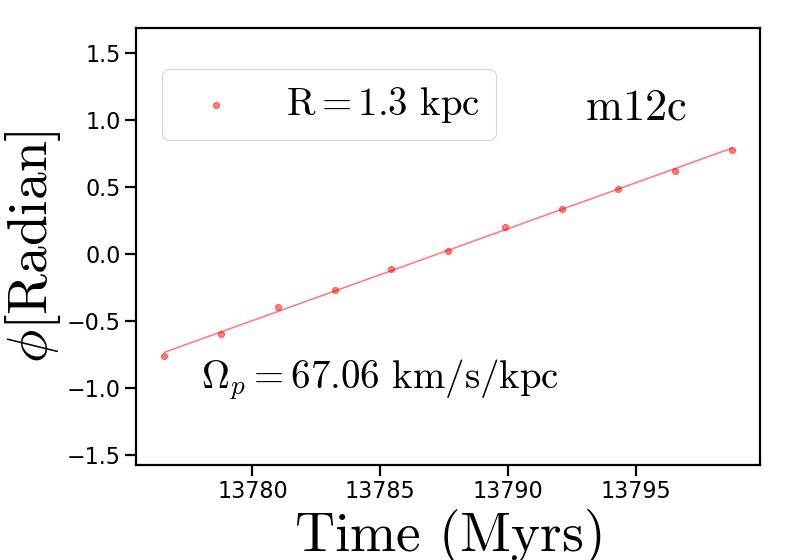} 
\caption{{\bf Direct measurement of pattern speed from high-cadence snapshots ($0.0016<\redshift<0.0$) in the last 22 Myr of evolution of the FIRE-2 galaxies.} We measure the pattern speed from the rate of change of Fourier phase angle $\phi=\frac{1}{2}\tan^{-1}(b_{2}/a_{2})$. }
    \label{fig:appendix:bar_pattern_speed_phi2}
\end{figure*}

In Figure \ref{fig:appendix:bar_pattern_speed_phi2}, we present the evolution of $\phi_{2}$ with time for bar in the galaxies \textsf{m12b}, \textsf{m12f}, \textsf{m12w}, \textsf{m12c} and \textsf{Remus}. The slope of the fitted lines is the pattern speed $\Omega_{p}=d\phi/dt$. We show the bar pattern speed $\Omega_{p}$ (in $\rm km s^{-1} kpc^{-1}$) at the bar length $a_s$ for all the barred galaxies mentioned above. For \textsf{m12b} and \textsf{m12f} we estimate the pattern speed at two radii close to bar length.

\begin{figure*}
\centering
\includegraphics[width=\textwidth]{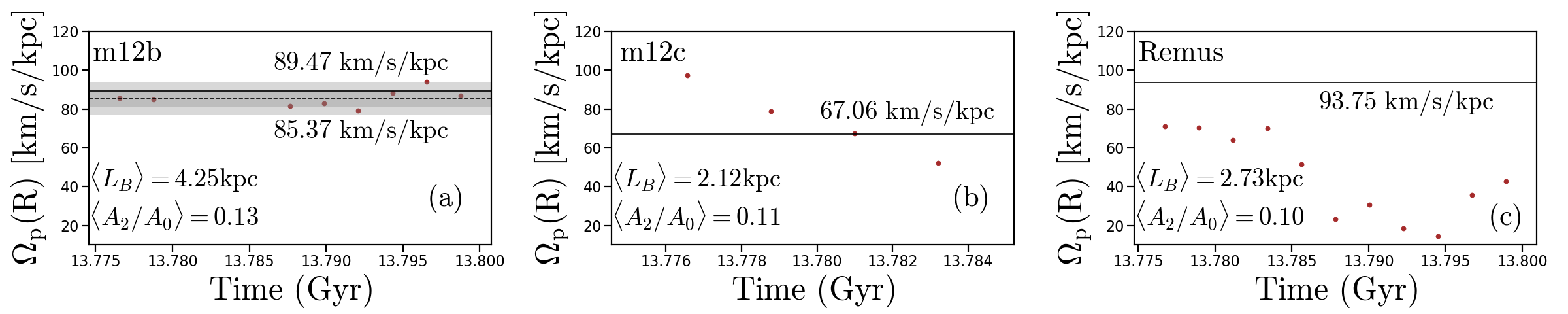} 
\caption{{\bf Comparison of bar pattern speed from direct measurement and using the TW method}. The black horizontal line indicates the pattern speed estimated from the direct method and the brown dots represent that from the TW method at different times between redshift $0.0016<\redshift<0.0$. TW measurements at some snapshots do not give a good fit and are not plotted. $\langle A_{2}/A_{0}\rangle$ and $\langle L_{B}\rangle$ are the mean bar strength and mean bar length for the same snapshot range. Bar length at each snapshot is measured using Method 4 in Section \ref{sec:bar_length_measurement}. TW method pattern speed matches with the direct method pattern speed for the bars with high strength and longer length like \textsf{m12b}, while it does not match for bars with lower strength and shorter length, like \textsf{m12c} and \textsf{Remus}.}
    \label{appendix:fig:bar_pattern_speed_phi2_TW_comparison}
\end{figure*}

\begin{figure*}
\centering
\includegraphics[width=\textwidth]{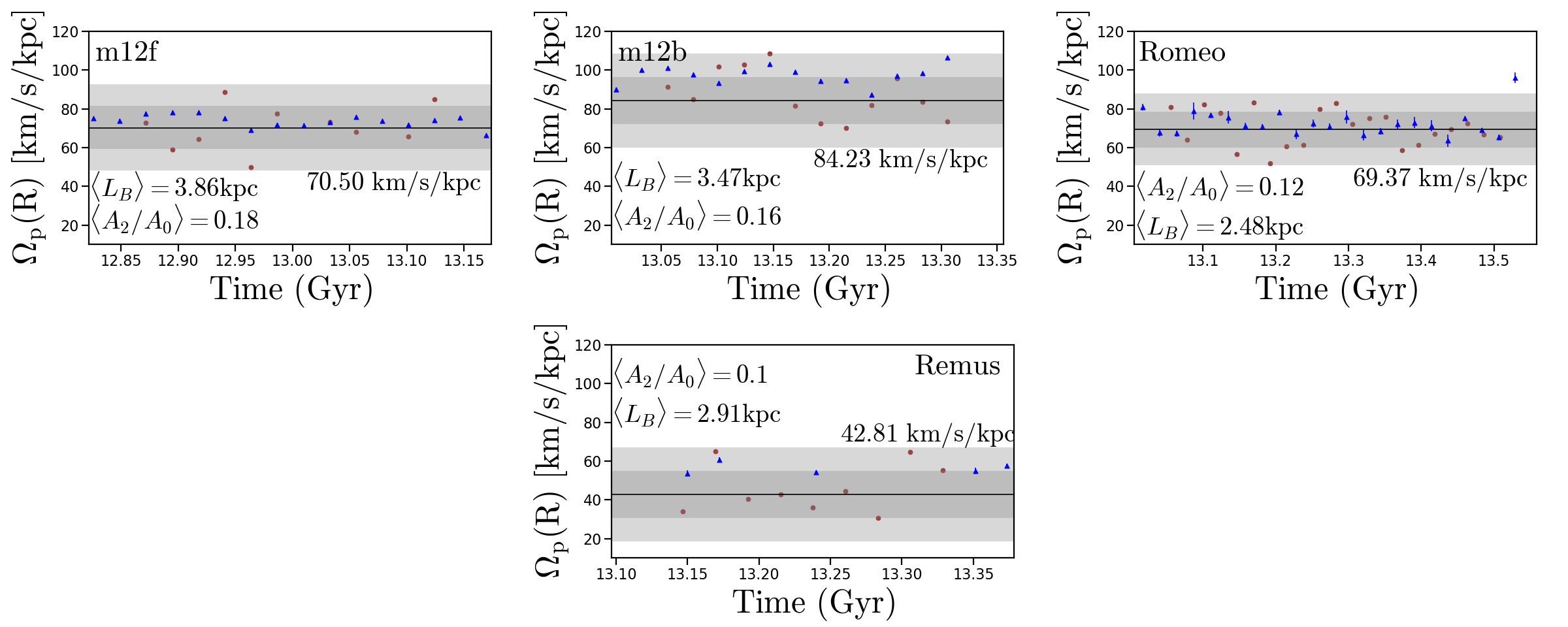} 
\caption{{\bf Median bar pattern speed measurement using the TW method from a range of time steps when the bar has maximum strength}. The pattern speed estimation through the TW method is possible for bars in \textsf{m12f}, \textsf{m12b}, \textsf{Romeo}, and \textsf{Remus} with high strength and long in length (see Figure \ref{appendix:fig:bar_pattern_speed_phi2_TW_comparison}). The black horizontal line indicates the median of the pattern speeds measured at each snapshot (brown dots) using the TW method and the median value lies between 70-84 km s$^{-1}$ kpc$^{-1}$ for the three galaxies. The dark grey shades indicate the $1\sigma$ and $2\sigma$ regions. $\langle A_{2}/A_{0}\rangle$ and $\langle L_{B}\rangle$ are the mean bar strength and mean bar length for the same snapshot range. Bar length at each snapshot is measured using Method 4 in Section \ref{sec:bar_length_measurement} from \cite{Aguerri.et.al.2000}. The bar in \textsf{Remus} is shorter in length and weaker in strength compared to \textsf{m12f}, \textsf{m12b}, \textsf{Romeo} and hence the pattern speed measurement is less reliable for \textsf{Remus}. The blue triangles, with blue errorbars in each panel show the pattern speed using the code \textsf{patternSpeed.py} from \citet{Dehnen2023}. Some errorbars are too small to be seen. We summarise the pattern speed estimations for all the FIRE-2 galaxies in Table \ref{table_bar_properties}.}
\label{appendix:fig:bar_pattern_speed_TW_m12b_m12f_romeo_remus}
\end{figure*}

We compare the bar pattern speed estimation by the direct method as described above with the pattern speed estimated using the TW method (see Section \ref{sec:bar_pattern_speed}). In Figure \ref{appendix:fig:bar_pattern_speed_phi2_TW_comparison} we present the comparison of bar pattern speeds in \textsf{m12b}, \textsf{m12c} and \textsf{Remus} for the last 22 Myrs. The pattern speed estimation through the TW method matches with the direct method when the bar is long and strong. For example, for \textsf{m12b} the pattern speed estimates closely match the value of 89.47 km s$^{-1}$ kpc$^{-1}$ from the direct method. While for \textsf{m12c} and \textsf{Remus} the bar length is much shorter. During applying the TW method an inclination $\sim 45^{o}$ of the galaxy plane with the line of sight makes the effective projected bar length further smaller, making it difficult to get a correct estimate of the slope for the $\langle X \rangle$ vs $\langle V_{\rm LOS} \rangle$ plot. In Figure \ref{appendix:fig:bar_pattern_speed_TW_m12b_m12f_romeo_remus} we present the median pattern speed estimation of the bars in \textsf{m12f}, \textsf{m12b}, \textsf{Romeo} and \textsf{Remus} for a range of snapshots where we otherwise cannot apply the direct estimation of pattern speed. We find that if either the bar length or the bar strength is low the TW method cannot be applied.
For example, for the bars in \textsf{m12m}, \textsf{m12c} and \textsf{m12w} the mean bar length is very short (2.49 kpc, 2.12 kpc, and 2.32 kpc respectively) and the bar strength is also less (0.08, 0.11 and 0.15 respectively), and the inspection of the fits of $\langle X \rangle$ vs $\langle V_{\rm LOS} \rangle$ in the TW method are not reliable in these cases. Thus, the TW method does not give correct estimates of pattern speeds for \textsf{m12m}, \textsf{m12c}, and \textsf{m12w}. We additionally compare our TW pattern speed measurements with those obtained using the code \textsf{patternSpeed.py} from \citet{Dehnen2023}, as shown in Figure \ref{appendix:fig:bar_pattern_speed_TW_m12b_m12f_romeo_remus}. In the figure, these measurements are represented by blue triangles and error bars, all of which fall within the shaded $2\sigma$ region.

\section{Evolution of bar pattern speed in \textsf{m12b} } \label{sec:appendix:m12b_patternspeed}
\begin{figure}
\centering
\includegraphics[width=\columnwidth]{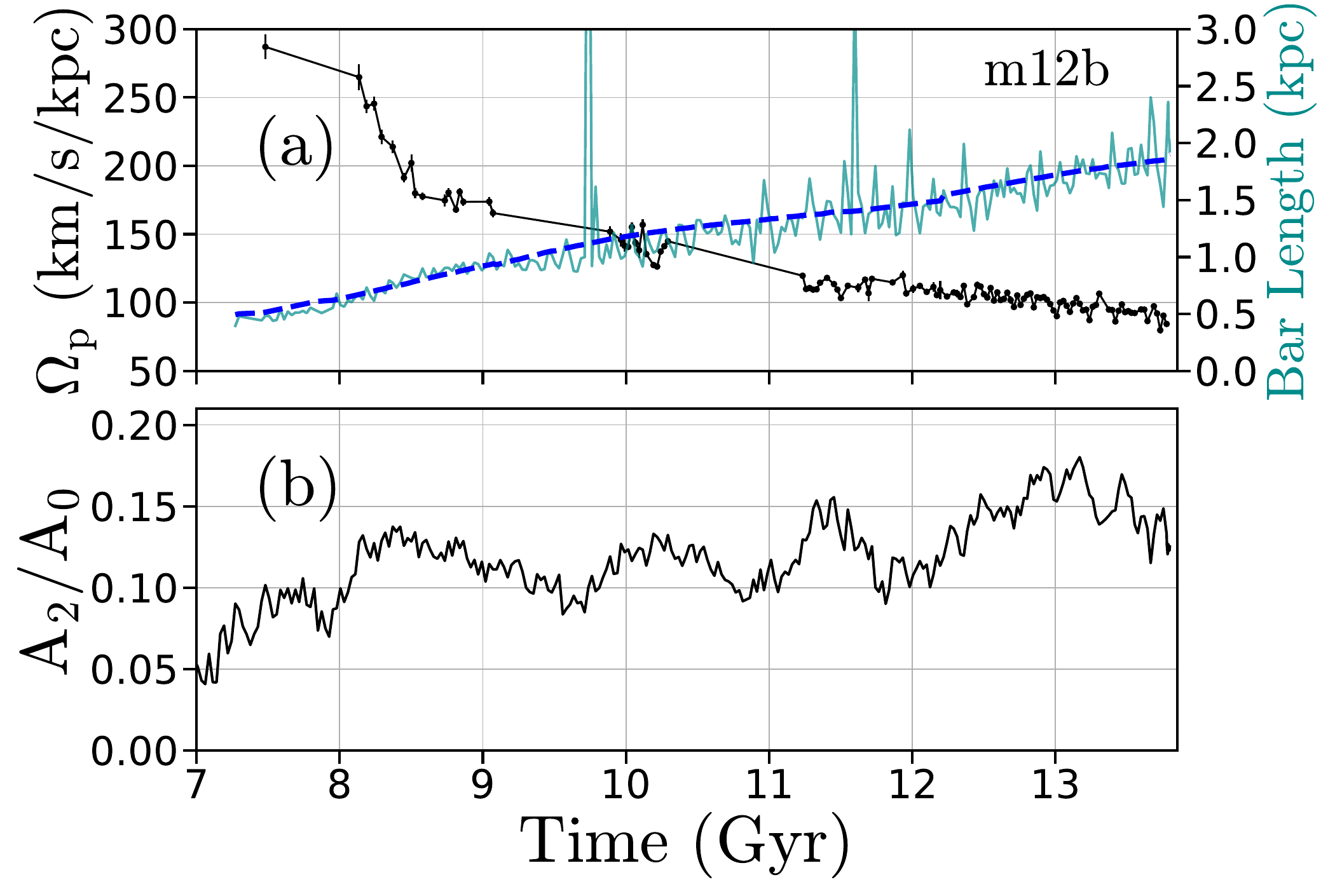} 
\caption{{\bf Pattern speed $\Omega_p$ decreases in \textsf{m12b} as the bar lengthens}. {\bf Panel a:} Evolution of bar pattern speed (black circles) and evolution of bar length (dark cyan), along with the smoothed bar length (blue dashed line). {\bf Panel b:} Evolution of bar strength in \textsf{m12b}. }
\label{appendix:fig:m12b_bar_patternspeed}
\end{figure}

Figure \ref{appendix:fig:m12b_bar_patternspeed} shows the evolution of bar pattern speed $\Omega_p$ measured for \textsf{m12b} using the code \textsf{patternSpeed.py} \citep{Dehnen2023}. The bar pattern speed in \textsf{m12b} decreases steadily as the bar grows stronger, lengthens and gains more stars that join the bar orbits. Decrease in bar pattern speed indicates loss of bar angular momentum \citep{Debattista.and.Sellwood.2000}, while the increase in bar length is due to the addition of more stars that can potentially increase the bar angular momentum. This complexity points to the challenge of disentangling the transfer of angular momentum between various components influenced by different processes in the galaxy's central region.

\section{Use of tidal index to search for satellites} \label{Appendix:tidal-index}
We investigate the FIRE-2 galaxy satellites that can induce bar formation in the disk. Previously many studies have looked into the formation of bars during galaxy encounters and flyby events \citep{Lang.et.al.2014, Lokas.et.al.2014, Lokas.2018} and found that during the encounter of two galaxies, bars form in both galaxies having similar masses. Bars also form in the minor galaxy as it interacts with the more massive one \citep{Lang.et.al.2014}. Bar formation depends on the orientation of the satellite galaxy orbit with respect to the direction of the total disk angular momentum of the more massive galaxy \citep{Lokas.2018, Cavanagh2020}. 

To find the satellites we investigate the merger trees of dark matter halos that are constructed using the codes ROCKSTAR \citep{Behroozi.et.al.2013a} and Consistent Trees \citep{Behroozi.et.al.2013b}, that identify dark matter halos and trace their evolutionary history. We read the merger trees using the python package {\it halo analysis} by \cite{2020ascl.soft02014Wetzel} to find out the instantaneous satellite mass and distance from the host center of mass. We use the python package {\it gizmo analysis} by \cite{2020ascl.soft02015Wetzel} to read the stellar particle positions of the satellite galaxies and find the retrograde or prograde orientation of the satellite orbit. With the motivation of finding the satellites that may induce bar formation, we search for all the satellites ($ M_{total}>10^{7}$ $\rm M_{\odot}$) in the FIRE-2 galaxies that undergo close tidal interactions with the host galaxy. 

\begin{figure*}
\centering
\includegraphics[width=\textwidth]{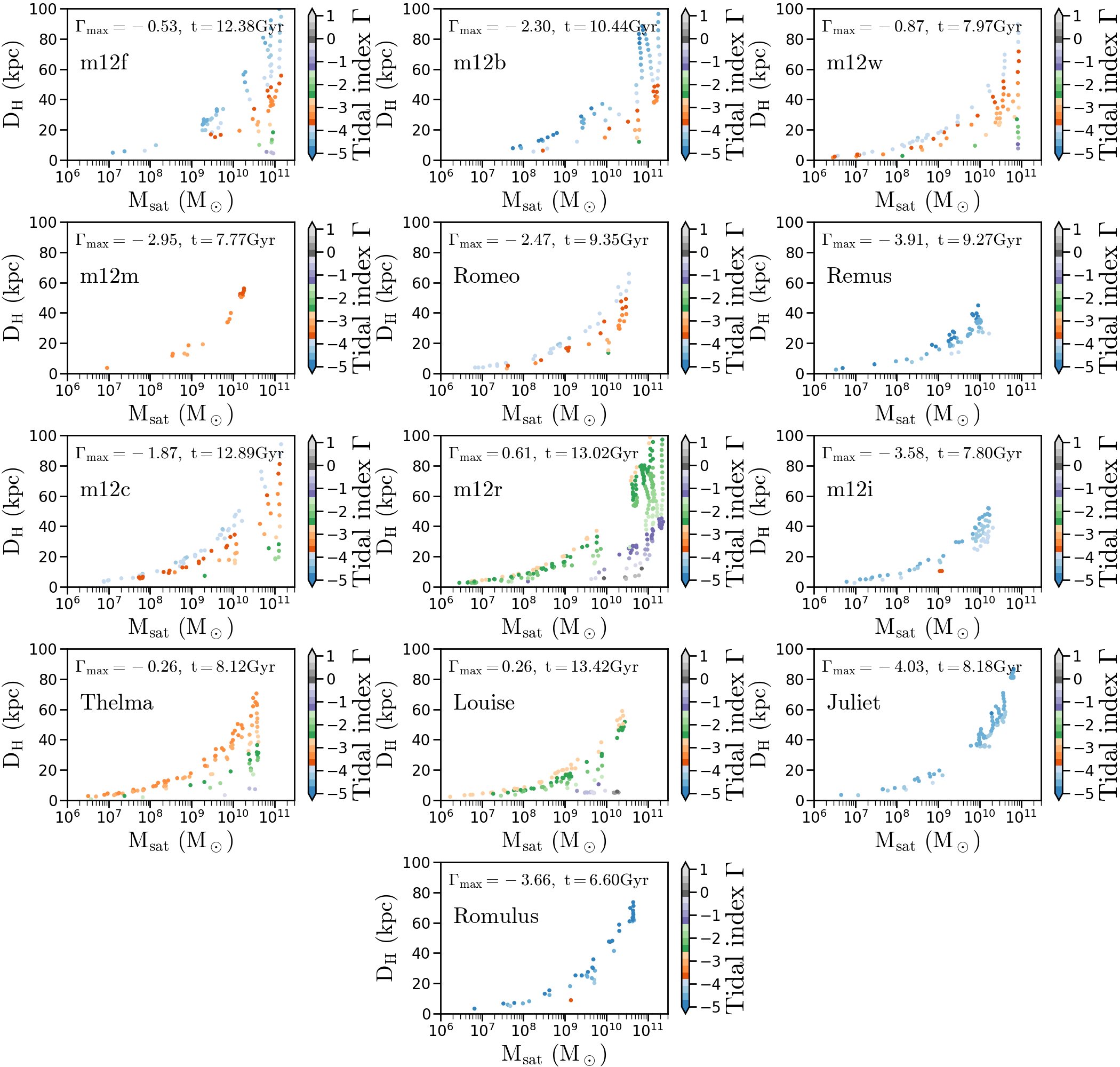}
\caption{{\bf The tidal index $\Gamma$ of all the satellites within a distance of 100 kpc from the host center, in the 13 FIRE-2 galaxies for a time period of 6.6-13.7 Gyrs of evolution}. The points in each panel represent the satellites at each snapshot with their instantaneous DM halo mass $M_{sat}$ (in $\rm M_{\odot}$) on the x-axis and the instantaneous total distance from the galaxy center $\rm D_{H}$ (in kpc) on the y-axis. The tidal index of the satellites for all the simulations, $\Gamma$ is shown with the color bar in the range $-5<\Gamma<1$. All satellites having $\Gamma\gtrsim-2.45$ may impart significant tidal force to the stellar disk. Often the time of the maximum tidal impact by a satellite (i.e., time of Max $\Gamma$ for each simulation) does not coincide with the time of bar formation.  }
    \label{fig:appendix_tidal_indices}
\end{figure*}

Figure \ref{fig:appendix_tidal_indices} shows the tidal index $\Gamma$ (Equation \ref{eq:tidal_index}) of the satellite galaxies in colored solid circles (color bar range: $-5<\Gamma <1$), along with their instantaneous DM halo mass (M$_{\rm sat}$) in x-axis and their instantaneous distance (D$_{\rm H}$) from the host center of mass (y-axis), at each time snapshot between 6.6-13.7 Gyrs. Note that $\Gamma$ encapsulates the properties of the satellites as well as the host disk properties (see Section \ref{sec:sat_interaction}). In Figure \ref{fig:appendix_tidal_indices} we present the different satellites in the FIRE-2 simulations, the maximum tidal index $\Gamma_{max}$ and the corresponding time (in Gyrs), for each simulation.

We find that the galaxies \textsf{m12f}, \textsf{m12c}, \textsf{m12r} and \textsf{Louise} undergo the highest tidal index interaction after 10 Gyrs of evolution when bar formation takes place, while the rest have their largest interaction in the early stages of evolution. One exception is \textsf{m12b} where the bar forms early ($\sim 8$ Gyrs). While studying bar formation in the FIRE-2 galaxies we search for the satellite interactions close to the bar formation time. We investigate the high $\Gamma$ satellites that are close to the bar formation time and enlist the satellites and their properties in Table \ref{table:satellite_properties}.

We test the effect of choosing a lower limit of $\Gamma$ ($\Gamma < -2.45$). We find that such a choice will increase the number of satellite interactions, but the new interactions will not be leading the bar formation process in these simulations. For example in the simulations \textsf{m12f}, \textsf{m12b} and \textsf{m12r}, we still have the strongest satellites (shown in blue in Figure \ref{fig:satellite_interaction_examples}) playing the major role in bar formation. A lower cutoff of tidal index will increase the number of satellite galaxies in \textsf{m12m}, \textsf{m12w}, \textsf{m12c}, \textsf{Romeo} and \textsf{Remus}, that we present in Section \ref{sec:secular_evolution} to be forming during the internal evolution of the disk. However the new satellites (see Figure \ref{fig:appendix-m12m-remus-romeo-satellites}) do not have their pericenter passages or mergers within 10 dynamical time of the bar formation in these galaxies. Hence, selecting a lower limit of $\Gamma$ does not seem to affect our results.

\section{Rare satellite interactions in \textsf{m12m}, \textsf{Remus}, \textsf{m12w}, \textsf{m12c} and \textsf{Romeo}}
\label{sec:appendix:m12m-satellite-interactions}
\begin{figure*}
\centering
	\includegraphics[width=0.45\textwidth]{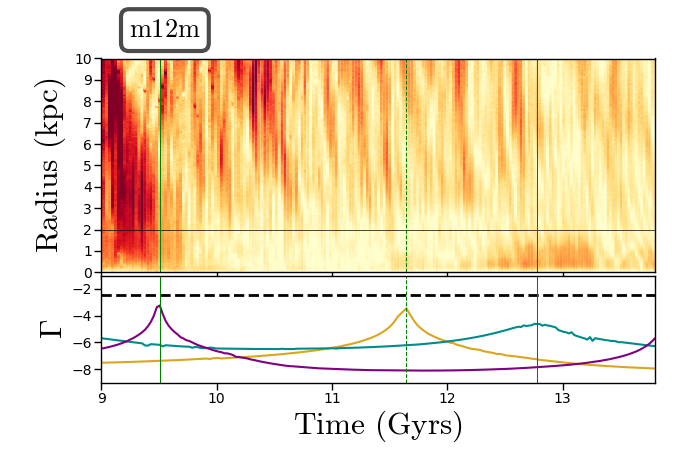}
	\hspace{1cm}
	\includegraphics[width=0.45\textwidth]{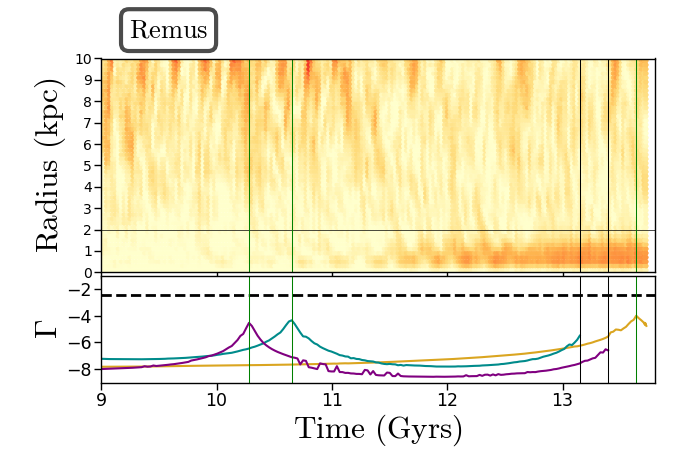} 
	\vspace{-0.2cm}\\
	\includegraphics[width=0.45\textwidth]{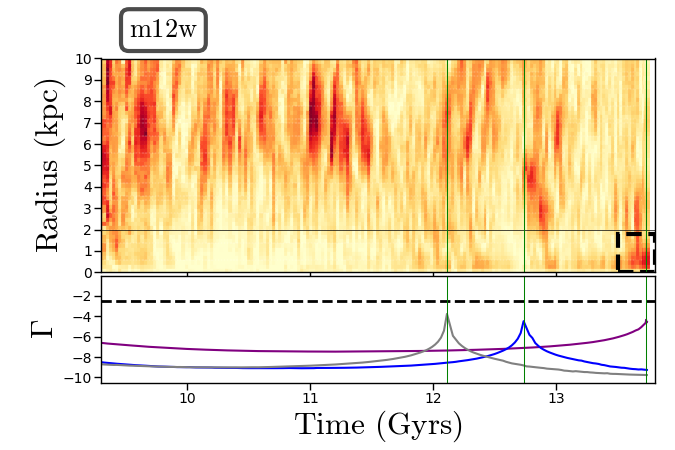}
	\hspace{1cm}
	\includegraphics[width=0.45\textwidth]{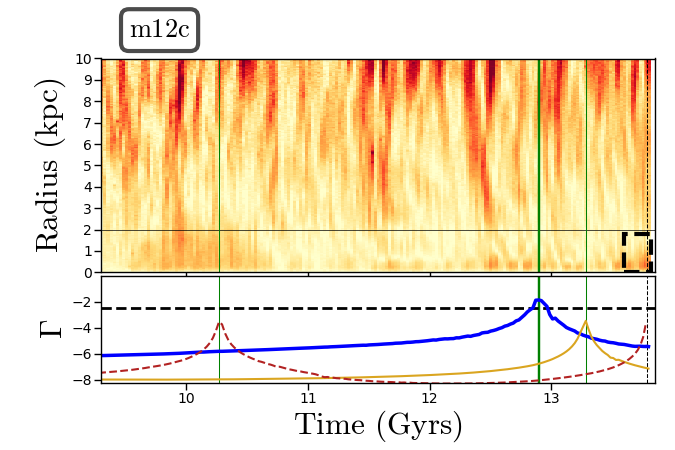} 
	\vspace{-0.2cm}\\
	\includegraphics[width=0.45\textwidth]{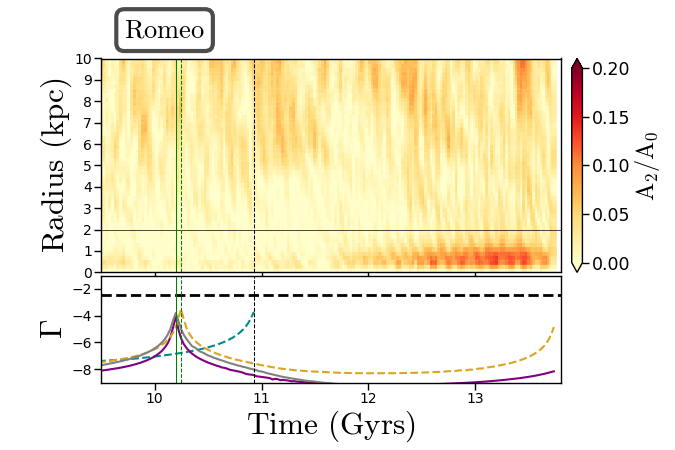}
    \caption{{\bf Low tidal index satellite interactions in \textsf{m12m}, \textsf{Remus}, \textsf{m12c}, \textsf{m12w} and \textsf{Romeo}}. {\bf Top panel for each simulation:} Evolution of bar strength (shown with color bar in the range: $0.0<A_2/A_0<0.2$). The dark reddish patch in the central region (radius $<2$ kpc) in the top panels shows the presence of the bar. {\bf Bottom panel for each simulation:} Evolution of scaled tidal index $\Gamma$ for the highest tidal index satellites close to bar formation. Colours and lines are same as Figures \ref{fig:satellite_interaction_examples}. We highlight the bar in \textsf{m12w} and \textsf{m12c} with dashed rectangular boxes as their duration in the disk is short. In the galaxies \textsf{m12m}, \textsf{m12c}, \textsf{Remus} and \textsf{Romeo}, the satellite interactions are less impactful with low tidal indices ($\Gamma<<-2.45$). Even though the satellite in \textsf{m12c} has high $\Gamma$ ($> -2.45$), it forms a bar after 22 dynamical times, too long to be causally related. Apart from the low tidal index, the time of maximum tidal impact of the satellites and the time of formation of the bar (around $A_{2}/A_{0}\sim 0.1$) are well separated such that satellite interactions cannot be the reason of the formation of the bars in these systems. }
    \label{fig:appendix-m12m-remus-romeo-satellites}
\end{figure*}

In this Section, we present a few cases of satellite interactions in \textsf{m12m}, \textsf{Remus}, \textsf{m12w}, \textsf{m12c} and \textsf{Romeo} that are either very weak with a low tidal index ($\Gamma< -2.45$) and/or the time of interaction with the host galaxy disk is a lot early to bar formation (more than 10 dynamical times), such that bar formation can not be causally related to the satellite interaction (see Figure \ref{fig:appendix-m12m-remus-romeo-satellites}). We present the quantitative measurements of the highest tidal index satellite interactions in \textsf{m12m}, \textsf{Remus}, \textsf{m12w}, \textsf{m12c} and \textsf{Romeo} in Table \ref{table:satellite_properties}. Two of the strongest satellite interactions for \textsf{Romeo} are in retrograde and polar orbits ($\cos\theta_{orbit}\sim -0.6$ and -0.04) and also more than 1.7 Gyr early ($>50$ dynamical times) to bar formation. The interactions in \textsf{Remus} are not only very weak, but the interactions happen either after the bar has already formed in the disk or more than 1.5 Gyr early ($>40$ dynamical times) to bar formation. The ineffectiveness of the satellite interactions is similar in the case of \textsf{m12m} and \textsf{m12w} as they impart weak tidal forces (with low $\Gamma=$-3.45, -4.61 for \textsf{m12m} and $\Gamma=$-4.47 for \textsf{m12w}) and/or the interactions are more than 10s of dynamical times before bar formation (33 $T_{dyn}$ for \textsf{m12m} and 25 $T_{dyn}$ for  \textsf{m12c}).  So the bars in these galaxies form as a result of the internal evolution of the disk.

\section{Stellar Surface density profiles}\label{sec:appendix:stellar_surfacedensity}
\begin{figure}
\centering
\includegraphics[width=\columnwidth]{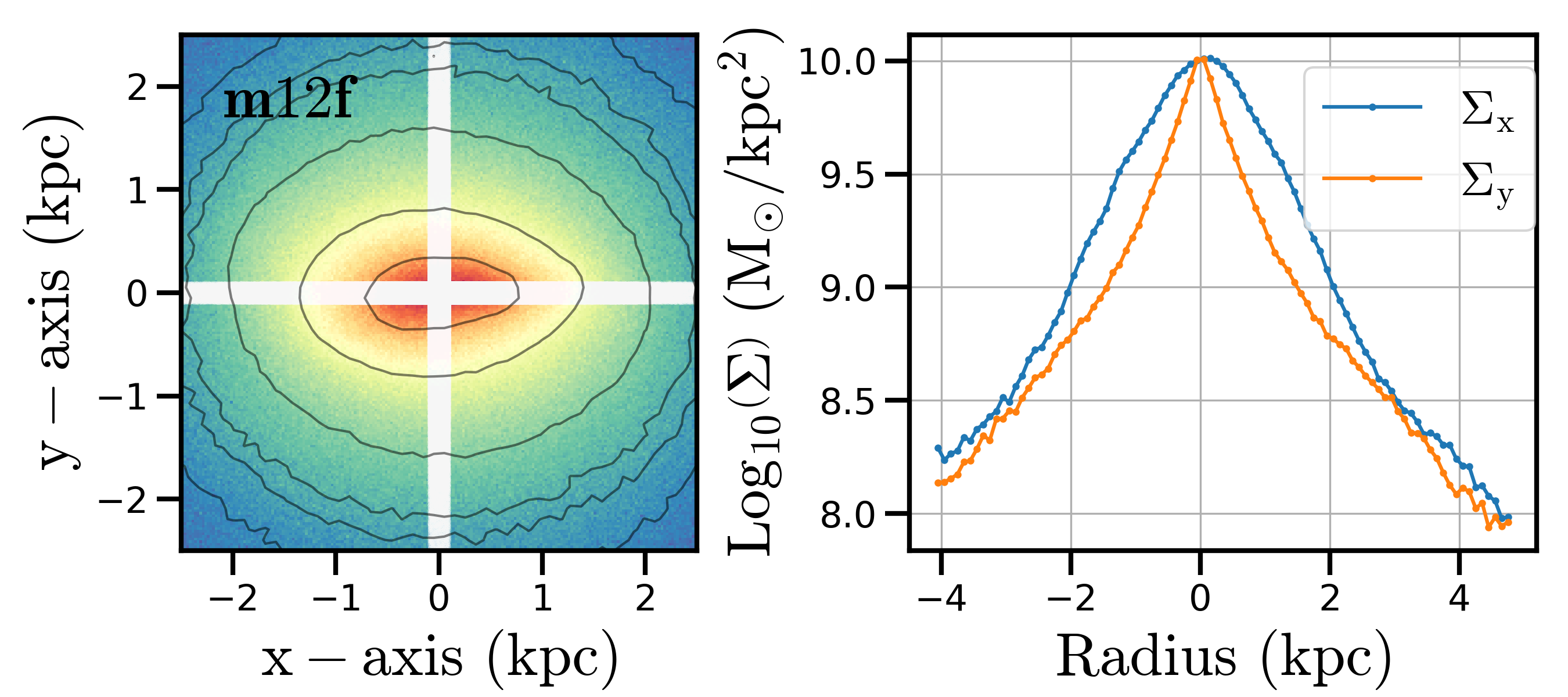}\\
\includegraphics[width=\columnwidth]{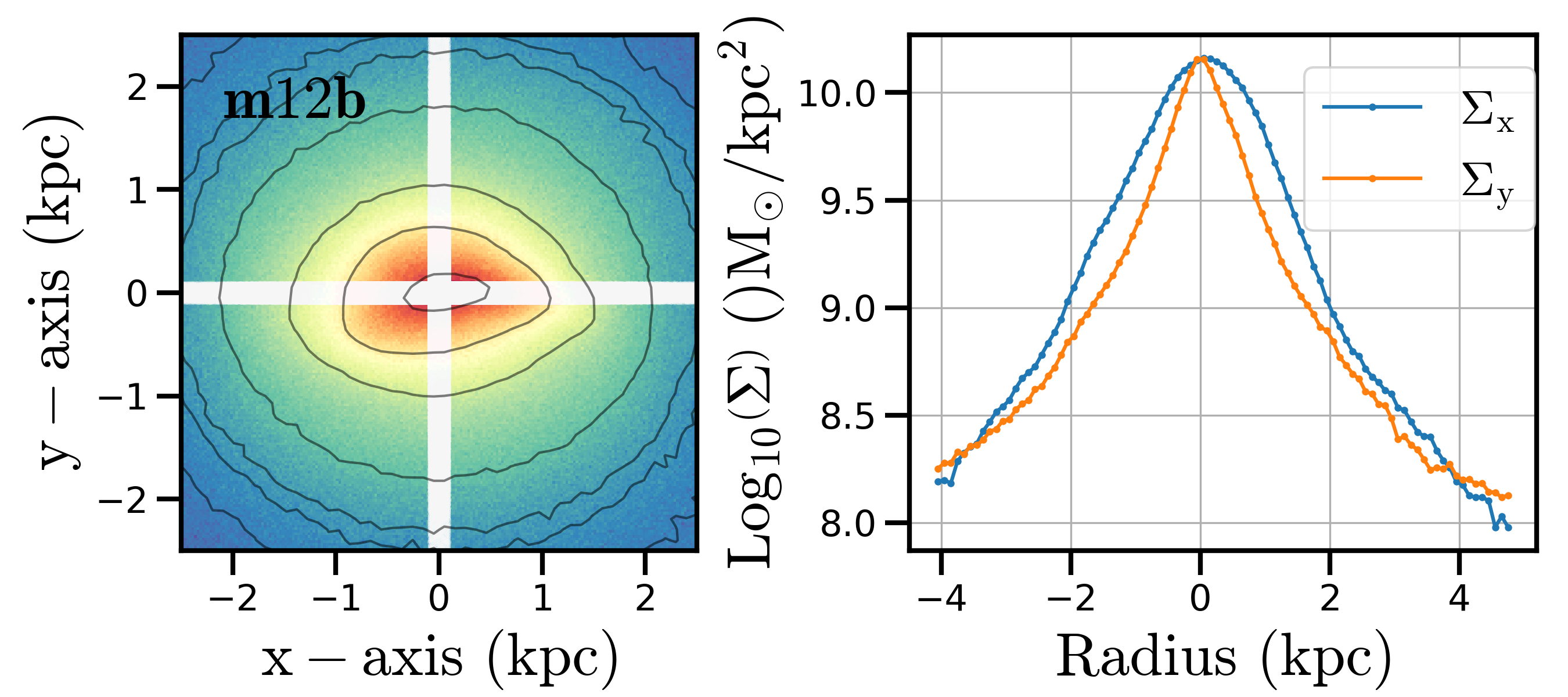}
\caption{ {\bf Stellar surface density profiles along bar major-axis and minor-axis in \textsf{m12f} (top row) and \textsf{m12b} (bottom row) at peak bar strength.} The left column shows the face-on images of central regions in \textsf{m12f} and \textsf{m12b}, with the slits (white) used to measure surface density in the right column. In the right column, the surface density $\Sigma_x$ (blue) is along the bar major axis, and $\Sigma_y$ (orange) is along the bar minor axis. }
\label{appendix:fig:stellar_surfacedensity}
\end{figure}

We present the stellar surface density profile along the bar major axis and the minor axis in \textsf{m12f} and \textsf{m12b}, at their peak bar strength, in Figure \ref{appendix:fig:stellar_surfacedensity}.

\section{Kinematic coldness of disks in FIRE-2}
\label{sec:appendix:kinematic_coldness_history}
\begin{figure*}
\centering
\includegraphics[width=0.9\textwidth]{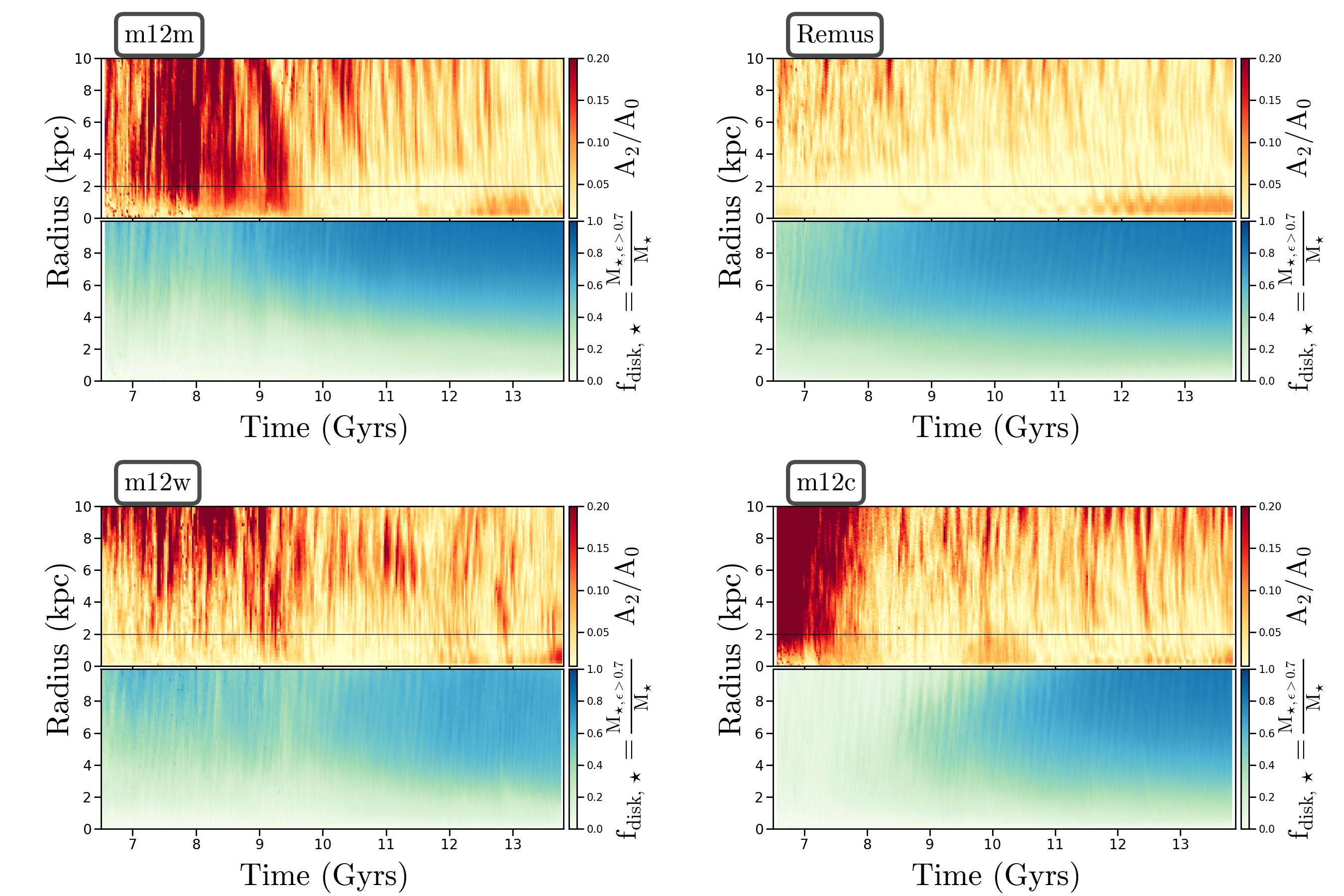}
\caption{ {\bf Weak bars and kinematic coldness of the stellar disk}. The history of bar strength ($A_{2}/A_{0}$) evolution (top panels for each simulation with color bar for bar strength: $0.0<A_{2}/A_{0}<0.2$) along with kinematic coldness parameter $f_{disk}(r)=M_{\star,\epsilon>0.7}/M_{\star}$ of the stellar disk (bottom panels for each simulation with color bar: $0.0<f_{disk}<1.0$) for all the barred galaxies in FIRE-2 (same as Figure \ref{fig:kinematic_coldness}). The bars in \textsf{m12m} and \textsf{Remus} form in kinematically hotter disks than \textsf{m12b} and \textsf{m12f}, during the internal evolution of the disks. \textsf{m12w} and \textsf{m12c} are short bars that arise in disks that are not so cold kinematically.}
    \label{appendix:fig:kinematic_coldness}
\end{figure*}

\begin{figure*}
\centering
\includegraphics[width=0.9\textwidth]{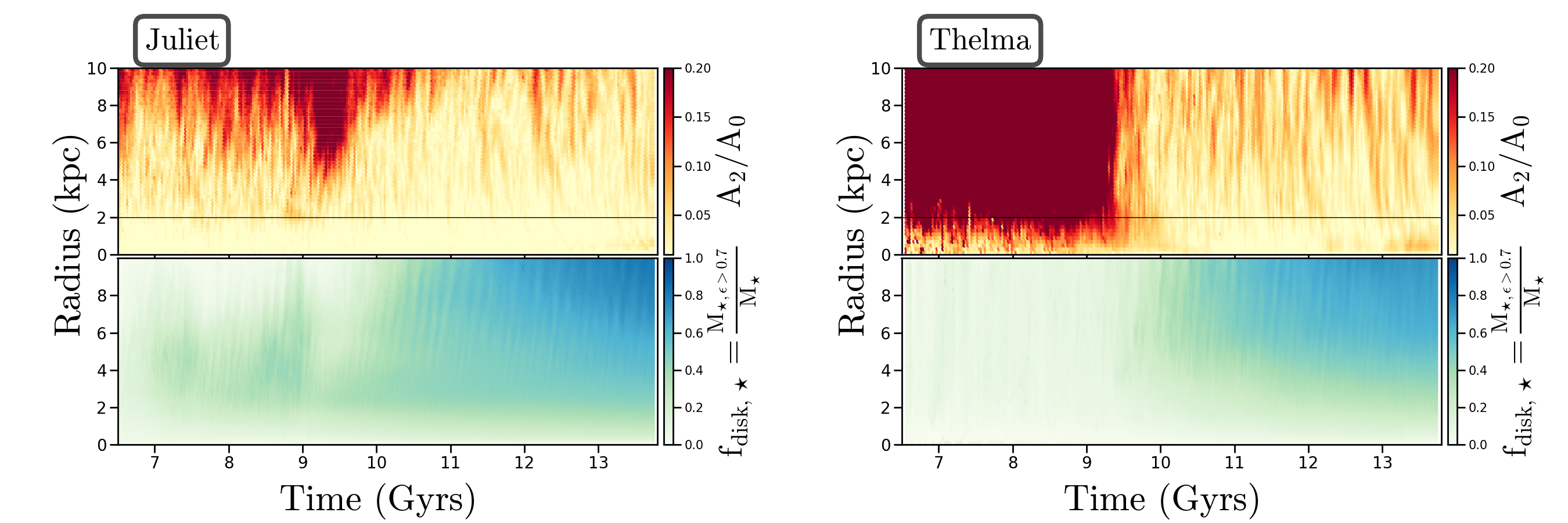}\\
\includegraphics[width=0.45\textwidth]{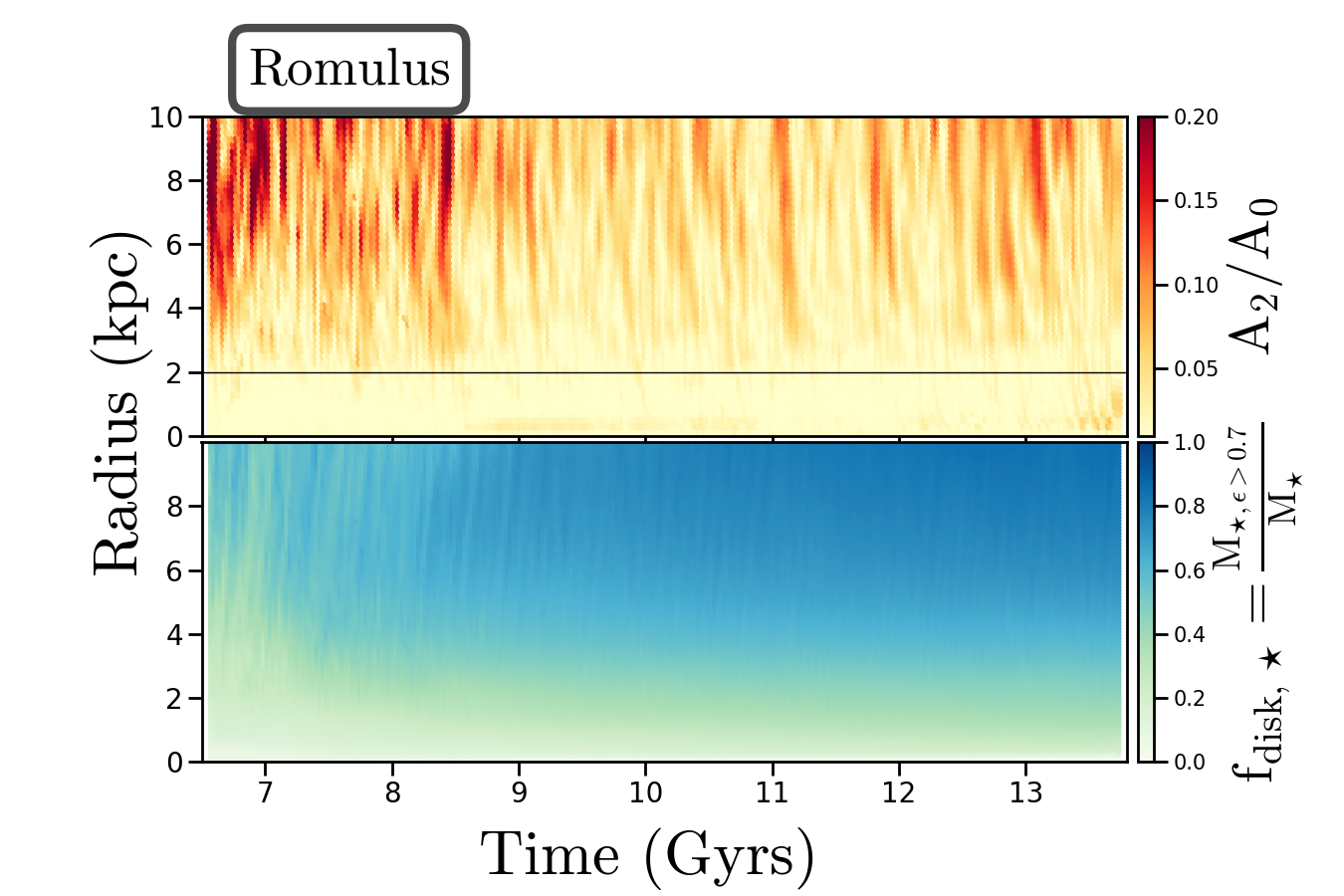}
\caption{ {\bf Unbarred galaxies and the kinematically hotness/coldness of their stellar disks}. The colors and axes are the same as Figure \ref{fig:kinematic_coldness}. \textsf{Thelma} is a failed bar with $A_{2}/A_{0}<0.1$. \textsf{Juliet} have a bulge at its center and a kinematically hot stellar disk. \textsf{Romulus} has a bulge at the center but the outer stellar disk is relatively cold kinematically.  }
    \label{appendix:fig:kinematic_coldness_unbarred}
\end{figure*}

In this Section, we present the history of the kinematic coldness of the stellar disks for the FIRE-2 barred galaxies (\textsf{m12m}, \textsf{m12w}, \textsf{m12c} and \textsf{Remus}) in Figure \ref{appendix:fig:kinematic_coldness} and the unbarred galaxies (\textsf{Juliet}, \textsf{Thelma} and \textsf{Romulus}) having central bulges in Figure \ref{appendix:fig:kinematic_coldness_unbarred}. 

We note that among the barred galaxies \textsf{m12m}, \textsf{Remus}, \textsf{m12w} and \textsf{m12c} have similar kinematically cold inner disk ($r<5$ kpc) at later stage of evolution ($>11$ Gyrs), even though the outer disk is kinematically hotter for \textsf{m12w} and \textsf{m12c} compared to the colder disk in \textsf{m12m} and \textsf{Remus}. The kinematic coldness of the inner disk is important for the bar instability to grow. However, we also note that a kinematically hotter disk does not always guarantee stability against bar formation \citep{Klypin2009, Aumer2017, Ghosh2023}, as seen for bar formation in \textsf{m12m} and \textsf{Remus} that have relatively hotter disks than \textsf{m12f}, \textsf{m12b}.

In Figure \ref{appendix:fig:kinematic_coldness_unbarred}, both the disks in \textsf{Juliet} and \textsf{Thelma} appear to be kinematically hotter compared to the disk in \textsf{Romulus}, however, none of them show a bar instability, probably due to the presence of stellar bulge.

\bibliography{sample631}{}
\bibliographystyle{aasjournal}

\end{document}